\documentclass[12pt]{article}
\usepackage{amsmath,amssymb,epsfig}
\makeatletter \@addtoreset{equation}{section} \makeatother
\renewcommand{\theequation}{\thesection.\arabic{equation}}
\newcommand{\ba}{\begin{array}}
\newcommand{\ea}{\end{array}}
\newcommand{\beq}{\begin{equation}}
\newcommand{\eeq}{\end{equation}}
\newcommand{\bea}{\begin{eqnarray}}
\newcommand{\eea}{\end{eqnarray}}

\def\bce{\begin{center}}
\def\ece{\end{center}}

\def\nonu{\nonumber}

\def\be{\beta}

\def\eps6{{\displaystyle \mathop{\epsilon}^{6}}{}}

\def\nab6{{\displaystyle \mathop{\nabla}^{6}}{}}

\def\0{{\sst{(0)}}}
\def\1{{\sst{(1)}}}
\def\2{{\sst{(2)}}}
\def\3{{\sst{(3)}}}
\def\4{{\sst{(4)}}}
\def\5{{\sst{(5)}}}
\def\6{{\sst{(6)}}}
\def\7{{\sst{(7)}}}
\def\8{{\sst{(8)}}}

\def\ba{\begin{array}}
\def\ea{\end{array}}
\def\beq{\begin{equation}}
\def\eeq{\end{equation}}
\def\be{\begin{equation}}
\def\ee{\end{equation}}

\def\eps{\epsilon}

\def\ba{\begin{array}}
\def\ea{\end{array}}
\def\beq{\begin{equation}}
\def\eeq{\end{equation}}
\def\be{\begin{equation}}
\def\ee{\end{equation}}

\def\eps{\epsilon}

\newcommand{\bean}{\begin{eqnarray*}}
\newcommand{\eean}{\end{eqnarray*}}

\begin{document}
\thispagestyle{empty} \addtocounter{page}{-1}
\begin{flushright}
\end{flushright}

\vspace*{1.3cm}

\centerline{ \Large \bf Meta-Stable Brane Configurations of   }
\vspace{.3cm} 
\centerline{ \Large \bf   Triple Product
  Gauge Groups } 
\vspace*{1.5cm}
\centerline{{\bf Changhyun Ahn} 
} 
\vspace*{1.0cm} 
\centerline{\it 
Department of Physics, Kyungpook National University, Taegu
702-701, Korea} 
\vspace*{0.8cm} 
\centerline{\tt ahn@knu.ac.kr} 
\vskip2cm

\centerline{\bf Abstract}
\vspace*{0.5cm}

From an ${\cal N}=1$ supersymmetric electric gauge theory 
with the gauge group $SU(N_c) \times SU(N_c') \times SU(N_c'')$ 
with fundamentals for each 
gauge group and the bifundamentals, we apply Seiberg dual 
to each gauge group  and obtain the ${\cal N}=1$
supersymmetric 
 dual magnetic gauge theories with dual matters including the
additional gauge singlets. 
By analyzing the F-term equations of the dual
magnetic 
superpotentials, we describe the intersecting brane configurations of
type 
IIA string theory corresponding to the meta-stable nonsupersymmetric 
vacua of this gauge theory.
We apply also to 
the case for ${\cal N}=1$ supersymmetric electric gauge theory 
with the gauge group $Sp(N_c) \times SO(2N_c') \times Sp(N_c'')$ 
with flavors for each 
gauge group and the bifundamentals. 
Finally, we describe the meta-stable brane configurations of 
multiple product gauge groups.  

\baselineskip=18pt
\newpage
\renewcommand{\theequation}
{\arabic{section}\mbox{.}\arabic{equation}}

\section{Introduction}

The nonsupersymmetric meta-stable vacua exist in ${\cal N}=1$ SQCD
with massive fundamental quarks where the  masses are 
much smaller than the dynamical
scale of the gauge sector \cite{ISS}.
The supersymmetry is 
broken by the rank condition and 
the classical flat directions can be 
lifted by quantum corrections which generate positive mass terms for the 
pseudomoduli leading to long-lived meta-stable vacua.   
When the direction 
which was not properly lifted by the one-loop potential occurs,
by adding a set of singlets, one gets meta-stable vacua
in the 
quiver gauge theory on fractional branes \cite{FU}. 
Other possible embedding by brane probes wrapping cycles of
local Calabi-Yau and corresponding quiver 
gauge theories have inequivalent meta-stable vacua \cite{OO}. 
See the review paper \cite{IS} for 
the recent developments of dynamical supersymmetry breaking.

The geometrical approach for these meta-stable vacua, using the brane 
configuration in type IIA string theory, 
has many interesting features \cite{OO1,FGU,BGHSS}. 
Turning on the quark masses in the 
electric theory corresponds to deform the superpotential by adding 
a term in linear in a singlet field in the dual magnetic theory. 
This is equivalent to move D6-branes in particular
directions in type IIA brane configuration. 
The misalignment of D4-branes 
connecting NS5'-brane can be 
analyzed as a nontrivial F-term conditions
providing nonzero vacuum expectation values of
dual quarks. The fact that some of D4-branes connecting NS5'-brane can move 
in other two directions freely is exactly the classical moduli 
space of nonsupersymmetric vacua. 
See
the review paper \cite{GK98} for the gauge theory and the brane dynamics.

When we add an adjoint matter field into 
above ${\cal N}=1$ SQCD with massive fundamental quarks,
a meta-stable 
nonsupersymmetric long-lived vacuum was found
by considering an addition of gauge singlet terms \cite{AGM}.
The corresponding brane configuration for these nonsupersymmetric 
meta-stable vacua was studied 
by realizing  these deformation terms in the magnetic dual theory
geometrically \cite{Ahn06}. 
The lifting of the tree-level supersymmetry breaking 
brane configuration to M-theory was described in \cite{BGHSS} 
by computing the equations of motion 
for the minimal area nonholomorphic 
curves.
The behavior at infinity of this nonsupersymmetric brane configuration 
was different from that of the standard 
supersymmetric ground state of MQCD. 
The M-theory lift for  symplectic and orthogonal gauge groups  
by introducing an orientifold 4-plane to the brane configuration for
unitary gauge group was presented in \cite{Ahn06-1}.

When we add a symmetric flavor and a conjugate symmetric flavor 
to an 
${\cal N}=1$ SQCD with massive fundamental quarks,
a meta-stable 
nonsupersymmetric long-lived vacuum was found
by adding an orientifold 6-plane to the above brane configuration \cite{Ahn07}.
Moreover, 
when an antisymmetric flavor, a conjugate symmetric flavor, 
and eight fundamental flavors are added to 
the 
${\cal N}=1$ SQCD with massive fundamental quarks,
the nonsupersymmetric minimal energy brane configuration was obtained
also in \cite{Ahn07-1}.
In these constructions, the precise brane motion was crucial to arrive
at the meta-stable brane configurations in order to truncate the
unwanted meson fields in the magnetic theory. 

Then it is natural to ask what happens when we increase the number of
gauge group.
When a bifundamental is added to 
the ${\cal N}=1$ product gauge theory with fundamentals for each gauge
group,
the nonsupersymmetric meta-stable brane configuration \cite{Ahn07-3} 
can be constructed.
Either the Seiberg dual for the first gauge group 
or for the second gauge group gave us 
the same meta-stable brane configuration.
Although the magnetic superpotential has many more terms, compared
with the ISS model \cite{ISS}, 
the F-term analysis enabled us to obtain the meta-stable vacua.
For the product of symplectic and orthogonal gauge groups, one needs
to add an orientifold 4-plane as well as three NS-branes, two kinds of
D4-branes
and two kinds of D6-branes \cite{Ahn07-2}.
On the other hand, when we add an orientifold 6-plane to the brane
configuration for product gauge theory with unitary groups, the matter
contents will be different in general and the extra NS-branes should
be included. In this case, the meta-stable brane configuration was
described in \cite{Ahn07-4}.     

When the D6-branes in above brane configurations 
are replaced by other NS5'-brane,
the meta-stable vacua of \cite{ISS} arises in some
region of parameter space when the D4-branes and
anti D4-branes can decay and the geometric misalignment of
flavor D4-branes occurs \cite{GK}. 
Adding an orientifold 4-plane or orientifold 6-plane 
to the brane configuration 
of \cite{GK} implies that the gauge group is a product of a symplectic
group and an orthogonal group or a product of unitary groups. Then 
the geometric misalignment of flavor 
D4-branes also occurs \cite{Ahn07-5}. Further generalization by using 
more NS-branes, orientifold 4-plane, or orientifold 6-plane 
was obtained from the recent works \cite{Ahn07-6} where 
there exists a triple product of gauge groups with bifundamentals 
and \cite{Ahn07-7} where other extra matter contents are present.   

In this paper, we continue to study for the meta-stable brane
configuration in the context of triple product gauge groups.
Compared with two product gauge groups studied in
\cite{Ahn07-3,Ahn07-2},  
the Seiberg dual for the
middle gauge group has a new feature in the sense that the ranks of
the first gauge group and the third gauge group appear in the number
of dual colors for the middle gauge group and there exist two possible
magnetic brane configurations where the magnetic superpotentials have
different form and also the different matter contents are present. 
On the other hand, the Seiberg duals for the first gauge group and the third
gauge group look similar to the ones in two product gauge groups
because when we take the Seiberg dual for the first(third) gauge group, the
fields for the third(first) gauge group in the magnetic theory are 
the same as those for the third(first) gauge group in the electric theory.  

One can easily generalize the meta-stable brane configuration 
to the ones corresponding to a multiple product of gauge groups.
Then one takes the Seiberg dual for the first gauge group factor, the
last gauge group factor and for any gauge group factor except the
first and last gauge group factors. One can write 
down the magnetic superpotentials in terms of the cubic interactions
between the gauge singlets and dual matters. 

We also add an orientifold 4-plane to the brane configurations
consisting of four NS-branes, three kinds of D4-branes and three kinds
of D6-branes in type IIA string theory. Then the gauge group will be a
product of symplectic and orthogonal gauge groups alternatively.
The meta-stable brane configuration for the 
multiple product gauge groups is also discussed.
 
In section 2,
we describe the type IIA brane configuration corresponding
to the electric theory based on the ${\cal N}=1$ $SU(N_c) \times
SU(N_c') \times SU(N_c'')$ 
gauge theory 
with fundamentals and bifundamentals, 
and deform this theory by adding the mass term
for the quarks for each gauge group. 
Then we construct the Seiberg dual magnetic theories for each gauge
group factor with corresponding dual
matters as well as additional gauge singlets, by brane motion and linking
number counting. 
Finally, the
nonsupersymmetric brane configurations are  found by recombination and
splitting for the flavor D4-branes.
One generalizes to the meta-stable brane configurations corresponding
to a multiple product of gauge groups and describe them
very briefly.

In section 3,
we describe the type IIA brane configuration corresponding
to the electric theory based on the ${\cal N}=1$ $Sp(N_c) \times
SO(2N_c') \times Sp(N_c'')$ 
gauge theory 
with fundamentals, vectors, and  bifundamentals, 
and deform this theory by adding the mass term
for the quarks for each gauge group.
In the brane configuration, this can be obtained from the brane
configuration given in section 2 by inserting the appropriate 
orientifold 4-planes.  
Then we construct the Seiberg dual magnetic theories for each gauge
group factor with corresponding dual
matters as well as extra gauge singlets, by brane motion and linking
number counting. 
Finally, the
nonsupersymmetric brane configurations are  found.
These can be seen from those brane configurations in section 2
by the action of orientifold 4-plane.
The generalization to a multiple product of gauge groups is described.
Compared with the ones in section 2, in this case, there exist more 
magnetic dual theories because the gauge group factors 
have both symplectic and orthogonal gauge groups.  

Finally, in section 4, 
we summarize what we have found in this paper and 
make some comments for the future directions
\footnote{Some different directions on the meta-stable vacua
are present in
recent relevant works \cite{ABF}-\cite{ABFK} where 
some of them are described in the type IIB string theory.
These are not complete list. It would be very interesting to find out
how the meta-stable brane configurations from 
type IIA string theory including the present work are related to
those brane configurations from type IIB string theory.}.

\section{Nonsupersymmetric meta-stable brane configurations
of $SU(N_c) \times SU(N_c') \times SU(N_c'')$ and 
its multiple product gauge theories}

For the meta-stable brane configurations, it is necessary
to have nonzero masses for the
quarks(corresponding to relative displacement between D6-branes and
D4-branes in type IIA string theory)
and to take the Seiberg dual theory(corresponding to the magnetic
theory obtained
from the electric theory via brane motion or field theory analysis) \cite{ISS}. 
We need to understand the brane configurations both from the
electric theory(where one has mass-deformed superpotential)
and from the magnetic theory(where the meta-stable states are long-lived 
parametrically). 
There exist four possible magnetic brane configurations for the triple
product gauge group $SU(N_c) \times SU(N_c') \times SU(N_c'')$, depending on
whether the dual gauge group we take is the first gauge group, the
second gauge group(in which there exist two magnetic brane configurations),  
or the third gauge group. 

After reviewing the electric brane configuration, 
we describe the four magnetic brane configurations, and then the
nonsupersymmetric brane configurations are  found by recombination
of some 
flavor D4-branes and color D4-branes and splitting 
procedure between those flavor D4-branes and the remnant of flavor D4-branes.

\subsection{Electric theory}

The gauge group is given by $SU(N_c) \times SU(N_c') 
\times SU(N_c'')$ and the matter 
contents \cite{BH,AT97} are given by 

$\bullet$
$N_f$ chiral multiplets $Q$ are 
in the fundamental representation under the $SU(N_c)$,
$N_f$ chiral multiplets $\widetilde{Q}$ are in the antifundamental
representation under the $SU(N_c)$ and then $Q$ are in the
representation $({\bf N_c,1, 1
})$ while $\widetilde{Q}$ are in the representation 
$({\bf \overline{N_c}, 1, 1})$
under the whole gauge group

$\bullet$
$N_f'$ chiral multiplets $Q'$ are 
in the fundamental representation under the $SU(N_c')$,
$N_f'$ chiral multiplets $\widetilde{Q}'$ are in the antifundamental
representation under the $SU(N_c')$ and then $Q'$ are in the
representation $({\bf 1, N_c', 1
})$ while $\widetilde{Q}'$ are in the representation $({\bf 1, 
\overline{N_c'}, 1})$
under the whole gauge group

$\bullet$
$N_f''$ chiral multiplets $Q''$ are 
in the fundamental representation under the $SU(N_c'')$,
$N_f''$ chiral multiplets $\widetilde{Q}''$ are in the antifundamental
representation under the $SU(N_c'')$ and then $Q''$ are in the
representation $({\bf 1, 1, N_c''
})$ while $\widetilde{Q}''$ are in the 
representation $({\bf 1, 1, \overline{N_c''}})$
under the whole gauge group

$\bullet$
The flavor-singlet field $F$ is 
in the bifundamental representation $({\bf N_c, \overline{N_c'}, 1 })$ 
under the whole gauge group and its complex conjugate field $\widetilde{F}$
 is 
in the bifundamental representation $({\bf \overline{N_c}, N_c', 1} )$ 
under the whole gauge group

$\bullet$
The flavor-singlet field $G$ is 
in the bifundamental representation $({\bf 1, N_c', \overline{N_c''} })$ 
under the whole gauge group and its complex conjugate field $\widetilde{G}$
 is 
in the bifundamental representation $({\bf 1, \overline{N_c'}, N_c''} )$ 
under the whole gauge group

This is a simple generalization to triple product gauge groups from
the two product gauge groups \cite{ILS,BIWW,BH,AT97}.
Then the gauge group and matter contents we consider 
are summarized as follows:
\bea
 & \mbox{gauge group}:& \;\;\;\;\;   SU(N_c) \times SU(N_c') \times
 SU(N_c'')  \nonu
\\
\mbox{matter}: 
 &Q_f \oplus \widetilde{Q}_{\widetilde{f}}& \;\;\;\;\;\;\;\;\; 
\;\;\; {(\bf \Box, 1, 1) \oplus (\overline{\Box}, 1, 1)}
\;\;\;\;\; (f, \widetilde{f}=1,  \cdots, N_f) 
\nonu \\
 &Q'_{f'} \oplus \widetilde{Q}'_{\widetilde{f}'}& \;\;\;\;\;\;\;\;\;\;
\;\; {(\bf 1, \Box, 1) \oplus ( 1, \overline{\Box}, 1)}
\;\;\;\;\; (f', \widetilde{f}' =1,  \cdots, N_f') 
\nonu \\
 &Q''_{f''} \oplus \widetilde{Q}''_{\widetilde{f}''}& \;\;\;\;\;\;\;\;\;\; 
\;\; {(\bf 1, 1, \Box) \oplus ( 1, 1, \overline{\Box})} 
\;\;\;\;\; (f'', \widetilde{f}'' =1,  \cdots, N_f'')
\nonu \\
 &F \oplus \widetilde{F}& \;\;\;\;\;\;\;\;\; 
\;\; {(\bf \Box, \overline{\Box},1) \oplus (\overline{\Box}, \Box, 1)} 
\nonu \\
 &G \oplus \widetilde{G}& \;\;\;\;\;\;\;\;\; 
\;\; {(\bf 1, \Box, \overline{\Box}) \oplus ( 1, \overline{\Box}, \Box)} 
\nonu 
\eea
In the electric theory, since there exist $N_f$ quarks $Q$, $N_f$
quarks $\widetilde{Q}$, one bifundamental field $F$(which will give
rise to the contribution of $N_c'$), and 
its complex conjugate field $\widetilde{F}$(which will give
rise to the contribution of $N_c'$), the coefficient of the beta function
of the first gauge group factor is
$
b_{SU(N_c)} = 3N_c -N_f-N_c'$.
Similarly,
since there exist $N_f'$ quarks $Q'$, $N_f'$
quarks $\widetilde{Q}'$, one bifundamental field $F$(which gives
the contribution of $N_c$), 
its complex conjugate field $\widetilde{F}$(which will give
the contribution of $N_c$),
one bifundamental field $G$(which will give
rise to the contribution of $N_c''$), and 
its complex conjugate field $\widetilde{G}$(which will give
the contribution of $N_c''$),
the coefficient of the beta function
for the second gauge group factor is
given by 
$
b_{SU(N_c')} = 3N_c'-N_f'-N_c-N_c''$.
Finally, since there exist $N_f''$ quarks $Q''$, $N_f''$
quarks $\widetilde{Q}''$, one bifundamental field $G$(which will give
rise to the contribution of $N_c'$), and 
its complex conjugate field $\widetilde{G}$(which will give
rise to the contribution of $N_c'$), the coefficient of the beta function
of the third gauge group factor can be read off
$
b_{SU(N_c'')} = 3N_c'' -N_f''-N_c'$.
There is a symmetry between $b_{SU(N_c)}$ and $b_{SU(N_c'')}$ such
that the former becomes the latter by replacing $N_c$ and $N_f$ with 
$N_c''$ and $N_f''$ respectively.

The anomaly free global symmetry contains $[SU(N_f) \times 
SU(N_f') \times SU(N_f'')]^2 \times
 U(1)_R$ \cite{BH} and let us denote the
strong coupling scales for $SU(N_c)$ as $\Lambda_1$, for $SU(N_c')$
as $\Lambda_2$ and for $SU(N_c'')$
as $\Lambda_3$ respectively.  
The electric theory is asymptotically free when $b_{SU(N_c)} >
0$ for the $SU(N_c)$ gauge theory,  when 
$b_{SU(N_c')} > 0$ for the $SU(N_c')$ gauge theory, and 
when 
$b_{SU(N_c'')} > 0$ for the $SU(N_c'')$ gauge theory. 

The classical electric superpotential is  
\bea
W_{elec} & = & \left( \mu A^2 + \lambda Q A \widetilde{Q} + \widetilde{F} A F +
\mu' A'^2 + \lambda' Q' A' \widetilde{Q}' + \widetilde{F} A' F +
\widetilde{G} A' G \right. \nonu \\
&+& \left.
 \mu'' A''^2 + \lambda'' Q'' A'' \widetilde{Q}'' + \widetilde{G} A'' G \right)
+ m Q \widetilde{Q} + m' Q' \widetilde{Q}' + m'' Q'' \widetilde{Q}''
\label{superpotential-1}
\eea
where the coefficient functions $\mu, \mu', \mu'', \lambda, \lambda'$
and $\lambda''$ are given by six rotation angles for the branes 
in type IIA string theory \cite{BH,BHKL}. 
We do not write down the dependences on these
angles explicitly. Each gauge group factor has two rotation
angles on NS-brane and D6-branes.
Here the adjoint field for $SU(N_c)$ gauge group is denoted by $A$, 
the adjoint field for $SU(N_c')$ gauge group is denoted by $A'$, and 
the adjoint field for $SU(N_c'')$ gauge group is denoted by $A''$.
The mass terms of these adjoint fields are related to the 
rotation angles of NS-branes. 
The couplings of fundamentals with these adjoint
fields are related also to the 
rotation angles of NS-branes as well as the rotation angles
of D6-branes.
We add the mass terms for each fundamental flavor.  
Setting the fields $Q'', \widetilde{Q}'', G, \widetilde{G}$ and $A''$ to zero, 
the superpotential becomes the one described in \cite{BH,BHKL,Ahn07-3}.

After integrating out the adjoint fields $A, A'$ and $A''$, 
this superpotential (\ref{superpotential-1}) 
at the particular orientations for branes, i.e., the case where 
any two neighboring NS-branes are perpendicular to each other, 
will reduce to the last three mass-deformed terms 
since the coefficient functions 
$\frac{1}{\mu}, \frac{1}{\mu'}$
and $\frac{1}{\mu''}$ vanish at these particular rotation angles for branes.
It does not matter whether $\lambda, \lambda'$ 
or $\lambda''$ vanishes because
even if these coefficient functions are not zero, 
$\lambda, \lambda'$- or $\lambda''$-dependent terms all vanish
because they contain 
$\frac{1}{\mu}$-, $\frac{1}{\mu'}$- or $\frac{1}{\mu''}$-prefactors.
Then the classical superpotential by deforming the massless case by adding the
mass terms for the quarks $Q(Q')[Q'']$ and
$\widetilde{Q}(\widetilde{Q}')[\widetilde{Q}'']$ 
is given by
\bea
W_{elec} = m Q \widetilde{Q} + m' Q' \widetilde{Q}' + m'' Q'' \widetilde{Q}''.
\label{superpotential}
\eea 
When we discuss the meta-stable brane configurations for the
particular dual gauge group with corresponding massive flavors, 
the other flavors corresponding to other two gauge groups will be massless.

The type IIA brane configuration for this mass-deformed theory 
can be described by as follows \footnote{Let
us introduce two complex coordinates $v \equiv x^4 + i x^5$ and $w \equiv
x^8 + i x^9$ for convenience. }.
The $N_c$-color 
D4-branes (01236) are suspended between the $NS5_L$-brane (012345) 
and the $NS5_L'$-brane (012389) along $x^6$
direction,
together with $N_f$ D6-branes (0123789) 
which are parallel to $NS5_L'$-brane and have nonzero $v$ direction.
The $NS5_R$-brane 
is located at the right hand side of
the $NS5_L'$-brane along the $x^6$ direction and 
there exist $N_c'$-color D4-branes
suspended 
between them, with  $N_f'$ D6-branes which have nonzero $v$ direction. 
Moreover, 
the $NS5_R'$-brane 
is located at the right hand side of
the $NS5_R$-brane along the $x^6$ direction and there 
exist $N_c''$-color D4-branes
suspended 
between them, with  $N_f''$ D6-branes which have nonzero $v$ direction
\footnote{Basically, if there are no $NS5_L$-brane, $N_c$ D4-branes and
$N_f$ D6-branes, this brane configuration is exactly the same as the
one in \cite{Ahn07-3}. One can consider the other brane configuration by
rotations of NS-branes in above brane configuration 
by $\frac{\pi}{2}$ angles. That is, $NS5_{L,R}$-branes go to
$NS5_{L,R}'$-branes and vice versa. In other words, this other brane
configuration can be obtained from the one in \cite{Ahn07-3}, by
adding  $NS5_R$-brane, $N_c''$ D4-branes and
$N_f''$ D6-branes to the right. 
Then we will see that 
all the meta-stable brane
configurations from this other brane configuration 
can be obtained by viewing the Figures 2B, 3B, 4B and 5B from the negative 
$w$ direction.
\label{rotation}  }.

We summarize the ${\cal N}=1$ supersymmetric electric brane
configuration in type IIA string theory as follows:

$\bullet$
$NS5_L(NS5_R)$-brane in (012345) directions. 

$\bullet$ 
$NS5_L'(NS5_R')$-brane in (012389) directions.

$\bullet$
$N_c(N_c')[N_c'']$-color D4-branes in (01236) directions. 
  
$\bullet$
$N_f(N_f')[N_f'']$ D6-branes in (0123789) directions. 

Now we draw this electric brane configuration in Figure 1 and we put
the coincident $N_f(N_f')[N_f'']$ D6-branes in 
the nonzero $v$ direction in general.
The quarks $Q(Q')[Q'']$ and
$\widetilde{Q}(\widetilde{Q}')[\widetilde{Q}'']$ 
correspond to strings
stretching between the
$N_c(N_c')[N_c'']$-color D4-branes with $N_f(N_f')[N_f'']$ D6-branes.
The bifundamentals $F(G)$ and $\widetilde{F}(\widetilde{G})$ 
correspond to   strings 
stretching between the
$N_c(N_c')$-color D4-branes with $N_c'(N_c'')$-color D4-branes. 

\begin{figure}[ht]
   \epsfxsize=3.0in 
\centerline{\epsffile{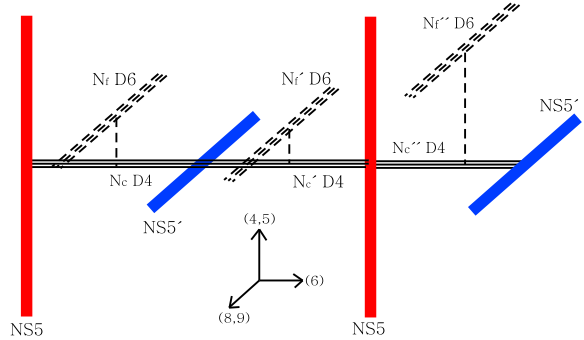}}
   \caption[FIG. \arabic{figure}.]{ 
The ${\cal N}=1$ supersymmetric electric brane configuration with
$SU(N_c) \times SU(N_c') \times SU(N_c'')$ gauge group with
fundamentals $Q(Q')[Q'']$ and
$\widetilde{Q}(\widetilde{Q}')[\widetilde{Q}'']$ 
for each gauge group and bifundamentals $F(G)$ and 
$\widetilde{F}(\widetilde{G})$.
The two NS5-branes can be written as $NS5_{L,R}$-branes while the two 
NS5'-branes can be denoted by $NS5_{L,R}'$-branes.}
\end{figure}


\subsection{Magnetic theory with dual for third gauge group}

One considers dualizing one of the gauge groups regarding as the
other gauge groups as a spectator.  
In this subsection,
one takes the Seiberg dual for the third gauge group factor $SU(N_c'')$
while remaining the first and second gauge group factors $SU(N_c)$ and  
$SU(N_c')$ unchanged. 
One ignores the dynamics of the other gauge groups. 
We consider the case where $\Lambda_3 >> \Lambda_1, \Lambda_2$, in other
words, the dualized group's dynamical scale is far above that of the
other spectator groups.

Let us move the $NS5_R$-brane to the right all the way past the  
$NS5'_R$-brane.  
After this brane motion, one arrives at the Figure 2A.
Note that there exists a creation of $N_f''$ D4-branes
connecting $N_f''$ D6-branes and $NS5'_R$-brane 
because the $N_f''$ D6-branes are perpendicular(or are not parallel) to the 
$NS5_R$-brane in Figure 1
and we consider
massless quarks for $Q(Q')$ and $\widetilde{Q}(\widetilde{Q}')$.
The linking number \cite{HW} of $NS5_R$-brane from Figure 2A
can be read off and is given by 
$
L_5 = \frac{N_f''}{2} -\widetilde{N}_c''$.
On the other hand, the linking number of $NS5_R$-brane from Figure 1
is
$
L_5 = -\frac{N_f''}{2} + N_c'' -N_c'$. 
Due to the connection of $N_c'$
D4-branes with $NS5_R$-brane in Figure 1, 
the presence of $N_c'$ in the linking
number arises. This $N_c'$ dependence will appear in many places later
and is common feature when we discuss about the product gauge group.
From these two relations, one obtains
the number of colors of dual magnetic theory
\bea
\widetilde{N}_c'' = N_f'' +N_c'-N_c''.
\label{number}
\eea

Let us draw this magnetic brane configuration in Figure 2A and recall
that we put
the coincident $N_f''$ D6-branes in the nonzero $v$-direction in the
electric theory and consider massless flavors for $Q$ and $Q'$ by
putting $N_f$ and $N_f'$ D6-branes at $v=0$.
The $N_f''$ created D4-branes connecting between
D6-branes and $NS5_R'$-brane can move freely in the $w$-direction.
Moreover, since $N_c'$ D4-branes are suspending between two equal
$NS5'_{L,R}$-branes located at different $x^6$ coordinate, these D4-branes
can slide along the $w$-direction also.
If we ignore the $NS5_L$-brane, $N_c$ D4-branes, $N_f$ 
D6-branes, the $NS5_L'$-brane, $N_c'$ D4-branes and $N_f'$ 
D6-branes(detaching these
branes from Figure 2A), 
then this brane configuration 
leads to the standard ${\cal N}=1$ SQCD with the magnetic gauge group 
$SU(\widetilde{N}_c''=N_f''-N_c'')$ with
$N_f''$ massive flavors \cite{OO,FGU,BGHSS}.

\begin{figure}[ht]
   \epsfxsize=4.0in 
\centerline{\epsffile{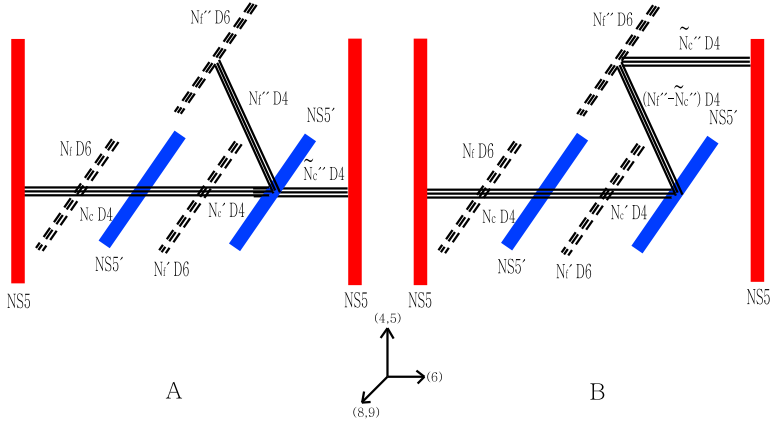}}
   \caption[FIG. \arabic{figure}.]{ 
The ${\cal N}=1$ supersymmetric magnetic brane configuration with
$SU(N_c) \times SU(N_c') \times SU(\widetilde{N}_c'')$ gauge group
with fundamentals $Q(Q')[q'']$ and 
$\widetilde{Q}(\widetilde{Q}')[\widetilde{q}'']$ 
for each gauge group, bifundamentals $F(g)$ and
$\widetilde{F}(\widetilde{g})$, and gauge singlets in Figure 2A. In
Figure 2B, the nonsupersymmetric minimal energy brane configuration
with the same gauge group and matter contents above 
for massless  $Q(Q')$ and
$\widetilde{Q}(\widetilde{Q}')$ is given. 
}
\end{figure}


The IR dynamics of the electric theory is described by 
Seiberg dual and one uses the variables of dual theory instead of the
original variables for the electric theory.  
Then the dual magnetic gauge group
is given by $SU(N_c) \times SU(N_c') 
\times SU(\widetilde{N}_c'')$ and the matter
contents are given by

$\bullet$
$N_f$ chiral multiplets $Q$ are 
in the fundamental representation under the $SU(N_c)$,
$N_f$ chiral multiplets $\widetilde{Q}$ are in the antifundamental
representation under the $SU(N_c)$ and then $Q$ are in the
representation $({\bf N_c, 1, 1
})$ while $\widetilde{Q}$ are in the representation $({\bf
  \overline{N_c}, 1, 1})$
under the whole gauge group

$\bullet$
$N_f'$ chiral multiplets $Q'$ are 
in the fundamental representation under the $SU(N_c')$,
$N_f'$ chiral multiplets $\widetilde{Q}'$ are in the antifundamental
representation under the $SU(N_c')$ and then $Q'$ are in the
representation $({\bf 1, N_c', 1
})$ while $\widetilde{Q}$ are in the representation $({\bf 1,
  \overline{N_c'}, 1})$
under the whole gauge group

$\bullet$ 
$N_f''$ chiral multiplets $q''$ are 
in the fundamental representation under the $SU(\widetilde{N}_c'')$,
$N_f''$ chiral multiplets $\widetilde{q}''$ are in the antifundamental
representation under the $SU(\widetilde{N}_c'')$ and then $q''$ are in the
representation $({\bf 1, 1, \widetilde{N}_c''})$ while 
$\widetilde{q}''$ are in the representation 
$({\bf 1, 1, \overline{\widetilde{N}_c''}})$
under the whole gauge group

$\bullet$
The flavor-singlet field $F$ is 
in the bifundamental representation $({\bf N_c, \overline{N_c'}, 1 })$ 
under the whole gauge group and its complex conjugate field $\widetilde{F}$
 is 
in the bifundamental representation $({\bf \overline{N_c}, N_c', 1})$ 
under the whole gauge group

$\bullet$
The flavor-singlet field $g$ is 
in the bifundamental representation $({\bf 1, N_c', 
\overline{\widetilde{N}_c''} })$ 
under the whole gauge group and its complex conjugate field $\widetilde{g}$
 is 
in the bifundamental representation $({\bf 1, \overline{N_c'}, 
\widetilde{N}_c''})$ 
under the whole gauge group

There are $(N_f''+N_c')^2$ gauge singlets in the third gauge group
factor
as follows:

$\bullet$
$N_f''$-fields $X'$ are in the fundamental representation under the
$SU(N_c')$, its complex conjugate
$N_f''$-fields $\widetilde{X}'$ are in the antifundamental
representation under the $SU(N_c')$ and then 
$X'$ are in the representation $({\bf 1, N_c', 1 })$ 
under the whole gauge group
while 
$\widetilde{X}'$ are in the representation $({\bf 1, \overline{N_c'}, 1 })$ 
under the whole gauge group

These additional $N_f''$-$SU(N_c')$ fundamentals and $N_f''$-$SU(N_c')$
antifundamentals 
are originating from 
the $SU(N_c'')$ chiral mesons $G Q''$ and $\widetilde{G}
\widetilde{Q}''$ 
respectively where the color indices are contracted each other. Therefore, 
there are free indices for a single color and a single flavor. 
Then the strings stretching between the $N_f''$ $D6$-branes and $N_c'$
D4-branes will give rise to these additional $N_f''$-$SU(N_c')$
fundamentals and  $N_f''$-$SU(N_c')$
antifundamentals.

$\bullet$
$N_f''^2$-fields $M''$ are in the representation $({\bf 1, 1, 1})$ under the
whole gauge group

This corresponds to the $SU(N_c'')$ chiral meson $Q'' \widetilde{Q}''$
where the color indices are contracted.
It is clear to see that since the
$N_f''$ D6-branes are parallel to the $NS5'_R$-brane from Figure 2A, 
the newly created $N_f''$-flavor D4-branes can slide along the plane
consisting of these $N_f''$ $D6$-branes and  $NS5'_R$-brane
freely.
The fluctuations of the gauge-singlet $M''$ correspond to the motion of $N_f''$
flavor D4-branes along (789) directions in Figure 2A.
As we will see later, for the nonsupersymmetric brane configuration,  
a misalignment for the $N_f''$-flavor D4-branes arises and some of the
vacuum expectation value of $M''$ is fixed and the remaining
components are arbitrary.

$\bullet$
The $N_c'^2$-fields 
$\Phi'$ is in the representation $({\bf 1, N_c'^2-1, 1}) \oplus ({\bf
  1, 1, 1
})$ 
under the whole gauge group  

This corresponds to the $SU(N_c'')$ chiral meson $G \widetilde{G}$
where the color indices for the third gauge group 
are contracted each other and
note that  
$G$  has a representation $({\bf 1, N_c', \overline{N_c''}})$ of an
electric theory while
$\widetilde{G}$
has a representation $({\bf 1, \overline{N_c'}, N_c''})$ of an
electric theory.
The fluctuations of the singlet $\Phi'$ correspond to the motion of
$N_c'$ D4-branes suspended two $NS5_{L,R}'$-branes along the (789)
directions in Figure 2A.  
The fact that these D4-branes
can slide along the $w$-direction implies that the vacuum expectation
value of $\Phi$ is arbitrary.

In the dual theory, 
since there exist $N_f''$ quarks $q''$, $N_f''$
quarks $\widetilde{q}''$, one bifundamental field $g$(which will give
rise to the contribution of $N_c'$), and 
its complex conjugate field $\widetilde{g}$(which will give
rise to the contribution of $N_c'$),
the coefficient of the beta function for the third 
gauge group factor 
is
$
b_{SU(\widetilde{N}_c'')}^{mag}
= 3\widetilde{N}_c''-N_f''-N_c'
=2N_f'' +2N_c'-3N_c''
$
where we inserted the number of colors given in (\ref{number}) in the
second equality. 
Since there exist $N_f'$ quarks $Q'$, $N_f'$
quarks $\widetilde{Q}'$, one bifundamental field $g$(which will give
rise to the contribution of $\widetilde{N}_c''$), 
its complex conjugate field $\widetilde{g}$(which will give
rise to the contribution of $\widetilde{N}_c''$),
$N_f''$-fields $X'$, its complex conjugate
$N_f''$-fields $\widetilde{X}'$, the singlet $\Phi'$(which will give
rise to $N_c'$), 
one bifundamental field $F$(which will give
rise to the contribution of $N_c$), and 
its complex conjugate field $\widetilde{F}$(which will give
rise to the contribution of $N_c$),
the coefficient of the beta function of 
second gauge group factor 
is
$
b_{SU(N_c')}^{mag} 
= 3N_c'-N_f'-\widetilde{N}_c''-N_f''-N_c'-N_c 
=N_c'+N_c''-N_c-2N_f''-N_f'$.
Similarly,
since there exist $N_f$ quarks $Q$, $N_f$
quarks $\widetilde{Q}$, one bifundamental field $F$(which will give
rise to the contribution of $N_c'$), and 
its complex conjugate field $\widetilde{F}$(which will give
rise to the contribution of $N_c'$),
the coefficient of the beta function of 
the first gauge group factor 
is
$
b_{SU(N_c)}^{mag} 
= 3N_c-N_f-N_c'=b_{SU(N_c)}$.
Note that the $SU(N_c)$ fields in the magnetic theory 
are the same as those of the electric theory.

Therefore, $SU(N_c)$, $SU(N_c')$ and 
$SU(\widetilde{N}_c'')$ gauge couplings are IR free
by requiring the negativeness of the coefficients of beta function.
One relies on the perturbative calculations at low energy 
for this magnetic IR free region $b_{SU(N_c)}^{mag} <
0$, $b_{SU(N_c')}^{mag} < 0$ and 
$b_{SU(\widetilde{N}_c'')}^{mag} < 0$.
Note that the $SU(N_c')$ fields in the magnetic theory 
are different from those of the electric theory.
Since $b_{SU(N_c')}-b_{SU(N_c')}^{mag} > 0$, $SU(N_c')$ is more
asymptotically free than $SU(N_c')^{mag}$.
Neglecting the $SU(N_c)$ and $SU(N_c')$ dynamics, 
the magnetic $SU(\widetilde{N}_c'')$
is IR free when 
\bea
N_c''-N_c' < N_f''< \frac{3}{2} N_c''-N_c'.
\nonu
\eea

Instead of $SU(N_c) \times SU(N_c') \times SU(N_c'')$ gauge theory,
by performing the dualizing the third gauge group,
we have an $SU(N_c) \times SU(N_c') \times SU(\widetilde{N}_c'')$
gauge theory with additional fields $X', \widetilde{X}', M''$ and $\Phi'$
and the gauge group and matter contents  
are summarized as follows:
\bea
 & \mbox{gauge group}:& \;\;\;\;\;   SU(N_c) \times SU(N_c') \times
 SU(\widetilde{N}_c'')  \nonu
\\
\mbox{matter}:  
&Q_f \oplus \widetilde{Q}_{\widetilde{f}}& \;\;\;\;\;\;\;\;\; 
\;\;\; {(\bf \Box, 1, 1) \oplus (\overline{\Box}, 1, 1)}
\;\; (f, \widetilde{f}=1,  \cdots, N_f) 
\nonu \\
 &Q'_{f'} \oplus \widetilde{Q}'_{\widetilde{f}'}& \;\;\;\;\;\;\;\;\;\;
\;\; {(\bf 1, \Box, 1) \oplus ( 1, \overline{\Box}, 1)}
\;\; (f', \widetilde{f}' =1,  \cdots, N_f') 
\nonu \\
 &q''_{f''} \oplus \widetilde{q}''_{\widetilde{f}''}& \;\;\;\;\;\;\;\;\;\; 
\;\; {(\bf 1, 1, \Box) \oplus ( 1, 1, \overline{\Box})} 
\;\; (f'', \widetilde{f}'' =1,  \cdots, N_f'')
\nonu \\
&F \oplus \widetilde{F}& \;\;\;\;\;\;\;\;\; 
\;\; {(\bf \Box, \overline{\Box},1) \oplus (\overline{\Box}, \Box, 1)} 
\nonu \\
 &g \oplus \widetilde{g}& \;\;\;\;\;\;\;\;\; 
\;\; {(\bf 1, \Box, \overline{\Box}) \oplus ( 1, \overline{\Box}, \Box)} 
\nonu \\
& (X_{n''}' \equiv) G Q'' \oplus \widetilde{G} \widetilde{Q}'' (\equiv 
\widetilde{X}_{\widetilde{n}''}') & 
 \;\;\;\;\;\;\;\;\;\;
\;\; {(\bf 1, \Box, 1) \oplus ( 1, \overline{\Box}, 1)}
\;\; (n'', \widetilde{n}'' =1,  \cdots, N_f'') 
\nonu \\
&  (M''_{f'',\widetilde{g}''} \equiv) Q'' \widetilde{Q}'' & 
 \;\;\;\;\;\;\;\;\;\;\;\;\;\;\;\;\;\;\;
\;\; {(\bf 1, 1, 1)} \;\;\;\;\;\;\;\;\;
\;\; (f'', \widetilde{g}'' =1,  \cdots, N_f'') 
\nonu \\
& (\Phi' \equiv) G \widetilde{G} & 
 \;\;\;\;\;\;\;\;\;\;
\;\; {(\bf 1, adj, 1) \oplus ( 1, 1, 1)}
\nonu
\eea

The dual magnetic superpotential, by taking the mass term 
(\ref{superpotential}) for $Q''$ and
$\widetilde{Q}''$ and massless flavors for $m=m'=0$
in the electric theory, which is equal to put a linear term in $M''$ in
the dual magnetic theory, in addition to the cubic interaction terms, 
takes the form  
\bea
W_{dual} = \left(M'' q'' \widetilde{q}'' + g X' q'' + 
\widetilde{g} \widetilde{q}''
\widetilde{X}' + \Phi' g \widetilde{g} \right) + m'' M''.
\nonu
\eea
Here $q''$ and $\widetilde{q}''$ are fundamental and antifundamental for
the third 
gauge group index respectively.
Then, $q'' \widetilde{q}''$ has rank $\widetilde{N}_c''$ while $m''$ has a
rank $N_f''$.  Therefore, the F-term condition, the derivative the 
superpotential $W_{dual}$ with respect to $M''$, cannot be satisfied 
if the rank $N_f''$ exceeds $\widetilde{N}_c''$. 
This is so-called rank condition and the supersymmetry is broken \cite{Ahn07-3}.    
Other F-term equations are satisfied by taking the vacuum expectation 
values of $g, \widetilde{g}, X'$ and $\widetilde{X}'$ to vanish.

More explicitly, the classical moduli space of vacua 
can be obtained from the following F-term
equations
\bea
 q''  \widetilde{q}'' +  m & = & 0, \qquad
\widetilde{q}'' M'' + X' g   =  0, \nonu \\
  M'' q''  + \widetilde{g} \widetilde{X}' & = & 0, \qquad
X' q'' +  \widetilde{g} \Phi'  =  0,
\nonu \\
q'' g & = & 0, \qquad
\widetilde{q}'' \widetilde{X}' +  \Phi' g  =  0, \nonu \\
\widetilde{g} \widetilde{q}'' & = & 0, \qquad
g \widetilde{g}  =  0. 
\nonu
\eea
Then, it is easy to see that there exist three reduced  equations
\bea
\widetilde{q}'' M'' =0= M'' q'', \qquad
 q''  \widetilde{q}'' +  m''  =  0
\nonu
\eea
and other F-term equations are satisfied if one takes the zero vacuum
expectation values for the fields $g, \widetilde{g}, X'$ and 
$\widetilde{X}'$.
Then the solutions for the equations  can be written as follows:
\bea
<q'' >  & = &  \left(
\begin{array}{c}
\sqrt{m} e^{\phi} {\bf 1}_{\widetilde{N}_c''}  \\
0
\end{array}
\right),  
< \widetilde{q}''> =
 \left(
\begin{array}{cc}
\sqrt{m} e^{-\phi}  {\bf 1}_{\widetilde{N}_c''}   &
0
\end{array}
\right), 
<M''>  =
 \left(
\begin{array}{cc}
0  & 0 
 \\
0 & M_0''  {\bf 1}_{N_f''-\widetilde{N}_c''} 
\end{array}
\right),
\nonu \\
<g> & = & <\widetilde{g}> = <X'> = <\widetilde{X}'>= 0.
\nonu
\eea
Let us expand around  a point on these vacua, as done in
\cite{ISS}. 
Then the remaining relevant terms of superpotential can be written as
\bea
W_{dual}^{rel} & = &  M_0'' \left( \delta \varphi  
\; \delta \widetilde{\varphi} + m'' \right) +
  \delta Z \; \delta \varphi  \; \widetilde{q}_0'' 
+ \delta \widetilde{Z} \; q_0''  
\delta \widetilde{\varphi}
\nonu
\eea
by following the same 
fluctuations for the various fields as in \cite{Ahn07}:
\bea
q''  & = &
\left(
\begin{array}{c}
q_0''  {\bf 1}_{\widetilde{N}_c''} +\frac{1}{\sqrt{2}}(\delta \chi_{+} + 
\delta \chi_{-})
 {\bf 1}_{\widetilde{N}_c''} \nonu \\
\delta \varphi
\end{array}
\right), 
\quad 
\widetilde{q}''   = 
 \left(
\begin{array}{cc}
\widetilde{q}_0''   {\bf 1}_{\widetilde{N}_c''} +
\frac{1}{\sqrt{2}}(\delta \chi_{+} - \delta \chi_{-})
  {\bf 1}_{\widetilde{N}_c''}   &
\delta \widetilde{\varphi}
\end{array}
\right),
\nonu \\
M''  & = &
 \left(
\begin{array}{cc}
\delta Y  & \delta Z
 \\
\delta \widetilde{Z} & M_0''  {\bf 1}_{N_f''-\widetilde{N}_c''} 
\end{array}
\right)
\nonu
\eea
as well as the fluctuations of $g, \widetilde{g}, X'$ and 
$\widetilde{X}'$.
Note that there exist also three kinds of terms, 
the vacuum  $<\widetilde{q}''>$ multiplied by 
$\delta \widetilde{g} \delta \widetilde{X}'$,  
the vacuum  $<q''>$ multiplied by $\delta X' 
\delta g$, and 
the vacuum  $<\Phi'>$ multiplied by $\delta g 
\delta \widetilde{g}$. However,
by redefining these, they do not enter the 
contributions for the one loop result, up to quadratic order. 
As done in \cite{Shih}, it turns out 
that $m_{M_0''}^2$ will contain $(\log 4 -1) > 0$ which implies that these
vacua are stable.

Now let us recombine $\widetilde{N}_c''$ flavor D4-branes among $N_f''$
flavor 
D4-branes(connecting between D6-branes and $NS5_R'$-brane) with the same number of 
color D4-branes(connecting between $NS5'_R$-brane and $NS5_R$-brane) and push
them in $+v$ direction from Figure 2A. We assume that $N_c'' \geq N_c'$. 
After this procedure, there are no color D4-branes between 
$NS5'_R$-brane and $NS5_R$-brane.
For the flavor D4-branes, we are left with only 
$(N_f''-\widetilde{N}_c'')=N_c''-N_c'$ flavor D4-branes
connecting between D6-branes and $NS5_R'$-brane.  

Then the minimal energy supersymmetry breaking brane configuration is
shown in Figure 2B.
If we ignore the $NS5'_L$-brane, $N_c'$ D4-branes,
$N_f'$ D6-branes, $NS5_L$-brane, $N_c$ D4-branes and $N_f$ 
D6-branes(detaching these from Figure 2B), 
as observed already, 
then this brane configuration 
corresponds to  the minimal energy supersymmetry breaking brane
configuration
for the ${\cal N}=1$ SQCD with the magnetic gauge group 
$SU(\widetilde{N}_c''=N_f''-N_c'')$ with
$N_f''$ massive flavors \cite{OO,FGU,BGHSS}.

The nonsupersymmetric minimal energy brane configuration in Figure 2B
leads to the Figure 3 of \cite{Ahn07-3} if we ignore the
$NS5_L$-brane, $N_f$ D6-branes and $N_c$ D4-branes.
Moreover, the Figure 2B with a replacement $N_f''$ D6-branes by 
the NS5'-brane(neglecting  the
$NS5_L$-brane, $N_f$ D6-branes and $N_c$ D4-branes)
will become the Figure 2B of \cite{Ahn07-6}
together with a reflection with respect to the $NS5_L$-brane and a
rotation of $NS5_R$-brane by $\frac{\pi}{2}$ angle.

When one moves the $NS5_R'$-brane to the left all the way past the 
$NS5_R$-brane, as long as the $N_f''$ D6-branes are not parallel to 
$NS5_R'$-brane in an electric theory, then  
one arrives at the other magnetic brane configuration similar to the Figure 2. 
The only difference is that the $N_f''$ D6-branes are located at the right
hand side of the $NS5_R$-brane.

Starting with $NS5_L'$-$NS5_L$-$NS5_R'$-$NS5_R$ branes configuration,
as in the footnote \ref{rotation}, 
and moving the $NS5_R$-brane to the left, 
one gets 
the nonsupersymmetric minimal energy brane configuration which is
exactly the Figure 5B(which will appear later) with a reflection with
respect to the NS5-brane. 
Furthermore, by 
moving the $NS5_L'$-brane to the right
(as long as the $N_f$ D6-branes are not parallel to 
$NS5_L'$-brane in an electric theory), 
one gets 
the nonsupersymmetric minimal energy brane configuration which is
exactly the new figure in previous paragraph with a reflection with
respect to the $NS5_L$-brane. 

After lifting 
the type IIA description to M-theory, the 
corresponding magnetic M5-brane configuration, 
in a 
background space of $x t = 
v^{N_f+N_f'} \prod_{k=1}^{N_f''} (v -e_k)$ where 
$e_k$ is the position of the $N_f''$ D6-branes in the $v$ direction and
this four dimensional space
replaces (45610) directions,
is described by \cite{Witten}
\bea
& & t^4 + ( v^{N_c }  + \cdots ) t^3 + 
( v^{N_c'+N_f} + \cdots) t^2  \nonu \\
&& +  (v^{\widetilde{N}_c''+2N_f+N_f'} + 
\cdots ) (v-m)^{N_f''} t + v^{3N_f+2N_f'}
(v -m)^{2N_f''} =0
\nonu
\eea
where 
we have ignored the lower power terms in $v$ in $t^3, t^2$ and $t$ 
and the scales for the gauge groups in front of the first term and the
last term, for simplicity. For fixed $x$, the coordinate $t$
corresponds to $y$.

From this curve 
of quartic equation for $t$ above, the asymptotic regions 
for four NS5-branes can be classified by looking at 
the first two terms providing $NS5_L$-brane asymptotic region, 
next two terms providing $NS5_L'$-brane asymptotic region,
next two terms providing $NS5_R'$-brane asymptotic region,
 and 
the last two terms giving $NS5_R$-brane asymptotic region
as follows:

1. $v \rightarrow \infty$ limit implies
\bea
w \rightarrow 0, && \quad y \sim    v^{N_c} + \cdots \quad
\mbox{$NS_L$ asymptotic region},   
\nonu \\
w \rightarrow 0, && \quad y \sim    
v^{N_f+N_f'+N_f''-\widetilde{N}_c''} + \cdots \quad
\mbox{$NS_R$ asymptotic region}.
\nonu  
\eea

2.  $w \rightarrow \infty$ limit implies
\bea
v  \rightarrow    m, && \quad 
y \sim  w^{N_c'-N_c+N_f}
 +\cdots
\quad \mbox{$NS_{L}'$ asymptotic region}, 
\nonu
\\
v  \rightarrow   m, && \quad  
y \sim w^{\widetilde{N}_c''-N_c'+N_f+N_f'+N_f''}
+\cdots
\quad \mbox{$NS_{R}'$ asymptotic region}. 
\nonu
\eea
Here the two $NS5'_{L,R}$-branes are moving in the
$+v$ direction holding everything else fixed instead of moving
$N_f''$ D6-branes in the $+v$ direction, as in \cite{BGHSS}. 
The harmonic function sourced by the D6-branes can be written
explicitly by summing over three contributions from the 
$N_f$, $N_f'$ and $N_f''$ D6-branes and similar analysis to both solve
the differential equation and find out the nonholomorphic curve can be
done. 
We expect that an instability from a new M5-brane mode arises.

\subsection{Magnetic theory  with dual for second gauge group}

Based on the procedure in previous subsection,
one continues to consider dualizing one of the gauge groups regarding as the
other gauge groups as a spectator. 
One takes the Seiberg dual for the second gauge group factor $SU(N_c')$
while remaining the first and third gauge group factor $SU(N_c)$ and  
$SU(N_c'')$ unchanged. 
Also, 
since the dualized group's dynamical scale is far above that of the
other spectator groups, 
we consider the case where $\Lambda_2 >> \Lambda_1, \Lambda_3$.

Let us move the $NS5_R$-brane in Figure 1
to the left all the way past the  
$NS5'_L$-brane.  
There is also a possibility to move $NS5_L'$-brane to the right instead. 
We'll come to this next subsection.
After this brane motion, one arrives at the Figure 3A.
The linking number of $NS5_R$-brane from Figure 3A
is 
$
L_5 = -\frac{N_f'}{2} +\widetilde{N}_c'-N_c
$
while
the linking number of $NS5_R$-brane from Figure 1
is
$
L_5 = \frac{N_f'}{2} + N_c'' -N_c'$. 
Due to the connection of $N_c''$
D4-branes with $NS5_R$-brane in Figure 1, 
the presence of $N_c''$ in the linking
number arises.
From these two relations, one obtains
the number of colors of dual magnetic theory
\bea
\widetilde{N}_c' = N_f' +N_c''+N_c-N_c'.
\label{number1}
\eea
Compared with the previous case given by (\ref{number}), 
the dependence on the ranks of 
both the first gauge group and the third gauge group occurs.

Let us draw this magnetic brane configuration in Figure 3A and recall
that we put
the coincident $N_f'$ D6-branes in the nonzero $v$-direction in the
electric theory as well as massless flavors for $Q(Q'')$ and
$\widetilde{Q}(
\widetilde{Q}'')$ for simplicity.
If we ignore the $NS5_L$-brane, $N_c$ D4-branes, $N_f$ D6-branes, $N_f''$ 
D6-branes, the $NS5_R'$-brane and $N_c''$ D4-branes(detaching these
branes from Figure 3A), 
then this brane configuration 
leads to the standard ${\cal N}=1$ SQCD with the magnetic gauge group 
$SU(\widetilde{N}_c'=N_f'-N_c')$ with
$N_f'$ massive flavors \cite{OO,FGU,BGHSS}.

\begin{figure}[ht]
   \epsfxsize=4.0in 
\centerline{\epsffile{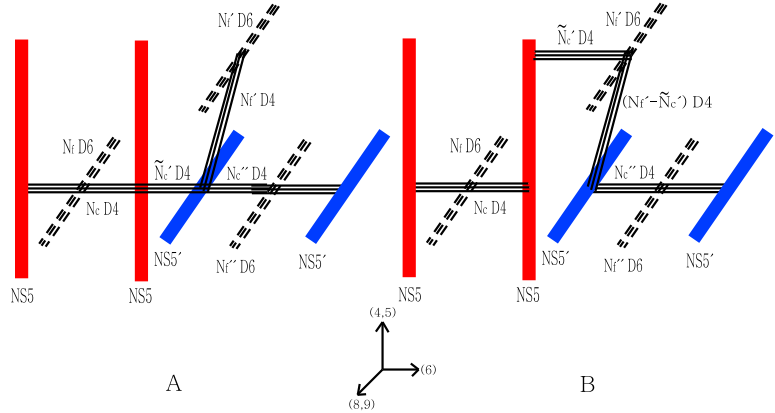}}
   \caption[FIG. \arabic{figure}.]{ 
The ${\cal N}=1$ supersymmetric magnetic brane configuration with
$SU(N_c) \times SU(\widetilde{N}_c') \times SU(N_c'')$ gauge group
with fundamentals $Q(q')[Q'']$ and
$\widetilde{Q}(\widetilde{q}')[\widetilde{Q}'']$ 
for each gauge group and bifundamentals $F(g)$ and
$\widetilde{F}(\widetilde{g})$ and gauge singlets in Figure 3A. In
Figure 3B, the nonsupersymmetric minimal energy brane configuration
with the same gauge group and matter contents above 
for massless  $Q(Q'')$ and
$\widetilde{Q}(\widetilde{Q}'')$ is given. 
 }
\end{figure}

In the dual theory, 
since there exist $N_f'$ quarks $q'$, $N_f'$
quarks $\widetilde{q}'$, one bifundamental field $g$(which gives
rise to the contribution of $N_c''$) and 
its complex conjugate field $\widetilde{g}$(which gives
the contribution of $N_c''$),
one bifundamental field $F$(which will give
rise to the contribution of $N_c$), and 
its complex conjugate field $\widetilde{F}$(which will give
rise to the contribution of $N_c$),
the coefficient of the beta function for the second 
gauge group factor 
is
$
b_{SU(\widetilde{N}_c')}^{mag}
= 3\widetilde{N}_c'-N_f'-N_c''-N_c
=2N_f' +2N_c''+2N_c-3N_c'
$
where we inserted the number of colors given in (\ref{number1}) in the
second equality. 
Since there exist $N_f''$ quarks $Q''$, $N_f''$
quarks $\widetilde{Q}''$, one bifundamental field $g$(which will give
rise to the contribution of $\widetilde{N}_c'$), 
its complex conjugate $\widetilde{g}$(which will give
rise to the contribution of $\widetilde{N}_c'$),
$N_f'$-fields $X''$, its complex conjugate
$N_f'$-fields $\widetilde{X}''$, and the singlet $\Phi''$(which will give
rise to $N_c''$), 
the coefficient of the beta function of 
second gauge group factor 
is
$
b_{SU(N_c'')}^{mag} 
= 3N_c''-N_f''-\widetilde{N}_c'-N_f'-N_c'' 
=N_c''+N_c'-N_c-2N_f'-N_f''$.
Similarly,
since there exist $N_f$ quarks $Q$, $N_f$
quarks $\widetilde{Q}$, one bifundamental field $F$(which will give
rise to the contribution of $\widetilde{N}_c'$), 
and its complex conjugate field $\widetilde{F}$(which will give
rise to the contribution of $\widetilde{N}_c'$),
the coefficient of the beta function of 
the first gauge group factor 
is
$
b_{SU(N_c)}^{mag} 
= 3N_c-N_f-\widetilde{N}_c'=2N_c-N_c''+N_c'-N_f-N_f'$.

Therefore, $SU(N_c)$, $SU(\widetilde{N}_c')$ and 
$SU(N_c'')$ gauge couplings are IR free
by requiring the negativeness of the coefficients of beta function.
One relies on the perturbative calculations at low energy 
for this magnetic IR free region $b_{SU(N_c)}^{mag} <
0$, $b_{SU(\widetilde{N}_c')}^{mag} < 0$ and 
$b_{SU(\widetilde{N}_c'')}^{mag} < 0$.
The $SU(N_c)$ fields in the magnetic theory
are different from those of electric theory and also
the $SU(N_c'')$ fields in the magnetic theory
are different from those of electric theory.
Neglecting the $SU(N_c)$ and $SU(N_c'')$ dynamics, 
the magnetic $SU(\widetilde{N}_c')$
is IR free when 
\bea
N_c'-N_c''-N_c < N_f' < \frac{3}{2} N_c'-N_c''-N_c.
\nonu
\eea
Here the $N_c$- and $N_c''$-dependence appears.

Then one can summarize the gauge group and matter contents where there
are additional fields  $X'', \widetilde{X}'', M'$ and $\Phi''$ as follows:
\bea
 & \mbox{gauge group}:& \;\;\;\;\;   SU(N_c) \times SU(\widetilde{N}_c') \times
 SU(N_c'')  \nonu
\\
\mbox{matter}:  
 &Q_f \oplus \widetilde{Q}_{\widetilde{f}}& \;\;\;\;\;\;\;\;\; 
\;\;\; {(\bf \Box, 1, 1) \oplus (\overline{\Box}, 1, 1)}
\;\;\;\;\; (f, \widetilde{f}=1,  \cdots, N_f) 
\nonu \\
 &q'_{f'} \oplus \widetilde{q}'_{\widetilde{f}'}& \;\;\;\;\;\;\;\;\;\;
\;\; {(\bf 1, \Box, 1) \oplus ( 1, \overline{\Box}, 1)}
\;\;\;\;\; (f', \widetilde{f}' =1,  \cdots, N_f') 
\nonu \\
 &Q''_{f''} \oplus \widetilde{Q}''_{\widetilde{f}''}& \;\;\;\;\;\;\;\;\;\; 
\;\; {(\bf 1, 1, \Box) \oplus ( 1, 1, \overline{\Box})} 
\;\;\;\;\; (f'', \widetilde{f}'' =1,  \cdots, N_f'')
\nonu \\
&F \oplus \widetilde{F}& \;\;\;\;\;\;\;\;\; 
\;\; {(\bf \Box, \overline{\Box},1) \oplus (\overline{\Box}, \Box, 1)} 
\nonu \\
 &g \oplus \widetilde{g}& \;\;\;\;\;\;\;\;\; 
\;\; {(\bf 1, \Box, \overline{\Box}) \oplus ( 1, \overline{\Box}, \Box)} 
\nonu \\
& (X_{n'}'' \equiv) \widetilde{G} Q' \oplus G \widetilde{Q}' (\equiv 
\widetilde{X}_{\widetilde{n}'}'') & 
 \;\;\;\;\;\;\;\;\;\;
\;\; {(\bf 1, 1, \Box) \oplus ( 1, 1, \overline{\Box})}
\;\;\;\;\; (n', \widetilde{n}' =1,  \cdots, N_f') 
\nonu \\
&  (M'_{f',\widetilde{g}'} \equiv) Q' \widetilde{Q}' & 
 \;\;\;\;\;\;\;\;\;\;\;\;\;\;\;\;\;\;\;
\;\; {(\bf 1, 1, 1)} \;\;\;\;\;\;\;\;\;
\;\;\;\;\; (f', \widetilde{g}' =1,  \cdots, N_f') 
\nonu \\
& (\Phi'' \equiv) G \widetilde{G} & 
 \;\;\;\;\;\;\;\;\;\;
\;\; {(\bf 1, 1, adj) \oplus ( 1, 1, 1)}
\nonu
\eea

The dual magnetic superpotential, by adding the mass term 
(\ref{superpotential}) for $Q'$ and
$\widetilde{Q}'$
in the electric theory which is equal to put a linear term in $M'$ in
the dual magnetic theory, is given by 
\bea
W_{dual} = \left(M' q' \widetilde{q}' + g \widetilde{X}'' \widetilde{q}' + 
\widetilde{g} q'
X'' + \Phi'' g \widetilde{g} \right) + m' M'.
\nonu
\eea
One sees also the usual cubic interaction terms between the additional
fields  $X'', \widetilde{X}'', M'$ and $\Phi''$
and their dual expressions, $\widetilde{g} q', g \widetilde{q}', 
q' \widetilde{q}'$
and $g \widetilde{g}$ respectively \cite{Ahn07-3}. 
Here $q'$ and $\widetilde{q}'$ are fundamental and antifundamental for
the gauge group index respectively.
Then, $q' \widetilde{q}'$ has rank $\widetilde{N}_c'$ while $m'$ has a
rank $N_f'$.  Therefore, the F-term condition, the derivative the 
superpotential $W_{dual}$ with respect to $M'$, cannot be satisfied 
if the rank $N_f'$ exceeds $\widetilde{N}_c'$. 
This is so-called rank condition and the supersymmetry is broken.    
Other F-term equations are satisfied by taking the vacuum expectation 
values of $g, \widetilde{g}, X''$ and $\widetilde{X}''$ to vanish.

Then the solutions for these equations can be written as follows:
\bea
<q' >  & = &  \left(
\begin{array}{c}
\sqrt{m} e^{\phi} {\bf 1}_{\widetilde{N}_c'}  \\
0
\end{array}
\right),  
< \widetilde{q}'> =
 \left(
\begin{array}{cc}
\sqrt{m} e^{-\phi}  {\bf 1}_{\widetilde{N}_c'}   &
0
\end{array}
\right), 
<M'>  =
 \left(
\begin{array}{cc}
0  & 0 
 \\
0 & M_0'  {\bf 1}_{N_f'-\widetilde{N}_c'} 
\end{array}
\right),
\nonu \\
<g> & = & <\widetilde{g}> = <X''> = <\widetilde{X}''>= 0.
\nonu
\eea
As we did in previous case, one can analyze the one loop 
computation by expanding the fields around the vacua and it will
lead to the fact that states are stable by realizing the mass of 
$m_{M_0'}^2$ positive.


Then the minimal energy supersymmetry breaking brane configuration is
shown in Figure 3B.
If we ignore the $NS5_L$-brane, $N_c$ D4-branes,
$N_f$ D6-branes, $NS5_R'$-brane, $N_c''$ D4-branes and $N_f''$ 
D6-branes(detaching these from Figure 3B), 
as observed already, 
then this brane configuration 
corresponds to  the minimal energy supersymmetry breaking brane
configuration
for the ${\cal N}=1$ SQCD with the magnetic gauge group 
$SU(\widetilde{N}_c'=N_f'-N_c')$ with
$N_f'$ massive flavors \cite{OO,FGU,BGHSS}.

The nonsupersymmetric minimal energy brane configuration Figure 3B
with vanishing $N_f''$ D6-branes
leads to the Figure 3 of \cite{Ahn07-3} with a reflection with
respect to the NS5-brane if we ignore the
$NS5_L$-brane, $N_f$ D6-branes and $N_c$ D4-branes.
Moreover, this Figure 3B with a replacement $N_f'$ D6-branes by 
the NS5'-brane(and neglecting  the
$NS5_L$-brane, $N_f$ D6-branes and $N_c$ D4-branes)
will become the Figure 2B of \cite{Ahn07-6}
with a
rotation of $NS5_R$-brane by $\frac{\pi}{2}$ angle or 
the Figure 4B of \cite{Ahn07-6} with a reflection with
respect to the $NS5_R$-brane 
if we ignore 
the
$NS5_R'$-brane, $N_f''$ D6-branes and $N_c''$ D4-branes from the
Figure 3B.

Starting with $NS5_L'$-$NS5_L$-$NS5_R'$-$NS5_R$ branes configuration,
as in the footnote \ref{rotation}, 
by moving the $NS5_L$-brane to the right, 
one gets 
the nonsupersymmetric minimal energy brane configuration which is
exactly the Figure 3 with a reflection with
respect to the NS5-brane. 

By lifting 
the type IIA description to M-theory, the 
corresponding magnetic M5-brane configuration, in 
a background space of $x t = 
v^{N_f+N_f''} \prod_{k=1}^{N_f'} (v -e_k)$
where $e_k$ is the position of the $N_f'$ D6-branes in the $v$ direction,
is described by 
\bea
t^4 + ( v^{N_c+N_f }  + \cdots ) t^3 + 
( v^{\widetilde{N}_c'+N_f} + \cdots) t^2 + (v^{N_c''+2N_f} + 
\cdots ) t + v^{3N_f+N_f''}
(v -m )^{N_f'} =0.
\nonu
\eea

From this curve  
of quartic equation for $t$ above, the asymptotic regions 
for four NS5-branes can be classified 
as follows:

1. $v \rightarrow \infty$ limit implies
\bea
w \rightarrow 0, && \quad y \sim    v^{N_c+N_f} + \cdots \quad
\mbox{$NS_L$ asymptotic region},   
\nonu \\
w \rightarrow 0, && \quad y \sim    
v^{\widetilde{N}_c'-N_c} + \cdots \quad
\mbox{$NS_R$ asymptotic region}.
\nonu  
\eea

2.  $w \rightarrow \infty$ limit implies
\bea
v  \rightarrow    m, && \quad 
y \sim  w^{N_c''-\widetilde{N}_c'+N_f}
 +\cdots
\quad \mbox{$NS_{L}'$ asymptotic region}, 
\nonu
\\
v  \rightarrow   m, && \quad  
y \sim w^{N_f+N_f'+N_f''-N_c''}
+\cdots
\quad \mbox{$NS_{R}'$ asymptotic region}. 
\nonu
\eea

\subsection{Magnetic theory  with dual for second gauge group}

For the dualizing the middle gauge group, there exists another
magnetic brane configuration. 
One considers dualizing one of the gauge groups regarding as the
other gauge groups as a spectator, as we did in previous subsections. 
Also we consider the case where $\Lambda_2 >> \Lambda_1, \Lambda_3$, in other
words, the dualized group's dynamical scale is far above that of the
other spectator groups.

Let us move the $NS5_L'$-brane in Figure 1
to the right all the way past the  
$NS5_R$-brane.  
After this brane motion, one arrives at the Figure 4A.
That is, we rotate $N_f'$ D6-branes a little bit(this
does not change the classical electric superpotential as we explained 
before) and 
after dualizing the second gauge group, 
we rotate those $N_f'$ D6-branes with opposite direction.
The linking number of $NS5'_L$-brane from Figure 4A
is 
$
L_5 = \frac{N_f'}{2} -\widetilde{N}_c'+N_c''$.
On the other hand, the linking number of $NS5'_L$-brane from modified 
Figure 1
is
$
L_5 = -\frac{N_f'}{2} + N_c' -N_c$. 
Due to the connection of $N_c$
D4-branes with $NS5_L'$-brane in Figure 1, the presence of $N_c$ in the linking
number arises.
From these two relations, one obtains
the number of colors of dual magnetic theory
\bea
\widetilde{N}_c' = N_f' +N_c''+N_c-N_c'
\label{Number1}
\eea
which is the same as before given in (\ref{number1}).

Let us draw this magnetic brane configuration in Figure 4A and recall
that we put
the coincident $N_f'$ D6-branes in the nonzero $v$-direction and 
the coincident $N_f$ and $N_f''$ D6-branes at $v=0$ in the
electric theory.

\begin{figure}[ht]
   \epsfxsize=4.0in 
\centerline{\epsffile{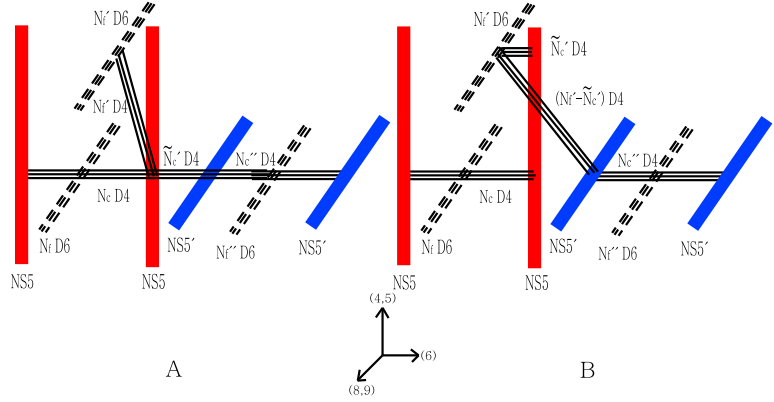}}
   \caption[FIG. \arabic{figure}.]{ 
The ${\cal N}=1$ supersymmetric magnetic brane configuration with
$SU(N_c) \times SU(\widetilde{N}_c') \times SU(N_c'')$ gauge group
with fundamentals $Q(q')[Q'']$ and
$\widetilde{Q}(\widetilde{q}')[\widetilde{Q}'']$ 
for each gauge group and bifundamentals $f(G)$ and
$\widetilde{f}(\widetilde{G})$ and gauge singlets in Figure 4A. In
Figure 4B, the nonsupersymmetric minimal energy brane configuration
with the same gauge group and matter contents above 
for massless  $Q(Q'')$ and
$\widetilde{Q}(\widetilde{Q}'')$ is given. 
 }
\end{figure}

In the dual theory, 
since there exist $N_f'$ quarks $q'$, $N_f'$
quarks $\widetilde{q}'$, one bifundamental field $G$(which will give
rise to the contribution of $N_c''$), 
its complex conjugate field $\widetilde{G}$(which will give
rise to the contribution of $N_c''$),
one bifundamental field $f$ which(will give
rise to the contribution of $N_c$), and 
its complex conjugate field $\widetilde{f}$(which will give
rise to the contribution of $N_c$),
the coefficient of the beta function for the second 
gauge group factor 
is
$
b_{SU(\widetilde{N}_c')}^{mag}
= 3\widetilde{N}_c'-N_f'-N_c''-N_c
=2N_f' +2N_c''+2N_c-3N_c'
$
where we inserted the number of colors given in (\ref{Number1}) in the
second equality and this is equal to the 
$b_{SU(\widetilde{N}_c')}^{mag}$
in previous subsection. 
Since there exist $N_f''$ quarks $Q''$, $N_f''$
quarks $\widetilde{Q}''$, one bifundamental field $G$(which will give
rise to the contribution of $\widetilde{N}_c'$), and
its complex conjugate field $\widetilde{G}$(which will give
rise to the contribution of $\widetilde{N}_c'$),
 the coefficient of the beta function of 
second gauge group factor 
is
$
b_{SU(N_c'')}^{mag} 
= 3N_c''-N_f''-\widetilde{N}_c' 
=2N_c''+N_c'-N_c-N_f'-N_f''$.
Similarly,
since there exist $N_f$ quarks $Q$, $N_f$
quarks $\widetilde{Q}$, one bifundamental field $f$(which will give
rise to the contribution of $\widetilde{N}_c'$), 
its complex conjugate field $\widetilde{f}$(which will give
rise to the contribution of $\widetilde{N}_c'$),
$N_f'$-fields $X$, its complex conjugate
$N_f'$-fields $\widetilde{X}$, and the singlet 
$\Phi$(which will give
rise to $N_c$),
the coefficient of the beta function of 
the first gauge group factor 
is
$
b_{SU(N_c)}^{mag} 
= 3N_c-N_f-\widetilde{N}_c'-N_f'-N_c=N_c-N_c''+N_c'-N_f-2N_f'$.

Therefore, $SU(N_c)$, $SU(\widetilde{N}_c')$ and 
$SU(N_c'')$ gauge couplings are IR free
by requiring the negativeness of the coefficients of beta function.
One relies on the perturbative calculations at low energy 
for this magnetic IR free region $b_{SU(N_c)}^{mag} <
0$, $b_{SU(\widetilde{N}_c')}^{mag} < 0$ and 
$b_{SU(\widetilde{N}_c'')}^{mag} < 0$.
Neglecting the $SU(N_c)$ and $SU(N_c'')$ dynamics, 
the magnetic $SU(\widetilde{N}_c')$
is IR free when 
\bea
N_c' -N_c''-N_c < N_f' < \frac{3}{2} N_c'-N_c''-N_c.
\nonu
\eea

Then one can summarize the gauge group and matter contents where there
are additional fields  $X, \widetilde{X}, M'$ and $\Phi$ as follows:
\bea
 & \mbox{gauge group}:& \;\;\;\;\;   SU(N_c) \times SU(\widetilde{N}_c') \times
 SU(N_c'')  \nonu
\\
\mbox{matter}:  
 &Q_f \oplus \widetilde{Q}_{\widetilde{f}}& \;\;\;\;\;\;\;\;\; 
\;\;\; {(\bf \Box, 1, 1) \oplus (\overline{\Box}, 1, 1)}
\;\;\;\;\; (f, \widetilde{f}=1,  \cdots, N_f) 
\nonu \\
 &q'_{f'} \oplus \widetilde{q}'_{\widetilde{f}'}& \;\;\;\;\;\;\;\;\;\;
\;\; {(\bf 1, \Box, 1) \oplus ( 1, \overline{\Box}, 1)}
\;\;\;\;\; (f', \widetilde{f}' =1,  \cdots, N_f') 
\nonu \\
 &Q''_{f''} \oplus \widetilde{Q}''_{\widetilde{f}''}& \;\;\;\;\;\;\;\;\;\; 
\;\; {(\bf 1, 1, \Box) \oplus ( 1, 1, \overline{\Box})} 
\;\;\;\;\; (f'', \widetilde{f}'' =1,  \cdots, N_f'')
\nonu \\
&f \oplus \widetilde{f}& \;\;\;\;\;\;\;\;\; 
\;\; {(\bf \Box, \overline{\Box},1) \oplus (\overline{\Box}, \Box, 1)} 
\nonu \\
 &G \oplus \widetilde{G}& \;\;\;\;\;\;\;\;\; 
\;\; {(\bf 1, \Box, \overline{\Box}) \oplus ( 1, \overline{\Box}, \Box)} 
\nonu \\
& (X_{n'} \equiv) F Q' \oplus \widetilde{F} \widetilde{Q}' (\equiv 
\widetilde{X}_{\widetilde{n}'}) & 
 \;\;\;\;\;\;\;\;\;\;
\;\; {(\bf \Box, 1, 1) \oplus ( \overline{\Box}, 1, 1)}
\;\;\;\;\; (n', \widetilde{n}' =1,  \cdots, N_f') 
\nonu \\
&  (M'_{f',\widetilde{g}'} \equiv) Q' \widetilde{Q}' & 
 \;\;\;\;\;\;\;\;\;\;\;\;\;\;\;\;\;\;\;
\;\; {(\bf 1, 1, 1)} \;\;\;\;\;\;\;\;\;
\;\;\;\;\; (f', \widetilde{g}' =1,  \cdots, N_f') 
\nonu \\
& (\Phi \equiv) F \widetilde{F} & 
 \;\;\;\;\;\;\;\;\;\;
\;\; {(\bf adj, 1, 1) \oplus ( 1, 1, 1)}
\nonu
\eea

The dual magnetic superpotential, by adding the mass term 
(\ref{superpotential}) for $Q'$ and
$\widetilde{Q}'$
in the electric theory, which is equal to put a linear term in $M'$ in
the dual magnetic theory, is given by 
\bea
W_{dual} = \left(M' q' \widetilde{q}' + f X q' + 
\widetilde{f} \widetilde{q}'
\widetilde{X} + \Phi f \widetilde{f} \right) + m' M'.
\nonu
\eea
Here $q'$ and $\widetilde{q'}$ are fundamental and antifundamental for
the gauge group index respectively \cite{Ahn07-3}.
Then, $q' \widetilde{q}'$ has rank $\widetilde{N}_c'$ while $m'$ has a
rank $N_f'$.  Therefore, the F-term condition, the derivative the 
superpotential $W_{dual}$ with respect to $M'$, cannot be satisfied 
if the rank $N_f'$ exceeds $\widetilde{N}_c'$. 
This is so-called rank condition and the supersymmetry is broken.    
Other F-term equations are satisfied by taking the vacuum expectation 
values of $f, \widetilde{f}, X$ and $\widetilde{X}$ to vanish.

Then the solutions can be written as follows:
\bea
<q' >  & = &  \left(
\begin{array}{c}
\sqrt{m} e^{\phi} {\bf 1}_{\widetilde{N}_c'}  \\
0
\end{array}
\right),  
< \widetilde{q}'> =
 \left(
\begin{array}{cc}
\sqrt{m} e^{-\phi}  {\bf 1}_{\widetilde{N}_c'}   &
0
\end{array}
\right), 
<M'>  =
 \left(
\begin{array}{cc}
0  & 0 
 \\
0 & M_0'  {\bf 1}_{N_f'-\widetilde{N}_c'} 
\end{array}
\right),
\nonu \\
<f> & = & <\widetilde{f}> = <X> = <\widetilde{X}>= 0.
\nonu
\eea
As we did in previous case, one can analyze the one loop 
computation by expanding the fields around the vacua and it will
lead to the fact that states are stable by realizing the mass of 
$m_{M_0'}^2$ positive.


Then the minimal energy supersymmetry breaking brane configuration is
shown in Figure 4B.
If we ignore the $NS5_L$-brane, $N_c$ D4-branes,
$N_f$ D6-branes, $NS5_R'$-brane, $N_c''$ D4-branes and $N_f''$ 
D6-branes(detaching these from Figure 4B), 
then this brane configuration 
looks similar to  the minimal energy supersymmetry breaking brane
configuration
for the ${\cal N}=1$ SQCD with the magnetic gauge group 
$SU(\widetilde{N}_c'=N_f'-N_c')$ with
$N_f'$ massive flavors. The difference occurs in the position of D6-branes. 

The nonsupersymmetric minimal energy brane configuration Figure 4B
with a replacement $N_f'$ D6-branes by 
the NS5'-brane(neglecting  the
$NS5_R'$-brane, $N_f''$ D6-branes and $N_c''$ D4-branes 
with vanishing $N_f$ D6-branes)
leads to 
the Figure 5B of \cite{Ahn07-6}
with a
rotation of $NS5_L'$-brane by $\frac{\pi}{2}$ angle. 

Starting with $NS5_L'$-$NS5_L$-$NS5_R'$-$NS5_R$ branes configuration,
as in the footnote \ref{rotation}, 
by moving the $NS5_R'$-brane to the left, 
one gets 
the nonsupersymmetric minimal energy brane configuration which is
exactly the Figure 4 with a reflection with
respect to the NS5-brane. 

After lifting 
the type IIA description to M-theory, the 
magnetic M5-brane configuration, in a 
background space of $x t = 
v^{N_f+N_f''} \prod_{k=1}^{N_f'} (v -e_k)$,
is described by 
\bea
&& t^4 + ( v^{N_c }  + \cdots ) t^3 + 
\left[ v^{\widetilde{N}_c'+N_f}(v-m)^{N_f'} + \cdots \right] t^2 \nonu \\
&& + 
\left[ v^{N_c''+2N_f}(v-m)^{2N_f'} + \cdots \right] t + v^{3N_f+N_f''}
(v -m )^{3N_f'} =0.
\nonu
\eea

From this curve  
of quartic equation for $t$ above, the asymptotic regions 
for four NS5-branes can be classified 
as follows

1. $v \rightarrow \infty$ limit implies
\bea
w \rightarrow 0, && \quad y \sim    v^{N_c} + \cdots \quad
\mbox{$NS_L$ asymptotic region},   
\nonu \\
w \rightarrow 0, && \quad y \sim    
v^{\widetilde{N}_c'-N_c+N_f+N_f'} + \cdots \quad
\mbox{$NS_R$ asymptotic region}.
\nonu  
\eea

2.  $w \rightarrow \infty$ limit implies
\bea
v  \rightarrow    m, && \quad 
y \sim  w^{N_c''-\widetilde{N}_c'+N_f+N_f'}
 +\cdots
\quad \mbox{$NS_{L}'$ asymptotic region}, 
\nonu
\\
v  \rightarrow   m, && \quad  
y \sim w^{N_f+N_f'+N_f''-N_c''}
+\cdots
\quad \mbox{$NS_{R}'$ asymptotic region}. 
\nonu
\eea

\subsection{Magnetic theory  with dual for first gauge group}

Now we turn to the last magnetic brane configuration.
One considers dualizing one of the gauge groups regarding as the
other gauge groups as a spectator, as done in previous subsections. 
Also we consider the case where $\Lambda_1 >> \Lambda_2, \Lambda_3$, in other
words, the dualized group's dynamical scale is far above that of the
other spectator groups.

Let us move the $NS5_L$-brane in Figure 1
to the right all the way past the  
$NS5'_L$-brane.  
After this brane motion, one arrives at the Figure 5A.
Recall that the $N_f$ D6-branes are perpendicular to the 
$NS5_L$-brane in Figure 1.
The linking number of $NS5_L$-brane from Figure 5A
is 
$
L_5 = \frac{N_f}{2} -\widetilde{N}_c +N_c'$.
 Due to the connection of $N_c'$
D4-branes with $NS5_L$-brane in Figure 5A, the presence of $N_c'$ in the linking
number arises.
On the other hand, the linking number of $NS5_L$-brane from Figure 1
is
$
L_5 = -\frac{N_f}{2} + N_c$.
From these two relations, one obtains
the number of colors of dual magnetic theory
\bea
\widetilde{N}_c = N_f +N_c'-N_c.
\label{number2}
\eea

Let us draw this magnetic brane configuration in Figure 5A and recall
that we put
the coincident $N_f$ D6-branes in the nonzero $v$-direction as well as
$N_f'$ and $N_f''$ D6-branes at $v=0$
in the
electric theory.
If we ignore the $NS5_R$-brane, $N_c'$ D4-branes, $N_f'$ 
D6-branes, the $NS5_R'$-brane, $N_f''$ D6-branes  
and $N_c''$ D4-branes(detaching these
branes from Figure 5A), 
then this brane configuration 
leads to the standard ${\cal N}=1$ SQCD with the magnetic gauge group 
$SU(\widetilde{N}_c=N_f-N_c)$ with
$N_f$ massive flavors \cite{OO,FGU,BGHSS}.

\begin{figure}[ht]
   \epsfxsize=4.0in 
\centerline{\epsffile{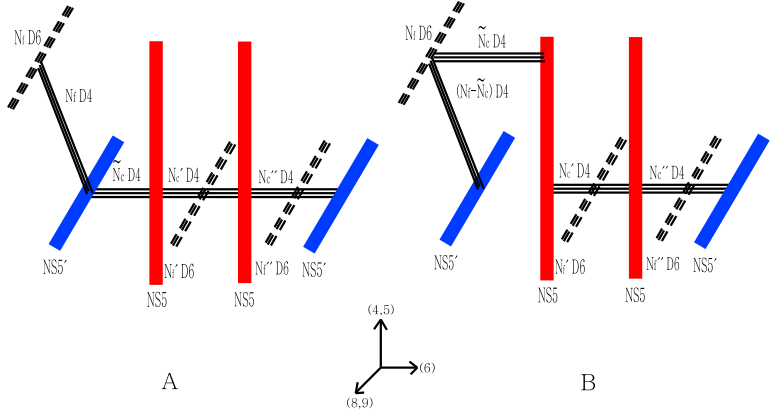}}
   \caption[FIG. \arabic{figure}.]{ 
The ${\cal N}=1$ supersymmetric magnetic brane configuration with
$SU(\widetilde{N}_c) \times SU(N_c') \times SU(N_c'')$ gauge group
with fundamentals $q(Q')[Q'']$ and
$\widetilde{q}(\widetilde{Q}')[\widetilde{Q}'']$ 
for each gauge group and bifundamentals $F(g)$ and
$\widetilde{F}(\widetilde{g})$ and gauge singlets in Figure 5A. In
Figure 5B, the nonsupersymmetric minimal energy brane configuration
with the same gauge group and matter contents above 
for massless  $Q'(Q'')$ and
$\widetilde{Q}'(\widetilde{Q}'')$ is given. 
 }
\end{figure}

In the dual theory, 
since there exist $N_f$ quarks $q$, $N_f$
quarks $\widetilde{q}$, one bifundamental field $f$(which will give
rise to the contribution of $N_c'$), and 
its complex conjugate $\widetilde{f}$(which will give
rise to the contribution of $N_c'$),
the coefficient of the beta function for the third 
gauge group factor 
is
$
b_{SU(\widetilde{N}_c)}^{mag}
= 3\widetilde{N}_c-N_f-N_c'
=2N_f +2N_c'-3N_c
$
where we inserted the number of colors given in (\ref{number2}) in the
second equality. 
Since there exist $N_f'$ quarks $Q'$, $N_f'$
quarks $\widetilde{Q}'$, one bifundamental field $f$(which will give
rise to the contribution of $\widetilde{N}_c$), 
its complex conjugate field $\widetilde{f}$(which will give
rise to the contribution of $\widetilde{N}_c$),
one bifundamental field $G$(which will give
rise to the contribution of $N_c''$), 
its complex conjugate $\widetilde{G}$(which will give
rise to the contribution of $N_c''$),
$N_f$ fields $X'$, its complex conjugate
$N_f$ fields $\widetilde{X}'$, and the singlet 
$\Phi'$(which will give
rise to $N_c'$), 
the coefficient of the beta function of 
second gauge group factor 
is
$
b_{SU(N_c')}^{mag} 
= 3N_c'-N_f'-\widetilde{N}_c-N_c''-N_f-N_c' 
=N_c'+N_c-N_c''-2N_f-N_f'$.
Similarly,
since there exist $N_f''$ quarks $Q''$, $N_f''$
quarks $\widetilde{Q}''$, one bifundamental field $G$(which will give
rise to the contribution of $N_c'$), 
and its complex conjugate field $\widetilde{G}$(which will give
rise to the contribution of $N_c'$),
the coefficient of the beta function of 
the third gauge group factor 
is
$
b_{SU(N_c'')}^{mag} 
= 3N_c''-N_f''-N_c'=b_{SU(N_c'')}$.

Therefore, $SU(\widetilde{N}_c)$, $SU(N_c')$ and 
$SU(N_c'')$ gauge couplings are IR free
by requiring the negativeness of the coefficients of beta function.
One relies on the perturbative calculations at low energy 
for this magnetic IR free region $b_{SU(\widetilde{N}_c)}^{mag} <
0$, $b_{SU(N_c)}^{mag} < 0$ and 
$b_{SU(N_c'')}^{mag} < 0$.
Note that the $SU(N_c')$ fields in the magnetic theory 
are different from those of the electric theory.
Since $b_{SU(N_c')}-b_{SU(N_c')}^{mag} > 0$, $SU(N_c')$ is more
asymptotically free than $SU(N_c')^{mag}$.
Neglecting the $SU(N_c')$ and $SU(N_c'')$ dynamics, 
the magnetic $SU(\widetilde{N}_c)$
is IR free when 
\bea
N_c -N_c' < N_f < \frac{3}{2} N_c-N_c'.
\nonu
\eea

Then one can summarize the gauge group and matter contents where there
are additional fields  $X', \widetilde{X}', M$ and $\Phi'$ as follows:
\bea
 & \mbox{gauge group}:& \;\;\;\;\;   SU(\widetilde{N}_c) \times SU(N_c') \times
 SU(N_c'')  \nonu
\\
\mbox{matter}:  
 &q_f \oplus \widetilde{q}_{\widetilde{f}}& \;\;\;\;\;\;\;\;\; 
\;\;\; {(\bf \Box, 1, 1) \oplus (\overline{\Box}, 1, 1)}
\;\;\;\;\; (f, \widetilde{f}=1,  \cdots, N_f) 
\nonu \\
 &Q'_{f'} \oplus \widetilde{Q}'_{\widetilde{f}'}& \;\;\;\;\;\;\;\;\;\;
\;\; {(\bf 1, \Box, 1) \oplus ( 1, \overline{\Box}, 1)}
\;\;\;\;\; (f', \widetilde{f}' =1,  \cdots, N_f') 
\nonu \\
 &Q''_{f''} \oplus \widetilde{Q}''_{\widetilde{f}''}& \;\;\;\;\;\;\;\;\;\; 
\;\; {(\bf 1, 1, \Box) \oplus ( 1, 1, \overline{\Box})} 
\;\;\;\;\; (f'', \widetilde{f}'' =1,  \cdots, N_f'')
\nonu \\
&f \oplus \widetilde{f}& \;\;\;\;\;\;\;\;\; 
\;\; {(\bf \Box, \overline{\Box},1) \oplus (\overline{\Box}, \Box, 1)} 
\nonu \\
 &G \oplus \widetilde{G}& \;\;\;\;\;\;\;\;\; 
\;\; {(\bf 1, \Box, \overline{\Box}) \oplus ( 1, \overline{\Box}, \Box)} 
\nonu \\
& (X_{n}' \equiv) \widetilde{F} Q \oplus F \widetilde{Q} (\equiv 
\widetilde{X}_{\widetilde{n}}') & 
 \;\;\;\;\;\;\;\;\;\;
\;\; {(\bf 1, \Box, 1) \oplus ( 1, \overline{\Box}, 1)}
\;\;\;\;\; (n, \widetilde{n} =1,  \cdots, N_f) 
\nonu \\
&  (M_{f,\widetilde{g}} \equiv) Q \widetilde{Q} & 
 \;\;\;\;\;\;\;\;\;\;\;\;\;\;\;\;\;\;\;
\;\; {(\bf 1, 1, 1)} \;\;\;\;\;\;\;\;\;
\;\;\;\;\; (f, \widetilde{g} =1,  \cdots, N_f) 
\nonu \\
& (\Phi' \equiv) F \widetilde{F} & 
 \;\;\;\;\;\;\;\;\;\;
\;\; {(\bf 1, adj, 1) \oplus ( 1, 1, 1)}
\nonu
\eea

The dual magnetic superpotential, by adding the mass term 
(\ref{superpotential}) for $Q$ and
$\widetilde{Q}$
in the electric theory, which is equal to put a linear term in $M$ in
the dual magnetic theory, is given by \cite{Ahn07-3}
\bea
W_{dual} = \left(M q \widetilde{q} + f \widetilde{X}' \widetilde{q} + 
\widetilde{f} q
X' + \Phi' f \widetilde{f} \right) + m M.
\nonu
\eea
Here $q$ and $\widetilde{q}$ are fundamental and antifundamental for
the gauge group index respectively.
Then, $q \widetilde{q}$ has rank $\widetilde{N}_c$ while $m$ has a
rank $N_f$.  Therefore, the F-term condition, the derivative the 
superpotential $W_{dual}$ with respect to $M$, cannot be satisfied 
if the rank $N_f$ exceeds $\widetilde{N}_c$. 
This is so-called rank condition and the supersymmetry is broken.    
Other F-term equations are satisfied by taking the vacuum expectation 
values of $f, \widetilde{f}, X'$ and $\widetilde{X}'$ to vanish.

Then the solutions can be written as follows:
\bea
<q >  & = &  \left(
\begin{array}{c}
\sqrt{m} e^{\phi} {\bf 1}_{\widetilde{N}_c}  \\
0
\end{array}
\right),  
< \widetilde{q}> =
 \left(
\begin{array}{cc}
\sqrt{m} e^{-\phi}  {\bf 1}_{\widetilde{N}_c}   &
0
\end{array}
\right), 
<M>  =
 \left(
\begin{array}{cc}
0  & 0 
 \\
0 & M_0  {\bf 1}_{N_f-\widetilde{N}_c} 
\end{array}
\right),
\nonu \\
<f> & = & <\widetilde{f}> = <X'> = <\widetilde{X}'>= 0.
\nonu
\eea
As we did in previous case, one can analyze the one loop 
computation by expanding the fields around the vacua and it will
lead to the fact that states are stable by realizing the mass of 
$m_{M_0}^2$ positive.


Then the minimal energy supersymmetry breaking brane configuration is
shown in Figure 5B.
If we ignore the $NS5_R$-brane, $N_c'$ D4-branes,
$N_f'$ D6-branes, $NS5_R'$-brane, $N_c''$ D4-branes and $N_f''$ 
D6-branes(detaching these from Figure 5B), 
as observed already, 
then this brane configuration 
is the minimal energy supersymmetry breaking brane
configuration
for the ${\cal N}=1$ SQCD with the magnetic gauge group 
$SU(\widetilde{N}_c=N_f-N_c)$ with
$N_f$ massive flavors \cite{OO,FGU,BGHSS}.  

The nonsupersymmetric minimal energy brane configuration Figure 5B
with a replacement $N_f$ D6-branes by 
the NS5'-brane(neglecting  the
$NS5_R'$-brane, $N_f''$ D6-branes and $N_c''$ D4-branes 
and $N_f'$ D6-branes)
leads to 
the Figure 4B of \cite{Ahn07-6}.

When we move the $NS5_L'$-brane in Figure 1 to the left all the way past the 
$NS5_L$-brane, then  
one arrives at the magnetic brane configuration similar to the Figure 5.
The only difference is that the $N_f$ D6-branes are located at the right
hand side of the $NS5_L$-brane.
Then this nonsupersymmetric minimal energy brane configuration 
 with a replacement $N_f$ D6-branes by 
the NS5'-brane(by neglecting  the
$NS5_R'$-brane, $N_f''$ D6-branes and $N_c''$ D4-branes) leads to 
the Figure 5B of \cite{Ahn07-6} with a reflection with respect to the
NS5-brane
and an extra rotation of
$NS5'_L$-brane by $\frac{\pi}{2}$ angle.

Starting with $NS5_L'$-$NS5_L$-$NS5_R'$-$NS5_R$ branes configuration,
as in the footnote \ref{rotation}, 
by moving the $NS5_L$-brane to the left, 
one gets 
the nonsupersymmetric minimal energy brane configuration which is
exactly the Figure 2 with a reflection with
respect to the NS5-brane. 
Furthermore, by 
moving the $NS5_R'$-brane to the right, 
one gets 
the nonsupersymmetric minimal energy brane configuration which is
exactly the new figure in previous paragraph with a reflection with
respect to the $NS5_L$-brane. 

The 
corresponding magnetic M5-brane configuration, in a 
background space of $x t = 
v^{N_f'+N_f''} \prod_{k=1}^{N_f} (v -e_k)$ where this four dimensional space
replaces (45610) directions,
is described by 
\bea
&& t^4 + ( v^{\widetilde{N}_c }  + \cdots ) t^3 + 
\left[ v^{N_c'}(v-m)^{N_f} + \cdots \right] t^2 \nonu \\
&& + 
\left[ v^{N_c''+N_f'}(v-m)^{2N_f} + \cdots \right] t + (v-m)^{3N_f} v^{2N_f'+N_f''}
=0.
\nonu
\eea 

From this curve  
of quartic equation for $t$ above, the asymptotic regions 
for four NS5-branes can be classified 
as follows:

1. $v \rightarrow \infty$ limit implies
\bea
w \rightarrow 0, && \quad y \sim    v^{N_c'-\widetilde{N}_c+N_f} + \cdots \quad
\mbox{$NS_L$ asymptotic region},   
\nonu \\
w \rightarrow 0, && \quad y \sim    
v^{N_f+N_f'+N_c''-N_c'} + \cdots \quad
\mbox{$NS_R$ asymptotic region}.
\nonu  
\eea

2.  $w \rightarrow \infty$ limit implies
\bea
v  \rightarrow    m, && \quad 
y \sim  w^{\widetilde{N}_c}
 +\cdots
\quad \mbox{$NS_{L}'$ asymptotic region}, 
\nonu
\\
v  \rightarrow   m, && \quad  
y \sim w^{N_f+N_f'+N_f''-N_c''}
+\cdots
\quad \mbox{$NS_{R}'$ asymptotic region}. 
\nonu
\eea

\subsection{Magnetic theories 
for the multiple product gauge groups}

Now one can generalize the method for the triple product gauge groups
to the finite $n$-multiple product gauge groups characterized by 
\cite{BH,AT97}
\bea
SU(N_{c,1}) \times \cdots
\times SU(N_{c,n})
\nonu
\eea
with the matter, 
the $(n-1)$ bifundametals $({\bf \Box_1, \overline{\Box}_2, 1, \cdots,  1})$,
$\cdots$, and $({\bf 1, \cdots, 1, \Box_{n-1}, \overline{\Box}_{n}})$, their
complex conjugate $(n-1)$ fields $({\bf \overline{\Box}_1, \Box_2, 1, \cdots, 1})$,
$\cdots$, 
and $({\bf 1, \cdots, 1, \overline{\Box}_{n-1}, \Box}_n)$, linking the
gauge groups together,
$n$-fundamentals $({\bf \Box_1, 1, \cdots, 1})$, $\cdots$, and 
$({\bf 1, \cdots,  1, \Box}_n)$, and $n$-antifundamentals  
$({\bf \overline{\Box}_1, 1, \cdots, 1})$, $\cdots$, and 
$({\bf 1, \cdots,  1, \overline{\Box}_n})$.
Then the mass-deformed superpotential can be written as
$
W_{elec} = \sum_{i=1}^n m_i Q_i \widetilde{Q}_i$. 
The brane configuration can be constructed similarly and 
any two neighboring NS-branes are perpendicular to each other. 

There exist $(2n-2)$ magnetic theories and they can be classified as
four cases as follows.

$\bullet$ Case 1 

When the Seiberg dual is taken for the first gauge group factor
 by
assuming that $\Lambda_1 >> \Lambda_i$ where $i=2, \cdots, n$, 
one follows the procedure given in the subsection 2.5.
The gauge group is 
\bea
SU(\widetilde{N}_{c,1} \equiv N_{f,1} +N_{c,2}-N_{c,1}) \times 
SU(N_{c,2}) \times \cdots \times SU(N_{c,n})
\nonu
\eea
and the matter contents are given by 
the dual quarks $q_1$ $({\bf \Box_1, 1, \cdots,
1})$ 
and $\widetilde{q}_1$ in the 
representation $({\bf \overline{\Box}_1, 1, \cdots, 1})$ 
as well as $(n-1)$ 
quarks $Q_i$ and
$\widetilde{Q}_i$ where $i=2, \cdots, n$, 
the bifundamentals $f_1$ in the representation 
 $({\bf \Box_1, \overline{\Box}_2, 1, \cdots,  1})$ under the dual gauge group,
$\cdots$, and $\widetilde{f}_1$ in the representation
$({\bf \overline{\Box}_1, \Box_2, 1, \cdots, 1})$ under the dual gauge
group  in
addition to $(n-2)$ bifundamentals $G_i$ and $\widetilde{G}_i$, and
various gauge singlets $X_2, \widetilde{X}_2, M_1$ and $\Phi_2$.
The corresponding brane configuration can be 
obtained similarly and 
the extra $(n-3)$ NS-branes, $(n-3)$ sets of D6-branes and $(n-3)$
sets of D4-branes  
are present at the right hand side of the $NS5_R'$-brane
of Figure 5.
The magnetic superpotential can be written as
$
W_{dual} = \left(M_1 q_1 \widetilde{q}_1 + f_1 \widetilde{X}_2 \widetilde{q}_1 + 
\widetilde{f}_1 q_1
X_2 + \Phi_2 f_1 \widetilde{f}_1 \right) + m_1 M_1$.
By computing the contribution for the one loop as in the subsection
2.5, 
the vacua are stable and the asymptotic behavior of $(n+1)$ NS-branes
can be obtained also.  

$\bullet$ Case 2 

When the Seiberg dual is taken for the last gauge group factor by
assuming that $\Lambda_n >> \Lambda_i$ where $i=1,2, \cdots, (n-1)$, 
one follows the procedure given in the subsection 2.2.
The gauge group is given by
\bea
SU(N_{c,1}) \times \cdots \times 
SU(N_{c,n-1}) \times SU(\widetilde{N}_{c,n} \equiv N_{f,n} +N_{c,n-1}-N_{c,n}).
\nonu
\eea
The corresponding brane configuration can be 
obtained similarly and 
the extra $(n-3)$ NS-branes, $(n-3)$ sets of D6-branes and $(n-3)$
sets of D4-branes  
are present at the left hand side of the $NS5_L$-brane
of Figure 2.
The magnetic superpotential can be written as
$
W_{dual} = \left(M_n q_n \widetilde{q}_n + g_{n-1} X_{n-1} q_n + 
\widetilde{g}_{n-1} \widetilde{q}_n
\widetilde{X}_{n-1} + \Phi_{n-1} g_{n-1} \widetilde{g}_{n-1} \right) + m_n M_n$.

$\bullet$ Case 3

When the Seiberg dual is taken for the middle gauge group factor
by
assuming that $\Lambda_i >> \Lambda_j$ where $j=1,2, \cdots, i-1, i+1,
\cdots, n$, 
one follows the procedure given in the subsection 2.3.
The gauge group is given by
\bea
\cdots \times SU(N_{c,i-1}) \times 
SU(\widetilde{N}_{c,i} \equiv N_{f,i}+N_{c,i+1}+N_{c,i-1}-N_{c,i}) 
\times SU(N_{c,i+1}) \times \cdots
\nonu
\eea
where $ 2 \leq i \leq n-1$ implying that the number of 
possible magnetic gauge group is $(n-2)$.
The corresponding brane configuration can be 
obtained similarly and 
the extra $(i-2)$ NS-branes, $(i-2)$ sets of D6-branes and $(i-2)$
sets of D4-branes  
are present at the left hand side of the $NS5_L$-brane
and the extra $(n-i-1)$ NS-branes, $(n-i-1)$ sets of D6-branes and $(n-i-1)$
sets of D4-branes  are present at the right hand side of the $NS5_R'$-brane
of Figure 3.
The magnetic superpotential can be written as
$
W_{dual} = \left(M_{i} q_i \widetilde{q}_i + 
g_{i} \widetilde{X}_{i+1} \widetilde{q}_i + 
\widetilde{g}_{i} q_i
X_{i+1} + \Phi_{i+1} g_{i} \widetilde{g}_{i+1} \right) + m_i M_{i}$.

$\bullet$ Case 4

When the Seiberg dual is taken for the middle gauge group factor with
different brane motion
by
assuming that $\Lambda_i >> \Lambda_j$ where $j=1,2, \cdots, i-1, i+1,
\cdots, n$, 
one follows the procedure given in the subsection 2.4.
The gauge group is given by
\bea
\cdots \times SU(N_{c,i-1}) \times 
SU(\widetilde{N}_{c,i} \equiv N_{f,i}+N_{c,i+1}+N_{c,i-1}-N_{c,i}) 
\times SU(N_{c,i+1}) \times \cdots
\nonu
\eea
where $ 2 \leq i \leq n-1$ implying that the number of 
possible magnetic gauge group is given by $(n-2)$.
The corresponding brane configuration can be 
obtained similarly and 
the extra $(i-2)$ NS-branes, $(i-2)$ sets of D6-branes and $(i-2)$
sets of D4-branes  
are present at the left hand side of the $NS5_L$-brane
and the extra $(n-i-1)$ NS-branes, $(n-i-1)$ sets of D6-branes and $(n-i-1)$
sets of D4-branes  are present at the right hand side of the $NS5_R'$-brane
of Figure 4.
The magnetic superpotential can be written as
$
W_{dual} = \left(M_{i} q_i \widetilde{q}_i + f_{i-1} X_{i-1} q_i + 
\widetilde{f}_{i-1} \widetilde{q}_i
\widetilde{X}_{i-1} + \Phi_{i-1} f_{i-1} \widetilde{f}_{i-1} \right) + m_i M_{i}$.

\section{Nonsupersymmetric meta-stable brane configurations
of $Sp(N_c) \times SO(2N_c') \times Sp(N_c'')$ and its multiple
product gauge theories }


\subsection{Electric theory}

%
%
%
%
%

The gauge group and matter contents are summarized as follows:
\bea
 & \mbox{ gauge group}:& \;\;\;\;\;   Sp(N_c) \times SO(2N_c') \times
 Sp(N_c'')  \nonu
\\
\mbox{matter}: 
 &Q_f & \;\;\;\;\;\;\;\;\; 
\;\;\;\;\;\;\;\;\;\;\;\; {(\bf \Box, 1, 1)} 
\;\;\;\;\; (f=1,  \cdots, 2N_f) 
\nonu \\
 &Q'_{f'} & \;\;\;\;\;\;\;\;\;\;
\;\;\;\;\;\;\;\;\;\;\; {(\bf 1, \Box, 1) }
\;\;\;\;\; (f'=1,  \cdots, 2N_f') 
\nonu \\
 &Q''_{f''} & \;\;\;\;\;\;\;\;\;\; 
\;\;\;\;\;\;\;\;\;\;\; {(\bf 1, 1, \Box) } 
\;\;\;\;\; (f''=1,  \cdots, 2N_f'')
\nonu \\
 &F & \;\;\;\;\;\;\;\;\;\;\;\;\;\;\;\;\;\;\; 
\;\; {(\bf \Box, \Box,1) } 
\nonu \\
 &G & \;\;\;\;\;\;\;\;\;\;\;\;\;\;\;\;\;\;\; 
\;\; {(\bf 1, \Box, \Box) } 
\nonu 
\eea

In the electric theory \footnote{For $Sp(N_c) \times SO(2N_c'+1)
  \times Sp(N_c'')$ gauge theory, the analysis can be done similarily
and we do not present here.}, 
since there exist $2N_f$ quarks $Q$, and 
one bifundamental field $F$(which will give
rise to the contribution of $2N_c'$), the coefficient of the beta function
of the first gauge group factor is
$
b_{Sp(N_c)} = 3(2N_c+2) -2N_f-2N_c'$.
Similarly,
since there exist $2N_f'$ quarks $Q'$, 
one bifundamental field $F$(which will give
rise to the contribution of $2N_c$), and
one bifundamental field $G$(which will give
rise to the contribution of $2N_c''$),
the coefficient of the beta function
of the second gauge group factor is
$
b_{SO(2N_c')} = 3(2N_c'-2)-2N_f'-2N_c-2N_c''$.
Finally, since there exist $2N_f''$ quarks $Q''$, and 
one bifundamental field $G$(which will give
rise to the contribution of $2N_c'$), the coefficient of the beta function
of the third gauge group factor is
$
b_{Sp(N_c'')} = 3(2N_c''+2) -2N_f''-2N_c'$.

The anomaly free global symmetry contains $ SU(2N_f) \times 
SU(2N_f') \times SU(2N_f'') \times U(1)_R$  and let us denote the
strong coupling scales for $Sp(N_c)$ as $\Lambda_1$, for $SO(2N_c')$
as $\Lambda_2$ and for $Sp(N_c'')$
as $\Lambda_3$.  
The theory is asymptotically free when $b_{Sp(N_c)} >
0$ for the $Sp(N_c)$ gauge theory,  when 
$b_{SO(2N_c')} > 0$ for the $SO(2N_c')$ gauge theory, and 
when 
$b_{Sp(N_c'')} > 0$ for the $Sp(N_c'')$ gauge theory. 

The classical electric superpotential can be obtained    
from 
(\ref{superpotential-1})
by orientifolding 
\bea
W_{elec} & = & \left( \mu A^2 + \lambda Q A Q + F A F +
\mu' A'^2 + \lambda' Q' A' Q' + F A' F +
G A' G \right. \nonu \\
&+& \left.
 \mu'' A''^2 + \lambda'' Q'' A'' Q'' + G A'' G \right)
+ m Q Q + m' Q' Q' + m'' Q'' Q''
\label{superpotentialelectric}
\eea
where the coefficient functions $\mu, \mu', \mu'', \lambda, \lambda'$
and $\lambda''$ are given by six rotation angles.
Here the adjoint field for $Sp(N_c)$ gauge group is denoted by $A$, 
the adjoint field for $SO(2N_c')$ gauge group is denoted by $A'$, and 
the adjoint field for $Sp(N_c'')$ gauge group is denoted by $A''$.
The mass terms of these adjoint fields are related to the 
rotation angles of NS-branes in type IIA brane configuration. 
The couplings of flavors with these adjoint
fields are related also to the 
rotation angles of NS-branes as well as the rotation angles
of D6-branes.
We add the mass terms for each flavor.  
Setting the fields $Q'', G$, and $A''$ to zero, 
the superpotential becomes the one described in \cite{Ahn07-2}.

After integrating out the adjoint fields $A, A'$ and $A''$, 
this superpotential (\ref{superpotentialelectric}) 
at the particular orientations for branes, i.e., the case where 
any two neighboring NS-branes are perpendicular to each other, 
will reduce to the last three mass-deformed terms 
since the coefficient functions 
$\frac{1}{\mu}, \frac{1}{\mu'}$
and $\frac{1}{\mu''}$ vanish at this particular rotation angles for branes.
Then the classical superpotential by deforming this theory by adding the
mass terms for the quarks $Q(Q')[Q'']$ 
is given by
\bea
W_{elec} = m Q Q + m' Q' Q' + m'' Q'' Q''.
\label{superpotential1}
\eea 

The type IIA brane configuration for this mass-deformed theory 
can be described by as follows. 
The $2N_c$-color 
D4-branes (01236) are suspended between the $NS5_L$-brane (012345) 
and the $NS5_L'$-brane (012389) along $x^6$
direction,
together with $N_f$ D6-branes (0123789) 
which are parallel to $NS5_L'$-brane and have 
nonzero $+v$ direction(and their mirrors).
The $NS5_R$-brane 
is located at the right hand side of
the $NS5_L'$-brane along the $x^6$ direction and 
there exist $2N_c'$-color D4-branes
suspended 
between them, with  $N_f'$ D6-branes which 
have nonzero $+v$ direction(and their mirrors). 
Moreover, 
the $NS5_R'$-brane 
is located at the right hand side of
the $NS5_R$-brane along the $x^6$ direction and there 
exist $2N_c''$-color D4-branes
suspended 
between them, with  
$N_f''$ D6-branes which have nonzero $+v$ 
direction(and their mirrors).

One summarizes the brane configuration as follows:

$\bullet$
$NS5_L(NS5_R)$-brane in (012345) directions. 

$\bullet$ 
$NS5_L'(NS5_R')$-brane in (012389) directions.

$\bullet$
$2N_c(2N_c')[2N_c'']$-color D4-branes in (01236) directions. 
  
$\bullet$
$2N_f(2N_f')[2N_f'']$ D6-branes in (0123789) directions. 

$\bullet$
$O4^{\pm}$-planes in (01236) directions

Now we draw this electric brane configuration in Figure 6 and we put
the coincident $N_f(N_f')[N_f'']$ D6-branes in the nonzero $+v$
direction(and their mirrors).
The quarks $Q(Q')[Q'']$
correspond to strings
between the
$2N_c(2N_c')[2N_c'']$-color D4-branes with $2N_f(2N_f')[2N_f'']$ D6-branes.
The bifundamentals $F(G)$ correspond to   strings 
stretching between the
$2N_c(2N_c')$-color D4-branes with $2N_c'(2N_c'')$-color D4-branes. 

\begin{figure}[ht]
   \epsfxsize=3.0in 
\centerline{\epsffile{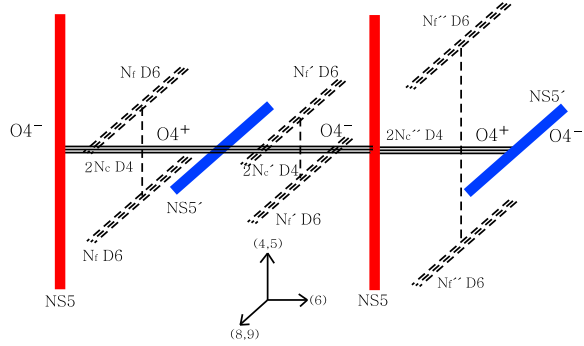}}
   \caption[FIG. \arabic{figure}.]{ 
The ${\cal N}=1$ supersymmetric electric brane configuration with
$Sp(N_c) \times SO(2N_c') \times Sp(N_c'')$ gauge group with
flavors $Q(Q')[Q'']$  
for each gauge group and bifundamentals $F(G)$.
}
\end{figure}


\subsection{Magnetic theory  with dual for third gauge group}

Let us take the Seiberg dual for the third gauge group factor $Sp(N_c'')$ 
while keeping the first and the second gauge group factors 
$Sp(N_c)$ and $SO(2N_c')$ untouched. 
Suppose that
$\Lambda_3 >> \Lambda_1, \Lambda_2$. 
This can be done by  
type IIA string theory side via brane motion. 
After we move  the $NS5_R$-brane in Figure 6 to the right all the way past
the $NS5_R'$-brane, 
we arrives at the Figure 7A.
Then the linking number 
of $NS5_R$-brane from Figure 7A is given by 
$
L_5 =\frac{(2N_f'')}{2}-1-(1)-2\widetilde{N}_c''$.
Note that $O4^{+}$-plane with RR charge $+1$ realizes a symplectic
gauge group while $O4^{-}$-plane with RR charge $-1$ does an
orthogonal gauge group. 
Originally, it was 
$
L_5=-\frac{(2N_f'')}{2}+ 1 -(-1)
+2N_c''-2N_c'
$
from Figure 6 before the brane motion.
Therefore, by the linking number conservation and equating these two
$L_5$'s each other, 
we are left with the number of colors in the magnetic
theory 
$
\widetilde{N}_c'' = N_f'' + N_c'-N_c'' -2$.
Note the dependence on $N_c'$ here and the constant piece $-2$ comes
from the presence of O4-plane.

Let us draw this magnetic brane configuration in Figure 7A
and we put
the coincident $N_f''$ D6-branes in 
the nonzero $+v$ direction(and its mirrors) as well as the massless $Q(Q')$.
If we ignore the $NS5_L$-brane, $2N_c$ D4-branes, $2N_f$ 
D6-branes, the $NS5_L'$-brane, $2N_c'$ D4-branes and $2N_f'$ 
D6-branes(detaching these
branes from Figure 7A), 
then this brane configuration 
leads to the standard ${\cal N}=1$ SQCD with the magnetic gauge group 
$Sp(\widetilde{N}_c''=N_f''-N_c''-2)$ with
$N_f''$ massive flavors \cite{FGU,Ahn06-1}.

\begin{figure}[ht]
   \epsfxsize=4.0in 
\centerline{\epsffile{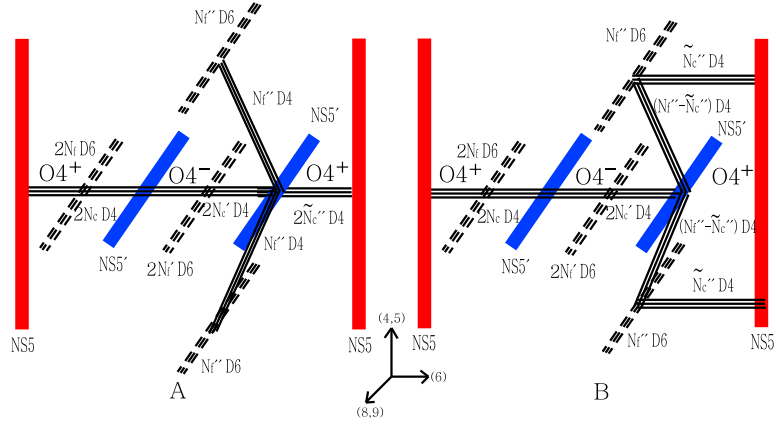}}
   \caption[FIG. \arabic{figure}.]{ 
The ${\cal N}=1$ supersymmetric magnetic brane configuration with
$Sp(N_c) \times SO(2N_c') \times Sp(\widetilde{N}_c'')$ gauge group
with flavors $Q(Q')[q'']$
for each gauge group, the  bifundamentals $F(g)$, and gauge singlets 
in Figure 7A. In
Figure 7B, the nonsupersymmetric minimal energy brane configuration
with the same gauge group and matter contents above 
for massless  $Q(Q')$ is given. 
 }
\end{figure}

In the dual theory, since 
there exist  $2N_f''$ fundamental fields $q''$, and  
one bifundamental field  
$g$(which will give rise to the contribution of $2N_c'$),
the coefficient of the beta function 
is
$
b_{Sp(\widetilde{N}_c'')}^{mag}= 3(2\widetilde{N}_c''+2)-2N_f''-2N_c'
$
and since there are  $2N_f'$ fundamental fields $Q'$,
$2\widetilde{N}_c''$ fundamental 
fields $g$ for the second factor,
 $2N_c$ fundamental 
fields $F$ for the first factor,
an antisymmetric tensor $\Phi'$(which will contribute to $(2N_c'-2)$), 
and $2N_f''$ fields $X'$, 
the coefficient of the beta function
 is
$
b_{SO(2N_c')}^{mag} = 3(2N_c' -2)-2N_f'-2\widetilde{N}_c''-
2N_c-(2N_c' -2)-2N_f''$.
Since there exist $2N_f$ quarks $Q$, and  
one bifundamental field $F$(which will give
rise to the contribution of $2N_c'$), the coefficient of the beta function
of the first gauge group factor is
$
b_{Sp(N_c)}^{mag} = 3(2N_c+2) -2N_f-2N_c'=b_{Sp(N_c)}$.
Note that the $Sp(N_c)$ fields in the magnetic theory are the same as 
those of the electric theory.

Therefore, both $Sp(\widetilde{N}_c'')$,  
$SO(2N_c')$, and $Sp(N_c)$ gauge couplings are IR free
by requiring the negativeness of the coefficients of beta function.
One can rely on the perturbative calculations at low energy 
for this magnetic IR free region $b_{Sp(\widetilde{N}_c'')}^{mag} < 0$, 
$b_{SO(2N_c')}^{mag} < 0$, and $b_{Sp(N_c)}^{mag} < 0$.
Note that the $SO(2N_c')$ fields in the magnetic theory 
are different from those of the electric theory.
Since $b_{SO(2N_c')}-b_{SO(2N_c')}^{mag} > 0$, $SO(2N_c')$ is more
asymptotically free than $SO(2N_c')^{mag}$.
Neglecting the $SO(2N_c')$ and $Sp(N_c)$ dynamics, 
the magnetic $Sp(\widetilde{N}_c'')$
is IR free when 
\bea
N_c'' +2 -N_c' < N_f'' < \frac{3}{2}(N_c'' +1)-N_c'.
\nonu
\eea

Then one can summarize the gauge group and matter contents where there
are additional fields  $X', M''$ and $\Phi'$ as follows:
\bea
 & \mbox{gauge group}:& \;\;\;\;\;   Sp(N_c) \times SO(2N_c') \times
 Sp(\widetilde{N}_c'')  \nonu
\\
\mbox{matter}: 
 &Q_f & \;\;\;\;\;\;\;\;\; 
\;\;\;\;\;\;\;\;\;\;\;\; {(\bf \Box, 1, 1)} 
\;\;\;\;\; (f=1,  \cdots, 2N_f) 
\nonu \\
 &Q'_{f'} & \;\;\;\;\;\;\;\;\;\;
\;\;\;\;\;\;\;\;\;\;\; {(\bf 1, \Box, 1) }
\;\;\;\;\; (f'=1,  \cdots, 2N_f') 
\nonu \\
 &q''_{f''} & \;\;\;\;\;\;\;\;\;\; 
\;\;\;\;\;\;\;\;\;\;\; {(\bf 1, 1, \Box) } 
\;\;\;\;\; (f''=1,  \cdots, 2N_f'')
\nonu \\
 &F & \;\;\;\;\;\;\;\;\;\;\;\;\;\;\;\;\;\;\; 
\;\; {(\bf \Box, \Box,1) } 
\nonu \\
 &g & \;\;\;\;\;\;\;\;\;\;\;\;\;\;\;\;\;\;\; 
\;\; {(\bf 1, \Box, \Box) } 
\nonu \\
 &(X_{n''}' \equiv) G Q'' & \;\;\;\;\;\;\;\;\;\;\;\;\;\;\;\;\;\;\; 
\;\; {(\bf 1, \Box, 1) } 
\;\;\;\;\; (n''=1,  \cdots, 2N_f'')
\nonu \\
 &(M_{f'', g''}'' \equiv) Q'' Q'' & \;\;\;\;\;\;\;\;\;\;\;\;\;\;\;\;\;\;\; 
\;\; {(\bf 1, 1, 1) } 
\;\;\;\;\; (f'', g''=1,  \cdots, 2N_f'')
\nonu \\
 &(\Phi' \equiv) G G & \;\;\;\;\;\;\;\;\;\;\;\;\;\;\;\;\;\;\; 
\;\; {(\bf 1, asymm, 1) } 
\nonu
\eea

The dual magnetic  tree level superpotential, by adding the mass term
for the $Q''$ in electric theory corresponding to add
a linear term in $M''$ in dual magnetic theory, 
is given by
\bea
W_{dual} = \left( M'' q'' q'' + g X' q'' + \Phi' g g \right)
+ m'' M''.
\nonu
\eea
Here $q''$ is fundamental for
the gauge group index \cite{Ahn07-2}.
Then, $q'' q''$ has rank $2\widetilde{N}_c''$ while $m''$ has a
rank $2N_f''$.  Therefore, the F-term condition, the derivative the 
superpotential $W_{dual}$ with respect to $M''$, cannot be satisfied 
if the $2N_f''$ exceeds $2\widetilde{N}_c''$. 
This is so-called rank condition and the 
supersymmetry is broken.    

More explicitly, 
the classical moduli space of vacua can be obtained from F-term
equations.
From the F-terms $F_{q''}$ and $F_{M''}$, one gets
$M'' q'' + g X'  =0=   q'' q''  +
m$.
Similarly, one obtains 
$\Phi' g +  X' q'' =0 = g g $ from 
the F-terms $F_{g}$ and $F_{\Phi'}$. 
Moreover, there is a relation 
$ q'' g =0$ from the F-term $F_{X'}$.
Then, one obtains 
the following solutions 
\bea
< q''> =  \left(
\begin{array}{c}
i \sqrt{m}  {\bf 1}_{2\widetilde{N}_c''}  \\
0
\end{array}
\right),  
<M''>  =
 \left(
\begin{array}{cc}
0  & 0 
 \\
0 & M_0''  {\bf 1}_{(N_f''-\widetilde{N}_c'')} \otimes i \sigma^2 
\end{array}
\right), 
<g>=0= <X'>
\nonu
\eea
where $M_0''  {\bf 1}_{(N_f''-\widetilde{N}_c'')} \otimes i \sigma^2$ 
is an arbitrary 
$2(N_f''-\widetilde{N}_c'') \times 2(N_f''-\widetilde{N}_c'')$ antisymmetric 
matrix and the zeros of 
$<q''>$  are $2(N_f''-\widetilde{N}_c'') \times 2\widetilde{N}_c'' $
zero matrices. 
Similarly, the zeros
of $2N_f'' \times 2N_f''$  matrix $M''$ are assumed also.

Then the superpotential becomes 
\bea
W_{dual}^{fluct}  =   M_0'' \left( \delta \varphi \;  
 \delta \varphi  + m \right) +
  \delta Z^T \; \delta \varphi   \; i \sqrt{m}
+  \delta Z \; i \sqrt{m}  \; 
\delta \varphi  
\nonu
\eea
by following the fluctuation for the fields  
\bea
q''   = 
\left(
\begin{array}{c}
i \sqrt{m}  {\bf 1}_{2\widetilde{N}_c''} +
(\delta \chi_{A} + \delta \chi_{S})
 {\bf 1}_{\widetilde{N}_c''} \otimes i \sigma^2 \nonu \\
\delta \varphi
\end{array}
\right), 
\qquad
M'' =
 \left(
\begin{array}{cc}
\delta Y  & \delta Z^T
 \\
-\delta Z & M_0''  {\bf 1}_{(N_f''-\widetilde{N}_c'')} \otimes i \sigma^2
\end{array}
\right)
\nonu
\eea
as well as the fluctuations for $g$ and $X'$.
There are two kinds of terms, 
the vacuum of $<q''>$ multiplied by 
$\delta X' \delta g$ and 
the vacuum of $<\Phi'>$ multiplied by $\delta g 
\delta g$.
By redefining these as 
$\delta \hat{X'} \delta \hat{g}$ and $\delta
\hat{g}
\delta \hat{g}$ respectively, they do not enter the 
contributions for the one loop result. 
At one loop, the effective potential $V_{eff}^{(1)}$ for $M_0''$ 
can be obtained 
from this superpotential which consists of 
the matrices $M$ and $N$ of \cite{Shih} 
where the defining function ${\cal F}(v^2)$ can be
computed.
Using the equation (2.14) of \cite{Shih} 
of $m_{M_0''}^2$ and ${\cal F}(v^2)$, one gets 
that $m_{M_0''}^2$ will contain $(\log 4 -1) > 0$.
This implies that these vacua are stable.


Then the minimal energy supersymmetry breaking brane configuration is
shown in Figure 7B.
If we ignore the $NS5_L$-brane, $2N_c$ D4-branes, $2N_f$ 
D6-branes, the $NS5_L'$-brane, $2N_c'$ D4-branes and $2N_f'$ 
D6-branes(detaching these
branes from Figure 7B), 
then this brane configuration 
corresponds to  the minimal energy supersymmetry breaking brane
configuration
for the ${\cal N}=1$ SQCD with the magnetic gauge group 
$Sp(\widetilde{N}_c'')$ with
$N_f''$ massive flavors \cite{FGU,Ahn06-1}.

The nonsupersymmetric minimal energy brane configuration Figure 7B
with vanishing $2N_f'$ D6-branes
leads to the Figure 3 of \cite{Ahn07-2} if we ignore the
$NS5_L$-brane, $2N_f$ D6-branes and $2N_c$ D4-branes.
Moreover, this Figure 7B with a replacement $N_f''$ D6-branes by 
the NS5'-brane(neglecting  the
$NS5_L$-brane, $2N_f$ D6-branes and $2N_c$ D4-branes)
will become the Figure 7B of \cite{Ahn07-6}
with a reflection with respect to the $NS5_L$-brane and a
rotation of $NS5_R$-brane by $\frac{\pi}{2}$ angle.


Starting with $NS5_L'$-$NS5_L$-$NS5_R'$-$NS5_R$ branes configuration,
as in the footnote \ref{rotation}, 
by moving the $NS5_R$-brane to the left, 
one gets 
the nonsupersymmetric minimal energy brane configuration which is
exactly the Figure 10B(which will appear later) with a reflection with
respect to the NS5-brane. 
Furthermore, by 
moving the $NS5_L'$-brane to the right, 
one gets 
the nonsupersymmetric minimal energy brane configuration which is
exactly the new figure in previous paragraph with a reflection with
respect to the $NS5_L$-brane. 

After lifting 
the type IIA description we explained so far to M-theory, the 
corresponding magnetic M5-brane configuration, in a background space 
of $x t = 
v^{2N_f+2N_f'} 
\prod_{k=1}^{N_f''} (v^2 -e_k^2)$ where this four dimensional space
replaces (45610) directions,
is characterized by \cite{LLL97}
\bea
& & t^4 + ( v^{2N_c+2 }  + \cdots ) t^3 + 
( v^{2N_c'+2N_f} + \cdots) t^2  \nonu \\
&& +  (v^{2\widetilde{N}_c''+2+4N_f+2N_f'} + 
\cdots ) (v^2-m^2)^{N_f''} t + v^{6N_f+4N_f'}
(v^2 -m^2)^{2N_f''} =0
\nonu
\eea
where 
we ignored both the lower power terms in $v$ 
and the scales for the gauge groups for simplicity.

From this curve 
of quartic equation for $t$ above, the asymptotic regions 
can be classified 
as follows:

1. $v \rightarrow \infty$ limit implies
\bea
w \rightarrow 0, && \quad y \sim    v^{2N_c+2} + \cdots \quad
\mbox{$NS_L$ asymptotic region},   
\nonu \\
w \rightarrow 0, && \quad y \sim    
v^{2N_f+2N_f'+2N_f''-2\widetilde{N}_c''-2} + \cdots \quad
\mbox{$NS_R$ asymptotic region}.
\nonu  
\eea

2.  $w \rightarrow \infty$ limit implies
\bea
v  \rightarrow    m, && \quad 
y \sim  w^{2N_c'-2N_c-2+2N_f}
 +\cdots
\quad \mbox{$NS_{L}'$ asymptotic region}, 
\nonu
\\
v  \rightarrow   m, && \quad  
y \sim w^{2\widetilde{N}_c''+2-2N_c'+2N_f+2N_f'+2N_f''}
+\cdots
\quad \mbox{$NS_{R}'$ asymptotic region}. 
\nonu
\eea
 

\subsection{Magnetic theory  with dual for second gauge group}

Let us take the Seiberg dual for the second gauge group factor $SO(2N_c')$ 
while keeping the first and the third gauge group factors 
$Sp(N_c)$ and $Sp(N_c'')$ untouched. 
Suppose that
$\Lambda_2 >> \Lambda_1, \Lambda_3$. 
After we move  the $NS5_R$-brane in Figure 6
to the left all the way past
the $NS5_L'$-brane, the linking number 
of $NS5_R$-brane from Figure 8A is given by 
$
L_5 =-\frac{(2N_f')}{2}-1-(1)+2\widetilde{N}_c'-2N_c$.
Originally, it was 
$
L_5=\frac{(2N_f')}{2}+ 1 -(-1)
+2N_c''-2N_c'
$
from Figure 6 in electric theory before the brane motion.
Therefore, by the linking number conservation and equating these two
$L_5$'s each other, 
we are left with the number of colors in the magnetic
theory 
$
\widetilde{N}_c' = N_f' + N_c''+N_c-N_c' +2$.
Here the dependence on $N_c$ and $N_c''$ arises.

Let us draw this magnetic brane configuration in Figure 8A 
and we put
the coincident $N_f'$ D6-branes in the nonzero $v$ direction(and its mirrors).
For the $N_f$ and $N_f''$ D6-branes we consider the massless flavors
and then they are located at $v=0$.
If we ignore the $NS5_L$-brane, $2N_c$ D4-branes, $2N_f$ 
D6-branes, the $NS5_R'$-brane, $2N_c''$ D4-branes and $2N_f''$ 
D6-branes(detaching these
branes from Figure 8A), 
then this brane configuration 
leads to the standard ${\cal N}=1$ SQCD with the magnetic gauge group 
$SO(2\widetilde{N}_c'=2(N_f'+N_c''+N_c-N_c'+2))$ with
$N_f'$ massive flavors \cite{FGU,Ahn06-1}.

\begin{figure}[ht]
   \epsfxsize=4.0in 
\centerline{\epsffile{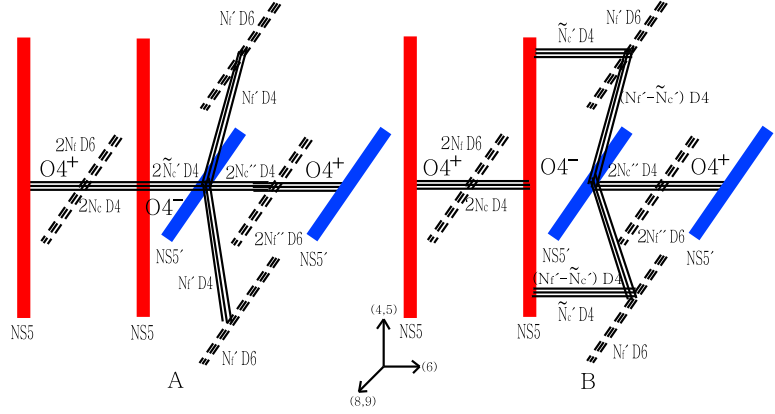}}
   \caption[FIG. \arabic{figure}.]{ 
The ${\cal N}=1$ supersymmetric magnetic brane configuration with
$Sp(N_c) \times SO(2\widetilde{N}_c') \times Sp(N_c'')$ gauge group
with flavors $Q(q')[Q'']$ 
for each gauge group, the bifundamentals $F(g)$, and gauge singlets 
in Figure 8A. In
Figure 8B, the nonsupersymmetric minimal energy brane configuration
with the same gauge group and matter contents above 
for massless  $Q(Q'')$ is given. 
 }
\end{figure}

In the dual theory, since 
there exist  $2N_f'$ fundamental fields $q'$,  
one bifundamental field $g$(which will give rise to 
the contribution of $2N_c''$), and
one bifundamental field  
$F$(which will give rise to the contribution of $2N_c$),
the coefficient of the beta function 
is
$
b_{SO(2\widetilde{N}_c')}^{mag}= 3(2\widetilde{N}_c'-2)-2N_f'-2N_c''-2N_c
$
and since there are $2\widetilde{N}_c'$ fundamental 
fields $g$ for the first factor,
 $2N_f''$ fundamental 
fields $Q''$,
an antisymmetric tensor $\Phi''$(which will contribute to $(2N_c''+2)$),
and a field $X''$(which will contribute to $2N_f'$), 
the coefficient of the beta function
 is
$
b_{Sp(N_c'')}^{mag} = 3(2N_c'' +2)-2\widetilde{N}_c'-2N_f''-(2N_c'' +2)-2N_f'$.
Since there exist $2N_f$ quarks $Q$, and
one bifundamental field $F$(which will give
rise to the contribution of $2\widetilde{N}_c'$), 
the coefficient of the beta function
of the first gauge group factor is
$
b_{Sp(N_c)}^{mag} = 3(2N_c+2) -2N_f-2\widetilde{N}_c'$.

Therefore, both $Sp(N_c)$,  
$SO(2\widetilde{N}_c')$, and $Sp(N_c'')$ gauge couplings are IR free
by requiring the negativeness of the coefficients of beta function.
One can rely on the perturbative calculations at low energy 
for this magnetic IR free region $b_{Sp(N_c)}^{mag} < 0$, 
$b_{SO(2\widetilde{N}_c')}^{mag} < 0$, and $b_{Sp(N_c'')}^{mag} < 0$.
Neglecting the $Sp(N_c)$ and $Sp(N_c'')$ dynamics, 
the magnetic $SO(2\widetilde{N}_c')$
is IR free when 
\bea
N_c'-N_c''-N_c -2 < N_f' < \frac{3}{2}(N_c' -1) -N_c''-N_c.
\nonu
\eea

Then the dual gauge group and matter contents 
with the additional fields $X'',M'$ and $\Phi''$ are 
\bea
 & \mbox{gauge group}:& \;\;\;\;\;   Sp(N_c) \times SO(2\widetilde{N}_c') \times
 Sp(N_c'')  \nonu
\\
\mbox{matter}: 
 &Q_f & \;\;\;\;\;\;\;\;\; 
\;\;\;\;\;\;\;\;\;\;\;\; {(\bf \Box, 1, 1)} 
\;\;\;\;\; (f=1,  \cdots, 2N_f) 
\nonu \\
 &q'_{f'} & \;\;\;\;\;\;\;\;\;\;
\;\;\;\;\;\;\;\;\;\;\; {(\bf 1, \Box, 1) }
\;\;\;\;\; (f'=1,  \cdots, 2N_f') 
\nonu \\
 &Q''_{f''} & \;\;\;\;\;\;\;\;\;\; 
\;\;\;\;\;\;\;\;\;\;\; {(\bf 1, 1, \Box) } 
\;\;\;\;\; (f''=1,  \cdots, 2N_f'')
\nonu \\
 &F & \;\;\;\;\;\;\;\;\;\;\;\;\;\;\;\;\;\;\; 
\;\; {(\bf \Box, \Box,1) } 
\nonu \\
 &g & \;\;\;\;\;\;\;\;\;\;\;\;\;\;\;\;\;\;\; 
\;\; {(\bf 1, \Box, \Box) } 
\nonu \\
 &(X_{n'}'' \equiv) G Q' & \;\;\;\;\;\;\;\;\;\;\;\;\;\;\;\;\;\;\; 
\;\; {(\bf 1, 1, \Box) } 
\;\;\;\;\; (n'=1,  \cdots, 2N_f')
\nonu \\
 &(M_{f',g'}' \equiv) Q' Q' & \;\;\;\;\;\;\;\;\;\;\;\;\;\;\;\;\;\;\; 
\;\; {(\bf 1, 1, 1) } 
\;\;\;\;\; (f', g'=1,  \cdots, 2N_f')
\nonu \\
 &(\Phi'' \equiv) G G & \;\;\;\;\;\;\;\;\;\;\;\;\;\;\;\;\;\;\; 
\;\; {(\bf 1, 1, symm) } 
\nonu
\eea

The dual magnetic  tree level superpotential, by adding the mass term
for the $Q'$ in electric theory corresponding to add
a linear term in $M'$ in dual magnetic theory, 
is given by
\bea
W_{dual} = \left( M' q' q' + g X'' q' + \Phi'' g g \right)
+ m' M'.
\nonu 
\eea
Here $q'$ is fundamental for
the gauge group index.
Then, $q' q'$ has rank $2\widetilde{N}_c'$ while $m'$ has a
rank $2N_f'$.  Therefore, the F-term condition, the derivative the 
superpotential $W_{dual}$ with respect to $M'$, cannot be satisfied 
if the rank $2N_f'$ exceeds $2\widetilde{N}_c'$. 
This is so-called rank condition and the supersymmetry is broken \cite{Ahn07-2}.  

The classical moduli space of vacua can be obtained from F-term
equations.
From the F-terms $F_{q'}$ and $F_{M'}$, one gets
$M' q' +  g X'' =0=   q' q'  +
m'$.
Similarly, one obtains 
$\Phi'' g +  X'' q'=0 = g g $ from 
the F-terms $F_{g}$ and $F_{\Phi''}$. 
Moreover, there is a relation 
$ q' g =0$ from the F-term $F_{X''}$.
Then, one obtains 
the following solutions 
\bea
< q'> =  \left(
\begin{array}{c}
i \sqrt{m}  {\bf 1}_{2\widetilde{N}_c'}  \\
0
\end{array}
\right),  \quad
<M'>  =
 \left(
\begin{array}{cc}
0  & 0 
 \\
0 & M_0'  {\bf 1}_{2(N_f'-\widetilde{N}_c')} 
\end{array}
\right), \quad
<g>=0= <X''>
\nonu
\eea
where $M_0'  {\bf 1}_{2(N_f'-\widetilde{N}_c')}$ 
is an arbitrary 
$2(N_f'-\widetilde{N}_c') \times 2(N_f'-\widetilde{N}_c')$ symmetric 
matrix and the zeros of 
$<q'>$  are $2(N_f'-\widetilde{N}_c') \times 2\widetilde{N}_c' $
zero matrices. 
Similarly, the zeros
of $2N_f' \times 2N_f'$  matrix $M'$ are assumed also.
As we did in previous case, one can analyze the one loop 
computation by expanding the fields around the vacua and it will
lead to the fact that states are stable by realizing the mass of 
$m_{M_0'}^2$ positive.


Then the minimal energy supersymmetry breaking brane configuration is
shown in Figure 8B.
If we ignore the $NS5_L$-brane, $2N_c$ D4-branes, $2N_f$ 
D6-branes, the $NS5_R'$-brane, $2N_c''$ D4-branes and $2N_f''$ 
D6-branes(detaching these
branes from Figure 8B), 
then this brane configuration 
corresponds to  the minimal energy supersymmetry breaking brane
configuration
for the ${\cal N}=1$ SQCD with the magnetic gauge group 
$SO(2\widetilde{N}_c')$ with
$N_f'$ massive flavors \cite{FGU,Ahn06-1}.

The nonsupersymmetric minimal energy brane configuration Figure 8B
with vanishing $2N_f''$ D6-branes
leads to the Figure 3 of \cite{Ahn07-2} with a reflection with
respect to the NS5-brane if we ignore the
$NS5_L$-brane, $2N_f$ D6-branes and $2N_c$ D4-branes.
Moreover, this Figure 8B with a replacement $N_f'$ D6-branes by 
the NS5'-brane(neglecting  the
$NS5_L$-brane, $N_f$ D6-branes and $N_c$ D4-branes)
will become the Figure 7B of \cite{Ahn07-6}
with a
rotation of $NS5_R$-brane by $\frac{\pi}{2}$ angle or 
the Figure 9B of \cite{Ahn07-6} with a reflection with
respect to the $NS5_R$-brane if we ignore 
the
$NS5_R'$-brane, $N_f''$ D6-branes and $N_c''$ D4-branes from the Figure 8B.

Starting with $NS5_L'$-$NS5_L$-$NS5_R'$-$NS5_R$ branes configuration,
as in the footnote \ref{rotation}, 
by moving the $NS5_L$-brane to the right, 
one gets 
the nonsupersymmetric minimal energy brane configuration which is
exactly the Figure 8 with a reflection with
respect to the NS5-brane.

After lifting 
the type IIA description to M-theory, the 
corresponding magnetic M5-brane configuration with equal mass for the
quarks, in a background space of $x t = 
v^{2N_f+2N_f''} \prod_{k=1}^{N_f'} (v^2 -e_k^2)$ where this four dimensional space
replaces (45610) directions,
is characterized by \cite{LLL97}
\bea
& & t^4 + ( v^{2N_c+2+2N_f }  + \cdots ) t^3 + 
( v^{2\widetilde{N}_c'+2N_f} + \cdots) t^2 \nonu \\
& & + (v^{2N_c''+2+4N_f} 
+ \cdots ) t + v^{6N_f+2N_f''}
(v^2 -m^2 )^{N_f'} =0
\nonu
\eea
where 
we ignored both the lower power terms in $v$ 
and the scales for the gauge groups for simplicity.

From this curve 
of quartic equation for $t$ above, the asymptotic regions 
can be classified by 
as follows:

1. $v \rightarrow \infty$ limit implies
\bea
w \rightarrow 0, && \quad y \sim    v^{2N_c+2+2N_f} + \cdots \quad
\mbox{$NS_L$ asymptotic region},   
\nonu \\
w \rightarrow 0, && \quad y \sim    
v^{2\widetilde{N}_c'-2N_c-2} + \cdots \quad
\mbox{$NS_R$ asymptotic region}.
\nonu  
\eea

2.  $w \rightarrow \infty$ limit implies
\bea
v  \rightarrow    m, && \quad 
y \sim  w^{2N_c''+2-2\widetilde{N}_c'+2N_f}
 +\cdots
\quad \mbox{$NS_{L}'$ asymptotic region}, 
\nonu
\\
v  \rightarrow   m, && \quad  
y \sim w^{2N_f+2N_f'+2N_f''-2N_c''-2}
+\cdots
\quad \mbox{$NS_{R}'$ asymptotic region}. 
\nonu
\eea


\subsection{Magnetic theory  with dual for second gauge group}

Let us take the Seiberg dual for the second gauge group factor $SO(2N_c')$ 
while keeping the first and the third gauge group factors 
$Sp(N_c)$ and $Sp(N_c'')$ untouched. 
Suppose that
$\Lambda_2 >> \Lambda_1, \Lambda_3$. 

Let us move the $NS5_L'$-brane in Figure 6 to the right all the way past the  
$NS5_R$-brane.  
After this brane motion, one arrives at the Figure 9A.
We need to change the Figure 6  a little bit such that 
the $N_f'$ D6-branes are not parallel to the 
$NS5_L'$-brane. That is, we rotate $N_f'$ D6-branes a little bit(this
does not change the classical electric superpotential as we explained 
before) and 
after dualizing the second gauge group, 
we rotate those $N_f'$ D6-branes with opposite direction.
The linking number of $NS5'_L$-brane from Figure 9A
is given by 
$
L_5 =\frac{(2N_f')}{2}+1-(-1)-2\widetilde{N}_c'+2N_c''$.
Originally, it was 
$
L_5=-\frac{(2N_f')}{2}+(-1) -(+1)
+2N_c'-2N_c
$
from Figure 6 before the brane motion.
Therefore, by the linking number conservation and equating these two
$L_5$'s each other, 
we are left with the number of colors in the magnetic
theory 
$
\widetilde{N}_c' = N_f' + N_c''+N_c-N_c'+2$.

Let us draw this magnetic brane configuration in Figure 9A 
and we put
the coincident $N_f'$ D6-branes in the nonzero 
$v$ direction(and its mirrors).

\begin{figure}[ht]
   \epsfxsize=4.0in 
\centerline{\epsffile{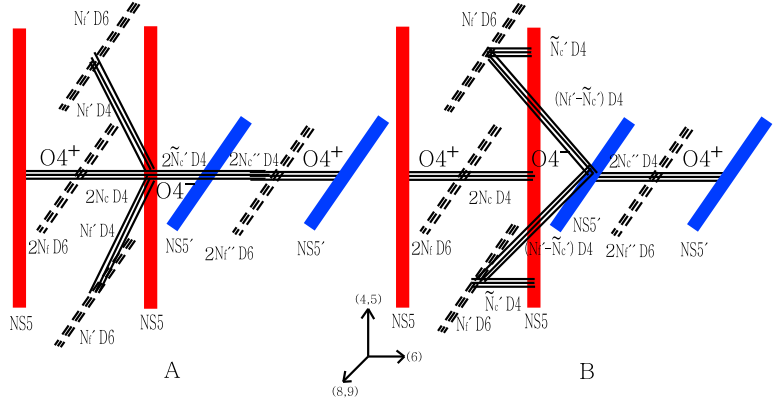}}
   \caption[FIG. \arabic{figure}.]{ 
The ${\cal N}=1$ supersymmetric magnetic brane configuration with
$Sp(N_c) \times SO(2\widetilde{N}_c') \times Sp(N_c'')$ gauge group
with flavors $Q(q')[Q'']$ 
for each gauge group, the bifundamentals $f(G)$, and gauge singlets 
in Figure 9A. In
Figure 9B, the nonsupersymmetric minimal energy brane configuration
with the same gauge group and matter contents above 
for massless  $Q(Q'')$ is given. 
 }
\end{figure}

Then the dual gauge group and matter contents 
with the additional fields $X,M'$ and $\Phi$ are 
\bea
 & \mbox{gauge group}:& \;\;\;\;\;   Sp(N_c) \times SO(2\widetilde{N}_c') \times
 Sp(N_c'')  \nonu
\\
\mbox{matter}: 
 &Q_f & \;\;\;\;\;\;\;\;\; 
\;\;\;\;\;\;\;\;\;\;\;\; {(\bf \Box, 1, 1)} 
\;\;\;\;\; (f=1,  \cdots, 2N_f) 
\nonu \\
 &q'_{f'} & \;\;\;\;\;\;\;\;\;\;
\;\;\;\;\;\;\;\;\;\;\; {(\bf 1, \Box, 1) }
\;\;\;\;\; (f'=1,  \cdots, 2N_f') 
\nonu \\
 &Q''_{f''} & \;\;\;\;\;\;\;\;\;\; 
\;\;\;\;\;\;\;\;\;\;\; {(\bf 1, 1, \Box) } 
\;\;\;\;\; (f''=1,  \cdots, 2N_f'')
\nonu \\
 &f & \;\;\;\;\;\;\;\;\;\;\;\;\;\;\;\;\;\;\; 
\;\; {(\bf \Box, \Box,1) } 
\nonu \\
 &G & \;\;\;\;\;\;\;\;\;\;\;\;\;\;\;\;\;\;\; 
\;\; {(\bf 1, \Box, \Box) } 
\nonu \\
 &(X_{n'} \equiv) F Q' & \;\;\;\;\;\;\;\;\;\;\;\;\;\;\;\;\;\;\; 
\;\; {(\bf \Box, 1, 1) } 
\;\;\;\;\; (n'=1,  \cdots, 2N_f')
\nonu \\
 &(M_{f',g'}' \equiv) Q' Q' & \;\;\;\;\;\;\;\;\;\;\;\;\;\;\;\;\;\;\; 
\;\; {(\bf 1, 1, 1) } 
\;\;\;\;\; (f', g'=1,  \cdots, 2N_f')
\nonu \\
 &(\Phi \equiv) F F & \;\;\;\;\;\;\;\;\;\;\;\;\;\;\;\;\;\;\; 
\;\; {(\bf symm, 1, 1) } 
\nonu
\eea

In the dual theory, since 
there exist  $2N_f'$ flavor fields $q'$ and  
one bifundamental field  
$G$(which will give rise to the contribution of $2N_c''$),
and one bifundamental field  
$f$(which will give rise to the contribution of $2N_c$),
the coefficient of the beta function 
is
$
b_{SO(\widetilde{N}_c')}^{mag}= 3(2\widetilde{N}_c'-2)-2N_f'-2N_c''-2N_c
$
which is the same as the $b_{SO(\widetilde{N}_c')}^{mag}$ 
in previous subsection
and since there are $2\widetilde{N}_c'$  
fields $G$ for the first factor,
and  $2N_f''$ fundamental 
fields $Q''$,
the coefficient of the beta function
 is
$
b_{Sp(N_c'')}^{mag} = 3(2N_c'' +2)-2\widetilde{N}_c'-2N_f''$.
Since there exist $2N_f$ quarks $Q$, 
one bifundamental field $f$(which will give
rise to the contribution of $2\widetilde{N}_c'$), 
an antisymmetric tensor $\Phi$(which will contribute to $(2N_c+2)$),
and a field $X$(which will contribute to $2N_f$),
the coefficient of the beta function
of the first gauge group factor is
$
b_{Sp(N_c)}^{mag} = 3(2N_c+2) -2N_f-2\widetilde{N}_c'-(2N_c+2)-2N_f$.

Therefore, both $Sp(N_c)$,  
$SO(2\widetilde{N}_c')$, and $Sp(N_c'')$ gauge couplings are IR free
by requiring the negativeness of the coefficients of beta function.
One can rely on the perturbative calculations at low energy 
for this magnetic IR free region $b_{Sp(N_c)}^{mag} < 0$, 
$b_{SO(2\widetilde{N}_c')}^{mag} < 0$, and $b_{Sp(N_c'')}^{mag} < 0$.
Neglecting the $Sp(N_c)$ and $Sp(N_c'')$ dynamics, 
the magnetic $SO(2\widetilde{N}_c')$
is IR free when 
\bea
 N_c' -N_c''-N_c-2 < N_f' < \frac{3}{2}(N_c' -1)-N_c''-N_c.
\nonu
\eea

The dual magnetic  tree level superpotential, by adding the mass term
for the $Q'$ in electric theory corresponding to add
a linear term in $M'$ in dual magnetic theory, 
is given by
\bea
W_{dual} = \left( M' q' q' + f X q' + \Phi f f  \right)
+ m' M'.
\nonu
\eea
Here $q'$ is fundamental for
the gauge group index.
Then, $q' q'$ has rank $2\widetilde{N}_c'$ while $m'$ has a
rank $2N_f'$.  Therefore, the F-term condition, the derivative the 
superpotential $W_{dual}$ with respect to $M'$, cannot be satisfied 
if the rank $2N_f'$ exceeds $2\widetilde{N}_c'$. 
This is so-called rank condition \cite{Ahn07-2} and the supersymmetry is broken.

The classical moduli space of vacua can be obtained from F-term
equations.
From the F-terms $F_{q'}$ and $F_{M'}$, one gets
$M' q' +  f X =0=   q' q'  +
m'$.
Similarly, one obtains 
$\Phi f +  X q' =0 = f f $ from 
the F-terms $F_{f}$ and $F_{\Phi}$. 
Moreover, there is a relation 
$ q' f =0$ from the F-term $F_{X}$.
Then, one obtains 
the following solutions 
\bea
< q'> =  \left(
\begin{array}{c}
i \sqrt{m}  {\bf 1}_{2\widetilde{N}_c'}  \\
0
\end{array}
\right),  \quad
<M'>  =
 \left(
\begin{array}{cc}
0  & 0 
 \\
0 & M_0'  {\bf 1}_{2(N_f'-\widetilde{N}_c')} 
\end{array}
\right), \quad
<f>=0= <X>
\nonu
\eea
where $M_0'  {\bf 1}_{2(N_f'-\widetilde{N}_c')}$ 
is an arbitrary 
$2(N_f'-\widetilde{N}_c') \times 2(N_f'-\widetilde{N}_c')$ symmetric 
matrix and the zeros of 
$<q'>$  are $2(N_f'-\widetilde{N}_c') \times 2\widetilde{N}_c' $
zero matrices. 
Similarly the zeros of $2N_f' \times 2N_f'$  matrix $M'$ are assumed also.
As we did in previous case, one can analyze the one loop 
computation by expanding the fields around the vacua and it will
lead to the fact that states are stable by realizing the mass of 
$m_{M_0'}^2$ positive.


Then the minimal energy supersymmetry breaking brane configuration is
shown in Figure 9B.
If we ignore the $NS5_L$-brane, $2N_c$ D4-branes,
$2N_f$ D6-branes, $NS5_R'$-brane, $2N_c''$ D4-branes and $2N_f''$ 
D6-branes(detaching these from Figure 9B), 
as observed already, 
then this brane configuration 
looks similar to  the minimal energy supersymmetry breaking brane
configuration
for the ${\cal N}=1$ SQCD with the magnetic gauge group 
$SO(2\widetilde{N}_c'=2(N_f'-N_c'+2))$ with
$N_f'$ massive flavors. 
The difference occurs in the position of D6-branes. 

The nonsupersymmetric minimal energy brane configuration Figure 9B
with a replacement $N_f'$ D6-branes by 
the NS5'-brane(neglecting  the
$NS5_R'$-brane, $2N_f''$ D6-branes and $2N_c''$ D4-branes 
with vanishing $2N_f$ D6-branes)
leads to 
the Figure 10B of \cite{Ahn07-6}
with a
rotation of $NS5_L'$-brane by $\frac{\pi}{2}$ angle. 

Starting with $NS5_L'$-$NS5_L$-$NS5_R'$-$NS5_R$ branes configuration,
as in the footnote \ref{rotation}, 
by moving the $NS5_R'$-brane to the left, 
one gets 
the nonsupersymmetric minimal energy brane configuration which is
exactly the Figure 9 with a reflection with
respect to the NS5-brane. 

After lifting 
the type IIA description we explained so far to M-theory, the 
corresponding magnetic M5-brane configuration, in a background space of $x t = 
v^{2N_f+2N_f''} \prod_{k=1}^{N_f'} (v^2 -e_k^2)$ where this four dimensional space
replaces (45610) directions,
is characterized by \cite{LLL97}
\bea
&& t^4 + ( v^{2N_c+2 }  + \cdots ) t^3 + 
\left[ v^{2\widetilde{N}_c'+2N_f}(v^2-m^2)^{N_f'} + \cdots \right] t^2 \nonu \\
&& + 
\left[ v^{2N_c''+2+4N_f}(v^2-m^2)^{2N_f'} + \cdots \right] t + v^{6N_f+2N_f''}
(v^2 -m^2 )^{3N_f'} =0.
\nonu
\eea

From this curve 
of quartic equation for $t$ above, the asymptotic regions 
can be classified 
as follows:

1. $v \rightarrow \infty$ limit implies
\bea
w \rightarrow 0, && \quad y \sim    v^{2N_c+2} + \cdots \quad
\mbox{$NS_L$ asymptotic region},   
\nonu \\
w \rightarrow 0, && \quad y \sim    
v^{2\widetilde{N}_c'-2N_c-2+2N_f+2N_f'} + \cdots \quad
\mbox{$NS_R$ asymptotic region}.
\nonu  
\eea

2.  $w \rightarrow \infty$ limit implies
\bea
v  \rightarrow    m, && \quad 
y \sim  w^{2N_c''+2-2\widetilde{N}_c'+2N_f+2N_f'}
 +\cdots
\quad \mbox{$NS_{L}'$ asymptotic region}, 
\nonu
\\
v  \rightarrow   m, && \quad  
y \sim w^{2N_f+2N_f'+2N_f''-2N_c''-2}
+\cdots
\quad \mbox{$NS_{R}'$ asymptotic region}. 
\nonu
\eea


\subsection{Magnetic theory  with dual for first gauge group}

Let us take the Seiberg dual for the first gauge group factor $Sp(N_c)$ 
while keeping the second and the third gauge group factors 
$SO(2N_c')$ and $Sp(N_c'')$ untouched. 
Suppose that
$\Lambda_1 >> \Lambda_2, \Lambda_3$. 
After we move  the $NS5_L$-brane in Figure 6
to the right all the way past
the $NS5_L'$-brane, one arrives at Figure 10A and the linking number
of $NS5_L$-brane from Figure 10A is given by 
$
L_5 =\frac{(2N_f)}{2}-1-(1)-2\widetilde{N}_c+ 2N_c'$.
Originally, it was 
$
L_5=-\frac{(2N_f)}{2}+ 1 -(-1)
+2N_c
$
from Figure 6 before the brane motion.
Therefore, by the linking number conservation and equating these two
$L_5$'s each other, 
we are left with the number of colors in the magnetic
theory 
$
\widetilde{N}_c = N_f + N_c'-N_c -2$.

Let us draw this magnetic brane configuration in Figure 10A 
and we put
the coincident $N_f$ D6-branes in the nonzero $v$ direction(and its mirrors).
If we ignore the $NS5_R$-brane, $N_c'$ D4-branes, $N_f'$ 
D6-branes, the $NS5_R'$-brane, $N_f''$ D6-branes  
and $N_c''$ D4-branes(detaching these
branes from Figure 10A), 
then this brane configuration 
leads to the standard ${\cal N}=1$ SQCD with the magnetic gauge group 
$Sp(\widetilde{N}_c=N_f-N_c-2)$ with
$N_f$ massive flavors \cite{FGU,Ahn06-1}.

\begin{figure}[ht]
   \epsfxsize=4.0in 
\centerline{\epsffile{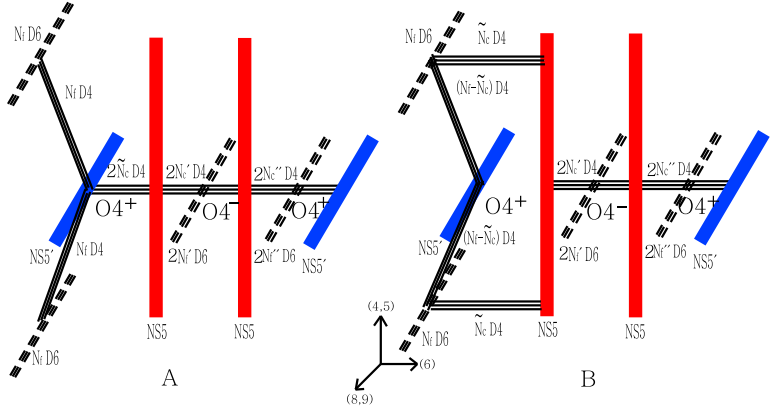}}
   \caption[FIG. \arabic{figure}.]{ 
The ${\cal N}=1$ supersymmetric magnetic brane configuration with
$Sp(\widetilde{N}_c) \times SO(2N_c') \times Sp(N_c'')$ gauge group
with fundamentals $q(Q')[Q'']$ 
for each gauge group, the bifundamentals $F(g)$, and gauge singlets 
in Figure 10A. In
Figure 10B, the nonsupersymmetric minimal energy brane configuration
with the same gauge group and matter contents above 
for massless  $Q'(Q'')$ is given. 
}
\end{figure}

In the dual theory, since 
there exist  $2N_f$ fundamental fields $q$, and  
one bifundamental field $f$(which will 
give rise to the contribution of $2N_c'$),
the coefficient of the beta function 
is
$
b_{Sp(\widetilde{N}_c)}^{mag}= 3(2\widetilde{N}_c+2)-2N_f-2N_c'
$
and since there are 
$2N_f'$  fundamental fields $Q'$,
$2\widetilde{N}_c$ fundamental 
fields $f$ for the second factor,
 $2N_c''$ fundamental 
fields $G$ for the 
flavor index of the first factor,
an antisymmetric tensor $\Phi'$(which will contribute to $(2N_c'-2)$), 
and $2N_f$ fields $X'$, 
the coefficient of the beta function
 is
$
b_{SO(2N_c')}^{mag} = 3(2N_c' -2)-2N_f'-2\widetilde{N}_c-2N_c''-(2N_c' -2)-2N_f$.
Since there exist $2N_f''$ quarks $Q''$, 
and one bifundamental field $G$(which will give
rise to the contribution of $2N_c'$), 
the coefficient of the beta function
of the third gauge group factor is
$
b_{Sp(N_c'')}^{mag} = 3(2N_c''+2) -2N_f''-2N_c'=b_{Sp(N_c'')}$.

Therefore, both $Sp(\widetilde{N}_c)$,  
$SO(2N_c')$, and $Sp(N_c'')$ gauge couplings are IR free
by requiring the negativeness of the coefficients of beta function.
One can rely on the perturbative calculations at low energy 
for this magnetic IR free region $b_{Sp(\widetilde{N}_c)}^{mag} < 0$, 
$b_{SO(2N_c')}^{mag} < 0$, and $b_{Sp(N_c'')}^{mag} < 0$.
Note that the $SO(2N_c')$ fields in the magnetic theory 
are different from those of the electric theory.
Since $b_{SO(2N_c')}-b_{SO(2N_c')}^{mag} > 0$, $SO(2N_c')$ is more
asymptotically free than $SO(2N_c')^{mag}$.
Neglecting the $SO(2N_c')$ and $Sp(N_c'')$ dynamics, 
the magnetic $Sp(\widetilde{N}_c)$
is IR free when 
\bea
N_c  -N_c'  +2 < N_f < \frac{3}{2}(N_c +1)-N_c'.
\nonu
\eea

Then the dual gauge group and matter contents 
with the additional fields $X',M$ and $\Phi'$ are 
\bea
 & \mbox{gauge group}:& \;\;\;\;\;   Sp(\widetilde{N}_c) \times SO(2N_c') \times
 Sp(N_c'')  \nonu
\\
\mbox{matter}: 
 &q_f & \;\;\;\;\;\;\;\;\; 
\;\;\;\;\;\;\;\;\;\;\;\; {(\bf \Box, 1, 1)} 
\;\;\;\;\; (f=1,  \cdots, 2N_f) 
\nonu \\
 &Q'_{f'} & \;\;\;\;\;\;\;\;\;\;
\;\;\;\;\;\;\;\;\;\;\; {(\bf 1, \Box, 1) }
\;\;\;\;\; (f'=1,  \cdots, 2N_f') 
\nonu \\
 &Q''_{f''} & \;\;\;\;\;\;\;\;\;\; 
\;\;\;\;\;\;\;\;\;\;\; {(\bf 1, 1, \Box) } 
\;\;\;\;\; (f''=1,  \cdots, 2N_f'')
\nonu \\
 &f & \;\;\;\;\;\;\;\;\;\;\;\;\;\;\;\;\;\;\; 
\;\; {(\bf \Box, \Box,1) } 
\nonu \\
 &G & \;\;\;\;\;\;\;\;\;\;\;\;\;\;\;\;\;\;\; 
\;\; {(\bf 1, \Box, \Box) } 
\nonu \\
 &(X_{n}' \equiv) F Q & \;\;\;\;\;\;\;\;\;\;\;\;\;\;\;\;\;\;\; 
\;\; {(\bf 1, \Box, 1) } 
\;\;\;\;\; (n=1,  \cdots, 2N_f)
\nonu \\
 &(M_{f, g} \equiv) Q Q & \;\;\;\;\;\;\;\;\;\;\;\;\;\;\;\;\;\;\; 
\;\; {(\bf 1, 1, 1) } 
\;\;\;\;\; (f, g=1,  \cdots, 2N_f)
\nonu \\
 &(\Phi' \equiv) F F & \;\;\;\;\;\;\;\;\;\;\;\;\;\;\;\;\;\;\; 
\;\; {(\bf 1, asymm, 1) } 
\nonu
\eea

The dual magnetic  tree level superpotential, by adding the mass term
for the $Q$ in electric theory corresponding to add
a linear term in $M$ in dual magnetic theory, 
is given by \cite{Ahn07-2}
\bea
W_{dual} = \left( M q q + f X' q + \Phi' f f \right)
+ m M.
\nonu
\eea
Here $q$ is fundamental for
the gauge group index.
Then, $q q$ has rank $2\widetilde{N}_c$ while $m$ has a
rank $2N_f$.  Therefore, the F-term condition, the derivative the 
superpotential $W_{dual}$ with respect to $M$, cannot be satisfied 
if the rank $2N_f$ exceeds $2\widetilde{N}_c$. 
This is so-called rank condition and the supersymmetry is broken.

The classical moduli space of vacua can be obtained from F-term
equations.
From the F-terms $F_{q}$ and $F_{M}$, one gets
$M q + f X'  =0=   q q  +
m$.
Similarly, one obtains 
$\Phi' f +  X' q =0 = f f $ from 
the F-terms $F_{f}$ and $F_{\Phi'}$. 
Moreover, there is a relation 
$ q f =0$ from the F-term $F_{X'}$.
Then, one obtains 
the following solutions 
\bea
< q> =  \left(
\begin{array}{c}
i \sqrt{m}  {\bf 1}_{2\widetilde{N}_c}  \\
0
\end{array}
\right),  
<M>  =
 \left(
\begin{array}{cc}
0  & 0 
 \\
0 & M_0  {\bf 1}_{(N_f-\widetilde{N}_c)} \otimes i \sigma^2 
\end{array}
\right), 
<f>=0= <X'>
\nonu
\eea
where $M_0  {\bf 1}_{(N_f-\widetilde{N}_c)} \otimes i \sigma^2$ 
is an arbitrary 
$2(N_f-\widetilde{N}_c) \times 2(N_f-\widetilde{N}_c)$ antisymmetric 
matrix and the zeros of 
$<q>$  are $2(N_f-\widetilde{N}_c) \times 2\widetilde{N}_c $
zero matrices. 
Similarly, the zeros
of $2N_f \times 2N_f$  matrix $M$ are assumed also.
As we did in previous case, one can analyze the one loop 
computation by expanding the fields around the vacua and it will
lead to the fact that states are stable by realizing the mass of 
$m_{M_0}^2$ positive.


Then the minimal energy supersymmetry breaking brane configuration is
shown in Figure 10B.
If we ignore the $NS5_R$-brane, $2N_c'$ D4-branes,
$2N_f'$ D6-branes, $NS5_R'$-brane, $2N_c''$ D4-branes and $2N_f''$ 
D6-branes(detaching these from Figure 10B), 
as observed already, 
then this brane configuration 
is the minimal energy supersymmetry breaking brane
configuration
for the ${\cal N}=1$ SQCD with the magnetic gauge group 
$Sp(\widetilde{N}_c=N_f-N_c-2)$ with
$N_f$ massive flavors \cite{FGU,Ahn06-1}.  

The nonsupersymmetric minimal energy brane 
configuration Figure 10B
with a replacement $N_f$ D6-branes by 
the NS5'-brane(neglecting  the
$NS5_R'$-brane, $2N_f''$ D6-branes and $2N_c''$ D4-branes 
with vanishing $2N_f'$ D6-branes)
leads to 
the Figure 10B of \cite{Ahn07-6}.

When we move the $NS5_L'$-brane in Figure 6 
to the left all the way past the 
$NS5_L$-brane, then  
one arrives at the magnetic brane configuration similar to the Figure 10.
The only difference is that the $N_f$ D6-branes are located at the right
hand side of the $NS5_L$-brane.
Then this nonsupersymmetric minimal energy brane configuration 
 with a replacement $N_f$ D6-branes by 
the NS5'-brane(neglecting  the
$NS5_R'$-brane, $2N_f''$ D6-branes and $2N_c''$ D4-branes) leads to 
the Figure 10B of \cite{Ahn07-6} with a reflection with respect to the
NS5-brane
and an extra rotation of
$NS5'_L$-brane by $\frac{\pi}{2}$ angle.

Starting with $NS5_L'$-$NS5_L$-$NS5_R'$-$NS5_R$ branes configuration,
as in the footnote \ref{rotation}, 
by moving the $NS5_L$-brane to the left, 
one gets 
the nonsupersymmetric minimal energy brane configuration which is
exactly the Figure 7 with a reflection with
respect to the NS5-brane. 
Furthermore, by 
moving the $NS5_R'$-brane to the right, 
one gets 
the nonsupersymmetric minimal energy brane configuration which is
exactly the new Figure in previous paragraph with a reflection with
respect to the $NS5_L$-brane.

After lifting 
the type IIA description we explained so far to M-theory, the 
corresponding magnetic M5-brane configuration, in a background space of $x t = 
v^{2N_f'+2N_f''} \prod_{k=1}^{N_f} (v^2 -e_k^2)$ where this four dimensional space
replaces (45610) directions,
is characterized by \cite{LLL97}
\bea
&& t^4 + ( v^{2\widetilde{N}_c+2 }  + \cdots ) t^3 + 
\left[ v^{2N_c'}(v^2-m^2)^{N_f} + \cdots \right] t^2 \nonu \\
&& + 
\left[ v^{2N_c''+2+2N_f'}(v^2-m^2)^{2N_f} + \cdots \right] t + 
(v^2-m^2)^{3N_f} v^{4N_f'+2N_f''}
=0.
\nonu
\eea

From this curve  
of quartic equation for $t$ above, the asymptotic regions 
can be classified 
as follows:

1. $v \rightarrow \infty$ limit implies
\bea
w \rightarrow 0, && \quad y \sim    v^{2N_c'-2\widetilde{N}_c-2+2N_f} + \cdots \quad
\mbox{$NS_L$ asymptotic region},   
\nonu \\
w \rightarrow 0, && \quad y \sim    
v^{2N_f+2N_f'+2N_c''+2-2N_c'} + \cdots \quad
\mbox{$NS_R$ asymptotic region}.
\nonu  
\eea

2.  $w \rightarrow \infty$ limit implies
\bea
v  \rightarrow    m, && \quad 
y \sim  w^{2\widetilde{N}_c+2}
 +\cdots
\quad \mbox{$NS_{L}'$ asymptotic region}, 
\nonu
\\
v  \rightarrow   m, && \quad  
y \sim w^{2N_f+2N_f'+2N_f''-2N_c''-2}
+\cdots
\quad \mbox{$NS_{R}'$ asymptotic region}. 
\nonu
\eea
 

\subsection{Other magnetic theories with opposite O4-plane charges}

By changing the charges of O4-plane in previous brane configuration given
in Figure 6, 
the type IIA brane configuration is realized by 
an ${\cal N}=1$ supersymmetric gauge theory with \cite{Ahn97}
\bea
SO(2N_c) \times Sp(N_c') \times SO(2N_c'')
\nonu
\eea
and corresponding matter contents. Then by deforming this electric theory by 
mass term for the quarks and taking the magnetic dual on
each gauge group factor, one gets the possible 
meta-stable brane configurations.

There exists an ${\cal N}=1$ magnetic supersymmetric gauge theory with 
\bea
SO(2N_c) \times Sp(N_c') \times SO(2\widetilde{N}_c''), \qquad
\widetilde{N}_c'' 
\equiv N_f''+N_c'-N_c''+2
\nonu
\eea 
with matters and the corresponding brane configuration is  given by
the Figure 7 with 
opposite O4-plane charges($O4^{+}$ goes to $O4^{-}$ and vice versa). 
The constant term $+2$ 
in the dual color has opposite sign, compared with the discussion of
subsection 3.2 where it has $-2$. This is due to the fact that the
linking number counting in this case has opposite sign for the
contribution of O4-plane since we have different charge in the
electric and magnetic theories.  
Since the dual gauge group is an orthogonal gauge group, the
corresponding gauge singlet $M''$ is a symmetric two index tensor in
the flavor indices(leading to different structure of 
vacuum expectation value, compared with the solution in the subsection
3.2) and the field $\Phi'$ is a symmetric under the
second gauge group.

Also 
there is an ${\cal N}=1$ magnetic supersymmetric gauge theory with 
\bea
SO(2N_c ) \times Sp(\widetilde{N}_c') \times
SO(2N_c''), \qquad
\widetilde{N}_c' \equiv N_f'+N_c''+N_c-N_c'-2
\nonu
\eea 
with matters and the corresponding brane configuration is  given by
the Figure 8 with 
opposite O4-plane charges($O4^{+}$ goes to $O4^{-}$ and vice versa). 
Since the dual gauge group is a symplectic gauge group, the
corresponding gauge singlet $M'$ is an  antisymmetric two index tensor in
the flavor indices(leading to different structure of 
vacuum expectation value, compared with the solution in the subsection
3.3) and the field $\Phi''$ is an antisymmetric under the
third gauge group.

Moreover, 
there is an ${\cal N}=1$ magnetic supersymmetric gauge theory with 
\bea
SO(2N_c ) \times Sp(\widetilde{N}_c') \times
SO(2N_c''), \qquad
\widetilde{N}_c' \equiv N_f'+N_c''+N_c-N_c'-2
\nonu
\eea 
with matters 
and the corresponding brane configuration is  given by
the Figure 9 with 
opposite O4-plane charges($O4^{+}$ goes to $O4^{-}$ and vice versa). 
Since the dual gauge group is a symplectic gauge group, the
corresponding gauge singlet $M'$ is an  antisymmetric two index tensor in
the flavor indices(leading to different structure of 
vacuum expectation value, compared with the solution in the subsection
3.4) and the field $\Phi$ is an antisymmetric under the
first gauge group.

Finally, 
there exists an ${\cal N}=1$ magnetic supersymmetric gauge theory with 
\bea
SO(2\widetilde{N}_c) \times Sp(N_c') \times
SO(2N_c''), \qquad
\widetilde{N}_c \equiv N_f+N_c'-N_c+2
\nonu
\eea 
with matters 
and the corresponding brane configuration is  given by
the Figure 10 with 
opposite O4-plane charges($O4^{+}$ goes to $O4^{-}$ and vice versa). 
Since the dual gauge group is a symplectic gauge group, the
corresponding gauge singlet $M$ is a  symmetric two index tensor in
the flavor indices(leading to different structure of 
vacuum expectation value, compared with the solution in the subsection
3.5) and the field $\Phi'$ is a symmetric under the
second gauge group.

The remaining analysis can be done easily
without any difficulty.

\subsection{Magnetic theories 
for the multiple product gauge groups}

\subsubsection{ The symplectic gauge groups both at the start and
end of the chain }

Now one can generalize the method for the triple product gauge groups
to the finite multiple product gauge groups characterized by \cite{Ahn97}
\bea
Sp(N_{c,1}) \times 
SO(2N_{c,2}) \times \cdots \times SO(2N_{c,n-1}) \times Sp(N_{c,n})
\nonu
\eea
with 
the $(n-1)$ bifundametals $({\bf \Box_1, \Box_2, 1, \cdots,  1})$,
$\cdots$, and $({\bf 1, \cdots, 1, \Box_{n-1}, \Box_n})$, and 
$n$-quarks $({\bf \Box_1, 1, \cdots, 1})$, $\cdots$, and 
$({\bf 1, \cdots,  1, \Box_n})$.
Note that there are the symplectic gauge groups both at the start and
end of the chain.  
Then the electric superpotential is 
$
W_{elec} = \sum_{i=1}^n Q_i Q_i$.

There exist $(2n-2)$ magnetic theories and they can be classified as
six cases as follows.

$\bullet$ Case 1 

When the Seiberg dual is taken for the first gauge group factor
 by
assuming that $\Lambda_1 >> \Lambda_i$ where $i=2, \cdots, n$, 
one follows the procedure given in the subsection 3.5.
The gauge group is 
\bea
Sp(\widetilde{N}_{c,1} \equiv N_{f,1} +N_{c,2}-N_{c,1}-2) \times 
SO(2N_{c,2}) \times \cdots \times Sp(N_{c,n})
\nonu
\eea
and the matter contents are given by 
the dual quark $q_1$ in the representation 
$({\bf \Box_1, 1, \cdots, 1 })$ 
as well as $(n-1)$ quarks $Q_i$ where $i=2, \cdots, n$, 
the bifundamentals $f_1$ in the representation
 $({\bf \Box_1, \Box_2, 1, \cdots,  1})$ under the gauge group,
$\cdots$ in
addition to $(n-2)$ bifundamentals $G_i$, and
various gauge singlets $X_2, M_1$ and $\Phi_2$.
The corresponding brane configuration can be 
obtained similarly and 
the extra $(n-3)$ NS-branes, $(n-3)$ sets of D6-branes and $(n-3)$
sets of D4-branes  
are present at the right hand side of the $NS5_R'$-brane
of Figure 10.
The magnetic superpotential is
$
W_{dual} = \left( M_1 q_1 q_1 + f_1 X_2 q_1 + \Phi_2 f_1 f_1 \right)
+ m_1 M_1$.
By computing the contribution for the one loop as in the subsection
3.5, 
the vacua are stable and the asymptotic behavior of $(n+1)$ NS-branes
can be obtained also. 

$\bullet$ Case 2 

When the Seiberg dual is taken for the last gauge group factor
 by
assuming that $\Lambda_n >> \Lambda_i$ where $i=1,2, \cdots, (n-1)$, 
one follows the procedure given in the subsection 3.2.
The gauge group is given by
\bea
Sp(N_{c,1}) \times \cdots \times 
SO(2N_{c,n-1}) \times Sp(\widetilde{N}_{c,n} \equiv N_{f,n} +N_{c,n-1}-N_{c,n}-2).
\nonu
\eea
The corresponding brane configuration can be 
obtained similarly and 
the extra $(n-3)$ NS-branes, $(n-3)$ sets of D6-branes and $(n-3)$
sets of D4-branes  
are present at the left hand side of the $NS5_L$-brane
of Figure 7.
The magnetic superpotential is
$
W_{dual} = \left( M_n q_n q_n + g_{n-1} X_{n-1} q_n + \Phi_{n-1}
  g_{n-1} g_{n-1} 
\right)
+ m_n M_n$.

$\bullet$ Case 3

When the Seiberg dual is taken for the middle gauge group factor
by
assuming that $\Lambda_i >> \Lambda_j$ where $j=1,2, \cdots, i-1, i+1,
\cdots, n$,  
one follows the procedure given in the subsection 3.3.
The gauge group is given by
\bea
\cdots \times Sp(N_{c,i-1}) \times 
SO(2\widetilde{N}_{c,i} \equiv 2(N_{f,i}+N_{c,i+1}+N_{c,i-1}-N_{c,i} +2)) 
\times Sp(N_{c,i+1}) \times \cdots
\nonu
\eea
where $ i=2, 4, \cdots, (n-1)$ implying that the number of 
possible magnetic gauge group is $\frac{n-1}{2}$.
The corresponding brane configuration can be 
obtained similarly and 
the extra $(i-2)$ NS-branes, $(i-2)$ sets of D6-branes and $(i-2)$
sets of D4-branes  
are present at the left hand side of the $NS5_L$-brane
and the extra $(n-i-1)$ NS-branes, $(n-i-1)$ sets of D6-branes and $(n-i-1)$
sets of D4-branes  are present at the right hand side of the $NS5_R'$-brane
of Figure 8.
The magnetic superpotential is
$
W_{dual} = \left( M_{i} q_i q_i + g_{i} X_{i+1} q_i + \Phi_{i+1}
  g_{i} g_{i} 
\right)
+ m_i M_{i}$.

$\bullet$ Case 4

When the Seiberg dual is taken for the middle gauge group factor
by
assuming that $\Lambda_i >> \Lambda_j$ where $j=1,2, \cdots, i-1, i+1,
\cdots, n$,  
one follows the procedure given in the subsection 3.4.
The gauge group is given by
\bea
\cdots \times Sp(N_{c,i-1}) \times 
SO(2\widetilde{N}_{c,i} \equiv 2(N_{f,i}+N_{c,i+1}+N_{c,i-1}-N_{c,i}+2)) 
\times Sp(N_{c,i+1}) \times \cdots
\nonu
\eea
where $ i=2, 4, \cdots, (n-1)$ implying that the number of 
possible magnetic gauge group is $\frac{n-1}{2}$.
The corresponding brane configuration can be 
obtained similarly and 
the extra $(i-2)$ NS-branes, $(i-2)$ sets of D6-branes and $(i-2)$
sets of D4-branes  
are present at the left hand side of the $NS5_L$-brane
and the extra $(n-i-2)$ NS-branes, $(n-i-2)$ sets of D6-branes and $(n-i-2)$
sets of D4-branes  are present at the right hand side of the $NS5_R'$-brane
of Figure 9.
The magnetic superpotential is
$
W_{dual} = \left( M_i q_i q_i + f_{i-1} X_{i-1} q_i + \Phi_{i-1}
  f_{i-1} f_{i-1} 
\right)
+ m_i M_i$.

$\bullet$ Case 5

When the Seiberg dual is taken for the symplectic gauge group factor
by
assuming that $\Lambda_i >> \Lambda_j$ where $j=1,2, \cdots, i-1, i+1,
\cdots, n$,   
one follows the procedure given in the subsection 3.6.
The gauge group is given by
\bea
\cdots \times SO(2N_{c,i-1}) \times 
Sp(\widetilde{N}_{c,i} \equiv N_{f,i}+N_{c,i+1}+N_{c,i-1}-N_{c,i} -2) 
\times SO(2N_{c,i+1}) \times \cdots
\nonu
\eea
where $ i=3, 5, \cdots, (n-2)$ implying that the number of 
possible magnetic gauge group is $\frac{n-3}{2}$.
The corresponding brane configuration can be 
obtained similarly and 
the extra $(i-2)$ NS-branes, $(i-2)$ sets of D6-branes and $(i-2)$
sets of D4-branes  
are present at the left hand side of the $NS5_L$-brane
and the extra $(n-i-1)$ NS-branes, $(n-i-1)$ sets of D6-branes and $(n-i-1)$
sets of D4-branes  are present at the right hand side of the $NS5_R'$-brane
of Figure 8.
The magnetic superpotential is
$
W_{dual} = \left( M_{i} q_i q_i + g_{i} X_{i+1} q_i + \Phi_{i+1}
  g_{i} g_{i} 
\right)
+ m_i M_{i}$.

$\bullet$ Case 6

When the Seiberg dual is taken for the  symplectic gauge group factor
by
assuming that $\Lambda_i >> \Lambda_j$ where $j=1,2, \cdots, i-1, i+1,
\cdots, n$,  
one follows the procedure given in the subsection 3.6.
The gauge group is given by
\bea
\cdots \times SO(2N_{c,i-1}) \times 
Sp(\widetilde{N}_{c,i} \equiv N_{f,i}+N_{c,i+1}+N_{c,i-1}-N_{c,i}-2) 
\times SO(2N_{c,i+1}) \times \cdots
\nonu
\eea
where $ i=3, 5, \cdots, (n-2)$ implying that the number of 
possible magnetic gauge group is $\frac{n-3}{2}$.
The corresponding brane configuration can be 
obtained similarly and 
the extra $(i-2)$ NS-branes, $(i-2)$ sets of D6-branes and $(i-2)$
sets of D4-branes  
are present at the left hand side of the $NS5_L$-brane
and the extra $(n-i-1)$ NS-branes, $(n-i-1)$ sets of D6-branes 
and $(n-i-1)$
sets of D4-branes  are present at the right hand side of the $NS5_R'$-brane
of Figure 9.
The magnetic superpotential is
$
W_{dual} = \left( M_i q_i q_i + f_{i-1} X_{i-1} q_i + \Phi_{i-1}
  f_{i-1} f_{i-1} 
\right)
+ m_i M_i$.

\subsubsection{ The orthogonal gauge groups both  at the start and
end of the chain }

When the electric theory is described by
\bea
SO(2N_{c,1}) \times 
Sp(N_{c,2}) \times \cdots \times Sp(N_{c,n-1}) \times SO(2N_{c,n})
\nonu
\eea
with 
the $(n-1)$ bifundametals $({\bf \Box_1, \Box_2, 1, \cdots,  1})$,
$\cdots$, and $({\bf 1, \cdots, 1, \Box_{n-1}, \Box_n})$, and 
$n$-quarks in the representation 
$({\bf \Box_1, 1, \cdots, 1})$, $\cdots$, and 
$({\bf 1, \cdots,  1, \Box_n})$,
there exist $(2n-2)$ magnetic theories and they can be classified as
six cases as follows.
Note that there are the orthogonal gauge groups both at the start and
end of the chain.

$\bullet$ Case $1'$ 

When the Seiberg dual is taken for the first gauge group factor
 by
assuming that $\Lambda_1 >> \Lambda_i$ where $i=2, \cdots, n$, 
one follows the procedure given in the subsection 3.5.
The gauge group is 
\bea
SO(2\widetilde{N}_{c,1} \equiv 2(N_{f,1} +N_{c,2}-N_{c,1}+2)) \times 
Sp(N_{c,2}) \times \cdots \times SO(2N_{c,n})
\nonu
\eea
and the matter contents are given by 
the dual quark $q_1$ in the representation 
$({\bf \Box_1, 1, \cdots, 1})$ 
as well as $(n-1)$ quarks $Q_i$ where $i=2, \cdots, n$, 
the bifundamentals $f_1$ in the representation
$({\bf \Box_1, \Box_2, 1, \cdots,  1})$,
$\cdots$ in
addition to $(n-2)$ bifundamentals $G_i$, and
various gauge singlets
$X_2, M_1$ and $\Phi_2$.
The corresponding brane configuration can be 
obtained similarly and 
the extra $(n-3)$ NS-branes, $(n-3)$ sets of D6-branes and $(n-3)$
sets of D4-branes  
are present at the right hand side of the $NS5_R'$-brane
of Figure 10.
The magnetic superpotential is
$
W_{dual} = \left( M_1 q_1 q_1 + f_1 X_2 q_1 + \Phi_2 f_1 f_1 \right)
+ m_1 M_1$.

$\bullet$ Case $2'$

When the Seiberg dual is taken for the last gauge group factor
 by
assuming that $\Lambda_n >> \Lambda_i$ where $i=1,2, \cdots, (n-1)$, 
one follows the procedure given in the subsection 3.2.
The gauge group is given by
\bea
SO(2N_{c,1}) \times \cdots \times 
Sp(N_{c,n-1}) \times SO(2\widetilde{N}_{c,n} 
\equiv 2(N_{f,n} +N_{c,n-1}-N_{c,n}+2)).
\nonu
\eea
The corresponding brane configuration can be 
obtained similarly and 
the extra $(n-3)$ NS-branes, $(n-3)$ sets of D6-branes and $(n-3)$
sets of D4-branes  
are present at the left hand side of the $NS5_L$-brane
of Figure 7.
The magnetic superpotential is
$
W_{dual} = \left( M_n q_n q_n + g_{n-1} X_{n-1} q_n + \Phi_{n-1}
  g_{n-1} g_{n-1} 
\right)
+ m_n M_n$.

$\bullet$ Case $3'$

When the Seiberg dual is taken for the middle gauge group factor
by
assuming that $\Lambda_i >> \Lambda_j$ where $j=1,2, \cdots, i-1, i+1,
\cdots, n$,  
one follows the procedure given in the subsection 3.3.
The gauge group is given by
\bea
\cdots \times Sp(N_{c,i-1}) \times 
SO(2\widetilde{N}_{c,i} \equiv 2(N_{f,i}+N_{c,i+1}+N_{c,i-1}-N_{c,i} +2)) 
\times Sp(N_{c,i+1}) \times \cdots
\nonu
\eea
where $ i=2, 4, \cdots, (n-1)$ implying that the number of 
possible magnetic gauge group is $\frac{n-1}{2}$.
The corresponding brane configuration can be 
obtained similarly and 
the extra $(i-2)$ NS-branes, $(i-2)$ sets of D6-branes and $(i-2)$
sets of D4-branes  
are present at the left hand side of the $NS5_L$-brane
and the extra $(n-i-1)$ NS-branes, $(n-i-1)$ sets of D6-branes and $(n-i-1)$
sets of D4-branes  are present at the right hand side of the $NS5_R'$-brane
of Figure 8.
The magnetic superpotential is
$
W_{dual} = \left( M_{i} q_i q_i + g_{i} X_{i+1} q_i + \Phi_{i+1}
  g_{i} g_{i} 
\right)
+ m_i M_{i}$.

$\bullet$ Case $4'$

When the Seiberg dual is taken for the middle gauge group factor
by
assuming that $\Lambda_i >> \Lambda_j$ where $j=1,2, \cdots, i-1, i+1,
\cdots, n$,  
one follows the procedure given in the subsection 3.4.
The gauge group is given by
\bea
\cdots \times Sp(N_{c,i-1}) \times 
SO(2\widetilde{N}_{c,i} \equiv 2(N_{f,i}+N_{c,i+1}+N_{c,i-1}-N_{c,i}+2)) 
\times Sp(N_{c,i+1}) \times \cdots
\nonu
\eea
where $ i=2, 4, \cdots, (n-1)$ implying that the number of 
possible magnetic gauge group is $\frac{n-1}{2}$.
The corresponding brane configuration can be 
obtained similarly and 
the extra $(i-2)$ NS-branes, $(i-2)$ sets of D6-branes and $(i-2)$
sets of D4-branes  
are present at the left hand side of the $NS5_L$-brane
and the extra $(n-i-1)$ NS-branes, $(n-i-1)$ 
sets of D6-branes and $(n-i-1)$
sets of D4-branes  are present at the right hand side of the $NS5_R'$-brane
of Figure 9.
The magnetic superpotential is
$
W_{dual} = \left( M_i q_i q_i + f_{i-1} X_{i-1} q_i + \Phi_{i-1}
  f_{i-1} f_{i-1} 
\right)
+ m_i M_i$.

$\bullet$ Case $5'$

When the Seiberg dual is taken for the symplectic gauge group factor
by
assuming that $\Lambda_i >> \Lambda_j$ where $j=1,2, \cdots, i-1, i+1,
\cdots, n$,   
one follows the procedure given in the subsection 3.6.
The gauge group is given by
\bea
\cdots \times SO(2N_{c,i-1}) \times 
Sp(\widetilde{N}_{c,i} \equiv N_{f,i}+N_{c,i+1}+N_{c,i-1}-N_{c,i} -2) 
\times SO(2N_{c,i+1}) \times \cdots
\nonu
\eea
where $ i=3, 5, \cdots, (n-2)$ implying that the number of 
possible magnetic gauge group is $\frac{n-3}{2}$.
The corresponding brane configuration can be 
obtained similarly and 
the extra $(i-2)$ NS-branes, $(i-2)$ sets of D6-branes and $(i-2)$
sets of D4-branes  
are present at the left hand side of the $NS5_L$-brane
and the extra $(n-i-1)$ NS-branes, $(n-i-1)$ sets of D6-branes and $(n-i-1)$
sets of D4-branes  are present at the right hand side of the $NS5_R'$-brane
of Figure 8.
The magnetic superpotential is
$
W_{dual} = \left( M_{i} q_i q_i + g_{i} X_{i+1} q_i + \Phi_{i+1}
  g_{i} g_{i} 
\right)
+ m_i M_{i}$.

$\bullet$ Case $6'$

When the Seiberg dual is taken for the  symplectic gauge group factor
by
assuming that $\Lambda_i >> \Lambda_j$ where $j=1,2, \cdots, i-1, i+1,
\cdots, n$,  
one follows the procedure given in the subsection 3.6.
The gauge group is given by
\bea
\cdots \times SO(2N_{c,i-1}) \times 
Sp(\widetilde{N}_{c,i} \equiv N_{f,i}+N_{c,i+1}+N_{c,i-1}-N_{c,i}-2) 
\times SO(2N_{c,i+1}) \times \cdots
\nonu
\eea
where $ i=3, 5, \cdots, (n-2)$ implying that the number of 
possible magnetic gauge group is $\frac{n-3}{2}$.
The corresponding brane configuration can be 
obtained similarly and 
the extra $(i-2)$ NS-branes, $(i-2)$ sets of D6-branes and $(i-2)$
sets of D4-branes  
are present at the left hand side of the $NS5_L$-brane
and the extra $(n-i-1)$ NS-branes, $(n-i-1)$ sets of D6-branes and $(n-i-1)$
sets of D4-branes  are present at the right hand side of the $NS5_R'$-brane
of Figure 9.
The magnetic superpotential is
$
W_{dual} = \left( M_i q_i q_i + f_{i-1} X_{i-1} q_i + \Phi_{i-1}
  f_{i-1} f_{i-1} 
\right)
+ m_i M_i$.

\subsubsection{ The symplectic gauge group at the start and
the orthogonal gauge group at end of the chain }

When the electric theory is described by
\bea
Sp(N_{c,1}) \times 
SO(2N_{c,2}) \times \cdots \times Sp(N_{c,n-1}) \times SO(2N_{c,n})
\nonu
\eea
with 
the $(n-1)$ bifundametals $({\bf \Box_1, \Box_2, 1, \cdots,  1})$,
$\cdots$, and $({\bf 1, \cdots, 1, \Box_{n-1}, \Box_n})$, and 
$n$-quarks in the representation 
$({\bf \Box_1, 1, \cdots, 1})$, $\cdots$, and 
$({\bf 1, \cdots,  1, \Box_n})$,
there exist $(2n-2)$ magnetic theories and they can be classified as
six cases as follows.
Note that there are the symplectic gauge group at the start and
the orthogonal gauge group at the
end of the chain.

$\bullet$ Case $1''$ 

When the Seiberg dual is taken for the first gauge group factor
 by
assuming that $\Lambda_1 >> \Lambda_i$ where $i=2, \cdots, n$, 
one follows the procedure given in the subsection 3.5.
The gauge group is 
\bea
Sp(\widetilde{N}_{c,1} \equiv N_{f,1} +N_{c,2}-N_{c,1}-2) \times 
SO(2N_{c,2}) \times \cdots \times SO(2N_{c,n})
\nonu
\eea
and the matter contents are given by 
the dual quark $q_1$ in the representation
$({\bf \Box_1, 1, \cdots, 1})$ 
as well as $(n-1)$ quarks $Q_i$ where $i=2, \cdots, n$, 
the bifundamentals $f_1$ in the representation
 $({\bf \Box_1, \Box_2, 1, \cdots,  1})$,
$\cdots$ in
addition to $(n-2)$ bifundamentals $G_i$, and
various gauge singlets
$X_2, M_1$ and $\Phi_2$.
The corresponding brane configuration can be 
obtained similarly and 
the extra $(n-3)$ NS-branes, $(n-3)$ sets of D6-branes and $(n-3)$
sets of D4-branes  
are present at the right hand side of the $NS5_R'$-brane
of Figure 10.
The magnetic superpotential is
$
W_{dual} = \left( M_1 q_1 q_1 + f_1 X_2 q_1 + \Phi_2 f_1 f_1 \right)
+ m_1 M_1$.

$\bullet$ Case $2''$

When the Seiberg dual is taken for the last gauge group factor
 by
assuming that $\Lambda_n >> \Lambda_i$ where $i=1,2, \cdots, (n-1)$, 
one follows the procedure given in the subsection 3.2.
The gauge group is given by
\bea
Sp(N_{c,1}) \times \cdots \times 
Sp(N_{c,n-1}) \times SO(2\widetilde{N}_{c,n} 
\equiv 2(N_{f,n} +N_{c,n-1}-N_{c,n}+2)).
\nonu
\eea
The corresponding brane configuration can be 
obtained similarly and 
the extra $(n-3)$ NS-branes, $(n-3)$ sets of D6-branes and $(n-3)$
sets of D4-branes  
are present at the left hand side of the $NS5_L$-brane
of Figure 7.
The magnetic superpotential is
$
W_{dual} = \left( M_n q_n q_n + g_{n-1} X_{n-1} q_n + \Phi_{n-1}
  g_{n-1} g_{n-1} 
\right)
+ m_n M_n$.

$\bullet$ Case $3''$

When the Seiberg dual is taken for the middle gauge group factor
by
assuming that $\Lambda_i >> \Lambda_j$ where $j=1,2, \cdots, i-1, i+1,
\cdots, n$, 
one follows the procedure given in the subsection 3.3.
The gauge group is given by
\bea
\cdots \times Sp(N_{c,i-1}) \times 
SO(2\widetilde{N}_{c,i} \equiv 2(N_{f,i}+N_{c,i+1}+N_{c,i-1}-N_{c,i} +2)) 
\times Sp(N_{c,i+1}) \times \cdots
\nonu
\eea
where $ i=2, 4, \cdots, (n-1)$ implying that the number of 
possible magnetic gauge group is $\frac{n-1}{2}$.
The corresponding brane configuration can be 
obtained similarly and 
the extra $(i-2)$ NS-branes, $(i-2)$ sets of D6-branes and $(i-2)$
sets of D4-branes  
are present at the left hand side of the $NS5_L$-brane
and the extra $(n-i-1)$ NS-branes, $(n-i-1)$ sets of D6-branes and $(n-i-1)$
sets of D4-branes  are present at the right hand side of the $NS5_R'$-brane
of Figure 8.
The magnetic superpotential is
$
W_{dual} = \left( M_{i} q_i q_i + g_{i} X_{i+1} q_i + \Phi_{i+1}
  g_{i} g_{i} 
\right)
+ m_i M_{i}$.

$\bullet$ Case $4''$

When the Seiberg dual is taken for the middle gauge group factor
by
assuming that $\Lambda_i >> \Lambda_j$ where $j=1,2, \cdots, i-1, i+1,
\cdots, n$,  
one follows the procedure given in the subsection 3.4.
The gauge group is given by
\bea
\cdots \times Sp(N_{c,i-1}) \times 
SO(2\widetilde{N}_{c,i} \equiv 2(N_{f,i}+N_{c,i+1}+N_{c,i-1}-N_{c,i}+2)) 
\times Sp(N_{c,i+1}) \times \cdots
\nonu
\eea
where $ i=2, 4, \cdots, (n-1)$ implying that the number of 
possible magnetic gauge group is $\frac{n-1}{2}$.
The corresponding brane configuration can be 
obtained similarly and 
the extra $(i-2)$ NS-branes, $(i-2)$ sets of D6-branes and $(i-2)$
sets of D4-branes  
are present at the left hand side of the $NS5_L$-brane
and the extra $(n-i-1)$ NS-branes, $(n-i-1)$ sets of D6-branes and $(n-i-1)$
sets of D4-branes  are present at the right hand side of the $NS5_R'$-brane
of Figure 9.
The magnetic superpotential is
$
W_{dual} = \left( M_i q_i q_i + f_{i-1} X_{i-1} q_i + \Phi_{i-1}
  f_{i-1} f_{i-1} 
\right)
+ m_i M_i$.

$\bullet$ Case $5''$

When the Seiberg dual is taken for the symplectic gauge group factor
by
assuming that $\Lambda_i >> \Lambda_j$ where $j=1,2, \cdots, i-1, i+1,
\cdots, n$,   
one follows the procedure given in the subsection 3.6.
The gauge group is given by
\bea
\cdots \times SO(2N_{c,i-1}) \times 
Sp(\widetilde{N}_{c,i} \equiv N_{f,i}+N_{c,i+1}+N_{c,i-1}-N_{c,i} -2) 
\times SO(2N_{c,i+1}) \times \cdots
\nonu
\eea
where $ i=3, 5, \cdots, (n-2)$ implying that the number of 
possible magnetic gauge group is $\frac{n-3}{2}$.
The corresponding brane configuration can be 
obtained similarly and 
the extra $(i-2)$ NS-branes, $(i-2)$ sets of D6-branes and $(i-2)$
sets of D4-branes  
are present at the left hand side of the $NS5_L$-brane
and the extra $(n-i-1)$ NS-branes, $(n-i-1)$ sets of D6-branes and $(n-i-1)$
sets of D4-branes  are present at the right hand side of the $NS5_R'$-brane
of Figure 8.
The magnetic superpotential is
$
W_{dual} = \left( M_{i} q_i q_i + g_{i} X_{i+1} q_i + \Phi_{i+1}
  g_{i} g_{i} 
\right)
+ m_i M_{i}$.

$\bullet$ Case $6''$

When the Seiberg dual is taken for the  symplectic gauge group factor
by
assuming that $\Lambda_i >> \Lambda_j$ where $j=1,2, \cdots, i-1, i+1,
\cdots, n$,  
one follows the procedure given in the subsection 3.6.
The gauge group is given by
\bea
\cdots \times SO(2N_{c,i-1}) \times 
Sp(\widetilde{N}_{c,i} \equiv N_{f,i}+N_{c,i+1}+N_{c,i-1}-N_{c,i}-2) 
\times SO(2N_{c,i+1}) \times \cdots
\nonu
\eea
where $ i=3, 5, \cdots, (n-2)$ implying that the number of 
possible magnetic gauge group is $\frac{n-3}{2}$.
The corresponding brane configuration can be 
obtained similarly and 
the extra $(i-2)$ NS-branes, $(i-2)$ sets of D6-branes and $(i-2)$
sets of D4-branes  
are present at the left hand side of the $NS5_L$-brane
and the extra $(n-i-1)$ NS-branes, $(n-i-1)$ sets of D6-branes and $(n-i-1)$
sets of D4-branes  are present at the right hand side of the $NS5_R'$-brane
of Figure 9.
The magnetic superpotential is
$
W_{dual} = \left( M_i q_i q_i + f_{i-1} X_{i-1} q_i + \Phi_{i-1}
  f_{i-1} f_{i-1} 
\right)
+ m_i M_i$.

\subsubsection{ The orthogonal gauge group at the start and
the symplectic gauge group at end of the chain }

When the electric theory is described by
\bea
SO(2N_{c,1}) \times 
Sp(N_{c,2}) \times \cdots \times SO(2N_{c,n-1}) \times Sp(N_{c,n})
\nonu
\eea
with 
the $(n-1)$ bifundametals 
$({\bf \Box_1, \Box_2, 1, \cdots,  1})$,
$\cdots$, and $({\bf 1, \cdots, 1, \Box_{n-1}, \Box_n})$, and 
$n$-quarks in the representation 
$({\bf \Box_1, 1, \cdots, 1})$, $\cdots$, and 
$({\bf 1, \cdots,  1, \Box_n})$,
there exist $(2n-2)$ magnetic theories and they can be classified as
six cases as follows.
Note that there are the orthogonal gauge group at the start and
the symplectic gauge group at the end of the chain.

$\bullet$ Case $1'''$ 

When the Seiberg dual is taken for the first gauge group factor
 by
assuming that $\Lambda_1 >> \Lambda_i$ where $i=2, \cdots, n$, 
one follows the procedure given in the subsection 3.5.
The gauge group is 
\bea
SO(2\widetilde{N}_{c,1} \equiv 2(N_{f,1} +N_{c,2}-N_{c,1}+2)) \times 
Sp(N_{c,2}) \times \cdots \times Sp(N_{c,n})
\nonu
\eea
and the matter contents are given by 
the dual quark $q_1$ 
in the representation 
$({\bf \Box_1, 1, \cdots, 1})$
as well as $(n-1)$ quarks $Q_i$ where $i=2, \cdots, n$, 
the bifundamentals $f_1$ in the representation
 $({\bf \Box_1, \Box_2, 1, \cdots,  1})$,
$\cdots$ in
addition to $(n-2)$ bifundamentals $G_i$, and
various gauge singlets
$X_2, M_1$ and $\Phi_2$.
The magnetic superpotential is
$
W_{dual} = \left( M_1 q_1 q_1 + f_1 X_2 q_1 + \Phi_2 f_1 f_1 \right)
+ m_1 M_1$.
The corresponding brane configuration can be 
obtained similarly and 
the extra $(n-3)$ NS-branes, $(n-3)$ sets of D6-branes and $(n-3)$
sets of D4-branes  
are present at the right hand side of the $NS5_R'$-brane
of Figure 10.

$\bullet$ Case $2'''$

When the Seiberg dual is taken for the last gauge group factor
 by
assuming that $\Lambda_n >> \Lambda_i$ where $i=1,2, \cdots, (n-1)$, 
one follows the procedure given in the subsection 3.2.
The gauge group is given by
\bea
SO(2N_{c,1}) \times \cdots \times 
SO(2N_{c,n-1}) \times Sp(\widetilde{N}_{c,n} \equiv 
N_{f,n} +N_{c,n-1}-N_{c,n}-2).
\nonu
\eea
The magnetic superpotential is
$
W_{dual} = \left( M_n q_n q_n + g_{n-1} X_{n-1} q_n + \Phi_{n-1}
  g_{n-1} g_{n-1} 
\right)
+ m_n M_n$.
The corresponding brane configuration can be 
obtained similarly and 
the extra $(n-3)$ NS-branes, $(n-3)$ sets of D6-branes and $(n-3)$
sets of D4-branes  
are present at the left hand side of the $NS5_L$-brane
of Figure 7.

$\bullet$ Case $3'''$

When the Seiberg dual is taken for the middle gauge group factor
by
assuming that $\Lambda_i >> \Lambda_j$ where $j=1,2, \cdots, i-1, i+1,
\cdots, n$,  
one follows the procedure given in the subsection 3.3.
The gauge group is given by
\bea
\cdots \times Sp(N_{c,i-1}) \times 
SO(2\widetilde{N}_{c,i} \equiv 2(N_{f,i}+N_{c,i+1}+N_{c,i-1}-N_{c,i} +2)) 
\times Sp(N_{c,i+1}) \times \cdots
\nonu
\eea
where $ i=2, 4, \cdots, (n-1)$ implying that the number of 
possible magnetic gauge group is $\frac{n-1}{2}$.
The magnetic superpotential is
$
W_{dual} = \left( M_{i} q_i q_i + g_{i} X_{i+1} q_i + \Phi_{i+1}
  g_{i} g_{i} 
\right)
+ m_i M_{i}$.
The corresponding brane configuration can be 
obtained similarly and 
the extra $(i-2)$ NS-branes, $(i-2)$ sets of D6-branes and $(i-2)$
sets of D4-branes  
are present at the left hand side of the $NS5_L$-brane
and the extra $(n-i-1)$ NS-branes, $(n-i-1)$ sets of D6-branes and $(n-i-1)$
sets of D4-branes  are present at the right hand side of the $NS5_R'$-brane
of Figure 8.

$\bullet$ Case $4'''$

When the Seiberg dual is taken for the middle gauge group factor
by
assuming that $\Lambda_i >> \Lambda_j$ where $j=1,2, \cdots, i-1, i+1,
\cdots, n$,  
one follows the procedure given in the subsection 3.4.
The gauge group is given by
\bea
\cdots \times Sp(N_{c,i-1}) \times 
SO(2\widetilde{N}_{c,i} \equiv 2(N_{f,i}+N_{c,i+1}+N_{c,i-1}-N_{c,i}+2)) 
\times Sp(N_{c,i+1}) \times \cdots
\nonu
\eea
where $ i=2, 4, \cdots, (n-1)$ implying that the number of 
possible magnetic gauge group is $\frac{n-1}{2}$.
The magnetic superpotential is
$
W_{dual} = \left( M_i q_i q_i + f_{i-1} X_{i-1} q_i + \Phi_{i-1}
  f_{i-1} f_{i-1} 
\right)
+ m_i M_i$.
The corresponding brane configuration can be 
obtained similarly and 
the extra $(i-2)$ NS-branes, $(i-2)$ sets of D6-branes and $(i-2)$
sets of D4-branes  
are present at the left hand side of the $NS5_L$-brane
and the extra $(n-i-1)$ NS-branes, $(n-i-1)$ sets of D6-branes and $(n-i-1)$
sets of D4-branes  are present at the right hand side of the $NS5_R'$-brane
of Figure 9.

$\bullet$ Case $5'''$

When the Seiberg dual is taken for the symplectic gauge group factor
by
assuming that $\Lambda_i >> \Lambda_j$ where $j=1,2, \cdots, i-1, i+1,
\cdots, n$,   
one follows the procedure given in the subsection 3.6.
The gauge group is given by
\bea
\cdots \times SO(2N_{c,i-1}) \times 
Sp(\widetilde{N}_{c,i} \equiv N_{f,i}+N_{c,i+1}+N_{c,i-1}-N_{c,i} -2) 
\times SO(2N_{c,i+1}) \times \cdots
\nonu
\eea
where $ i=3, 5, \cdots, (n-2)$ implying that the number of 
possible magnetic gauge group is $\frac{n-3}{2}$.
The magnetic superpotential is
$
W_{dual} = \left( M_{i} q_i q_i + g_{i} X_{i+1} q_i + \Phi_{i+1}
  g_{i} g_{i} 
\right)
+ m_i M_{i}$.
The corresponding brane configuration can be 
obtained similarly and 
the extra $(i-2)$ NS-branes, $(i-2)$ sets of D6-branes and $(i-2)$
sets of D4-branes  
are present at the left hand side of the $NS5_L$-brane
and the extra $(n-i-1)$ NS-branes, $(n-i-1)$ sets of D6-branes and $(n-i-1)$
sets of D4-branes  are present at the right hand side of the $NS5_R'$-brane
of Figure 8.

$\bullet$ Case $6'''$

When the Seiberg dual is taken for the  symplectic gauge group factor
by
assuming that $\Lambda_i >> \Lambda_j$ where $j=1,2, \cdots, i-1, i+1,
\cdots, n$,  
one follows the procedure given in the subsection 3.6.
The gauge group is given by
\bea
\cdots \times SO(2N_{c,i-1}) \times 
Sp(\widetilde{N}_{c,i} \equiv N_{f,i}+N_{c,i+1}+N_{c,i-1}-N_{c,i}-2) 
\times SO(2N_{c,i+1}) \times \cdots
\nonu
\eea
where $ i=3, 5, \cdots, (n-2)$ implying that the number of 
possible magnetic gauge group is $\frac{n-3}{2}$.
The magnetic superpotential is
$
W_{dual} = \left( M_i q_i q_i + f_{i-1} X_{i-1} q_i + \Phi_{i-1}
  f_{i-1} f_{i-1} 
\right)
+ m_i M_i$.
The corresponding brane configuration can be 
obtained similarly and 
the extra $(i-2)$ NS-branes, $(i-2)$ sets of D6-branes and $(i-2)$
sets of D4-branes  
are present at the left hand side of the $NS5_L$-brane
and the extra $(n-i-1)$ NS-branes, $(n-i-1)$ sets of D6-branes and $(n-i-1)$
sets of D4-branes  are present at the right hand side of the $NS5_R'$-brane
of Figure 9.

\section{Conclusions and outlook }

The meta-stable brane configurations we have found are summarized by
Figures 2B, 3B, 4B, 5B(and 7B, 8B, 9B, and 10B).
The nonsupersymmetric minimal energy brane configurations in Figure 2B
and 3B
lead to the Figure 3 of \cite{Ahn07-3} if we ignore the
$NS5_L$-brane, $N_f$ D6-branes and $N_c$ D4-branes.
Similarly the nonsupersymmetric minimal energy brane configurations in Figure 7B
and 8B
lead to the Figure 3 of \cite{Ahn07-2} if we ignore the
$NS5_L$-brane, $2N_f$ D6-branes and $2N_c$ D4-branes.

The Figures 2B and 7B with a replacement $N_f''$ D6-branes by 
the NS5'-brane by neglecting  the
$NS5_L$-brane, $N_f$ D6-branes and $N_c$ D4-branes
become the Figures 2B and 7B of \cite{Ahn07-6}
together with a reflection with respect to the $NS5_L$-brane and a
rotation of $NS5_R$-brane by $\frac{\pi}{2}$ angle respectively.
The Figures 3B and 8B with a replacement $N_f'$ D6-branes by 
the NS5'-brane by neglecting  the
$NS5_L$-brane, $N_f$ D6-branes and $N_c$ D4-branes
become the Figures 2B and 7B of \cite{Ahn07-6}
with a
rotation of $NS5_R$-brane by $\frac{\pi}{2}$ angle respectively or 
the Figures 4B and 9B of \cite{Ahn07-6} with a reflection with
respect to the $NS5_R$-brane 
if we ignore 
the
$NS5_R'$-brane, $N_f''$ D6-branes and $N_c''$ D4-branes from the
Figure 3B(8B) respectively.
The Figures 4B and 9B
with a replacement $N_f'$ D6-branes by 
the NS5'-brane by neglecting  the
$NS5_R'$-brane, $N_f''$ D6-branes and $N_c''$ D4-branes 
with vanishing $N_f$ D6-branes
lead to 
the Figures 5B and 10B of \cite{Ahn07-6}
with a
rotation of $NS5_L'$-brane by $\frac{\pi}{2}$ angle respectively. 

Finally, the Figures 5B and 10B
with a replacement $N_f$ D6-branes by 
the NS5'-brane, when we neglect  the
$NS5_R'$-brane, $N_f''$ D6-branes and $N_c''$ D4-branes 
and $N_f'$ D6-branes,
lead to 
the Figures 4B and 10B of \cite{Ahn07-6} respectively.

For the same gauge groups in this paper, one can add an orientifold 6-planes 
together with three more NS-branes. Totally there exist 
seven NS-branes. The relevant previous works are given in 
\cite{LLL,LL,LLL1,BHKL,EGKT}. 
Depending on the matter contents, there are two possibilities.
On the other hand, for the different triple product gauge groups, 
one can add an
orientifold 6-planes together with two more NS-branes. 
Totally, there exist six NS-branes.
It would be interesting to study these meta-stable brane
configurations in type IIA string theory.

\vspace{.7cm}

\centerline{\bf Acknowledgments}

I would like to thank 
Harvard High Energy
Theory Group, where this work was undertaken, for
the hospitality and the financial support and D. Shih for discussions. 
This work was supported by grant No.
R01-2006-000-10965-0 from the Basic Research Program of the Korea
Science \& Engineering Foundation.

\end{document}